\documentclass[twocolumn,twocolappendix]{aastex7}

\usepackage{amsmath}

\usepackage{booktabs}   
\usepackage{multirow} 
\shorttitle{The Ultra-long GRB 250702B}
\shortauthors{O'Connor et al.}
\submitjournal{ApJL}

\usepackage{siunitx}
\usepackage{booktabs}
\usepackage{longtable}
\usepackage{caption}

\definecolor{blazeorange}{rgb}{1.0, 0.4, 0.0}
\definecolor{seagreen}{rgb}{0.18, 0.55, 0.34}
\definecolor{darkgreen}{rgb}{0.08, 0.45, 0.2}
\definecolor{rufous}{rgb}{0.66, 0.11, 0.03}
\definecolor{royalfuchsia}{rgb}{0.79, 0.17, 0.57}
\definecolor{scarlet}{rgb}{1.0, 0.13, 0.0}
\definecolor{royalpurple}{rgb}{0.47, 0.32, 0.66}

\begin{document}

\title{

Comprehensive X-ray Observations of the Exceptional Ultra-long X-ray and Gamma-ray Transient GRB 250702B with Swift, NuSTAR, and Chandra: Insights from the X-ray Afterglow Properties 


}

\correspondingauthor{Brendan O'Connor}
\author[0000-0002-9700-0036]{Brendan O'Connor}
    \altaffiliation{McWilliams Fellow}
    \affiliation{McWilliams Center for Cosmology and Astrophysics, Department of Physics, Carnegie Mellon University, Pittsburgh, PA 15213, USA}
    \email[show]{boconno2@andrew.cmu.edu}

\author[0000-0003-0516-2968]{Ramandeep Gill}
    \affiliation{Instituto de Radioastronom\'ia y Astrof\'isica, Universidad Nacional Aut\'onoma de M\'exico, Antigua Carretera a P\'atzcuaro $\#$ 8701,  Ex-Hda. San Jos\'e de la Huerta, Morelia, Michoac\'an, C.P. 58089, M\'exico }
    \affiliation{Astrophysics Research Center of the Open university (ARCO), The Open University of Israel, P.O Box 808, Ra’anana 4353701, Israel}
    \email{rsgill.rg@gmail.com}

\author[0000-0001-5229-1995]{James DeLaunay}
	\affiliation{Department of Astronomy and Astrophysics, The Pennsylvania State University, 525 Davey Lab, University Park, PA 16802, USA}
    \affiliation{Institute for Gravitation and the Cosmos, The Pennsylvania State University, University Park, PA 16802, USA}
	\email{jjd330@psu.edu}

\author[0000-0002-8548-482X]{Jeremy Hare}
    \affiliation{Astrophysics Science Division, NASA Goddard Space Flight Center, 8800 Greenbelt Rd, Greenbelt, MD 20771, USA}
    \affiliation{Center for Research and Exploration in Space Science and Technology, NASA/GSFC, Greenbelt, Maryland 20771, USA}
    \affiliation{The Catholic University of America, 620 Michigan Ave., N.E. Washington, DC 20064, USA}
    \email{harej10@gmail.com}

\author[0000-0003-1386-7861]{Dheeraj Pasham}
	\affiliation{Eureka Scientific, 2452 Delmer Street Suite 100, Oakland, CA 94602-3017, USA}
    \affiliation{Department of Physics, The George Washington University, Washington, DC 20052, USA}
    \email{p.dheerajreddy@gmail.com}
    
\author[0000-0003-3765-6401]{Eric R. Coughlin}
	\affiliation{Department of Physics, Syracuse University, Syracuse, NY 13210, USA}
	\email{ecoughli@syr.edu}

\author[0000-0002-5116-844X]{Ananya Bandopadhyay}
	\affiliation{Department of Physics, Syracuse University, Syracuse, NY 13210, USA}
	\email{abandopa@syr.edu}

\author[0000-0002-8935-9882]{Akash Anumarlapudi}
	\affiliation{University of North Carolina at Chapel Hill, 120 E. Cameron Ave., Chapel Hill, NC 27514, USA}
	\email{akasha@unc.edu}

\author[0000-0001-7833-1043]{Paz Beniamini}
    \affiliation{Department of Natural Sciences, The Open University of Israel, P.O Box 808, Ra'anana 4353701, Israel}
    \affiliation{Astrophysics Research Center of the Open university (ARCO), The Open University of Israel, P.O Box 808, Ra’anana 4353701, Israel}
    \affiliation{Department of Physics, The George Washington University, Washington, DC 20052, USA}
    \email{paz.beniamini@gmail.com}

\author[0000-0001-8530-8941]{Jonathan Granot}
    \affiliation{Department of Natural Sciences, The Open University of Israel, P.O Box 808, Ra'anana 4353701, Israel}
    \affiliation{Astrophysics Research Center of the Open university (ARCO), The Open University of Israel, P.O Box 808, Ra’anana 4353701, Israel}
    \affiliation{Department of Physics, The George Washington University, Washington, DC 20052, USA}
    \email{granot@openu.ac.il}

\author[0000-0002-8977-1498]{Igor Andreoni}
	\affiliation{University of North Carolina at Chapel Hill, 120 E. Cameron Ave., Chapel Hill, NC 27514, USA}
	\email{igor.andreoni@unc.edu}

\author[0000-0001-8544-584X]{Jonathan Carney}
	\affiliation{University of North Carolina at Chapel Hill, 120 E. Cameron Ave., Chapel Hill, NC 27514, USA}
	\email{jcarney@unc.edu}

\author[0000-0002-1103-7082]{Michael J. Moss}
    \affiliation{NASA Postdoctoral Program Fellow, NASA Goddard Space Flight Center, Greenbelt, MD 20771, USA}
    \email{mikejmoss3@gmail.com}

\author[0000-0002-5274-6790]{Ersin G\"{o}\u{g}\"{u}\c{s}}
    \affiliation{Sabanc\i~University, Faculty of Engineering and Natural Sciences, \.Istanbul 34956 T\"urkiye}
    \email{ersing@sabanciuniv.edu}

\author[0000-0002-6745-4790]{Jamie A. Kennea}
	\affiliation{Department of Astronomy and Astrophysics, The Pennsylvania State University, 525 Davey Lab, University Park, PA 16802, USA}
	\email{jak51@psu.edu} 


\author[0009-0001-0574-2332]{Malte Busmann}
    \affiliation{University Observatory, Faculty of Physics, Ludwig-Maximilians-Universität München, Scheinerstr. 1, 81679 Munich, Germany}
    \email{m.busmann@physik.lmu.de}

\author[0000-0001-6849-1270]{Simone Dichiara}
    \affiliation{Department of Astronomy and Astrophysics, The Pennsylvania State University, 525 Davey Lab, University Park, PA 16802, USA}
    \email{sbd5667@psu.edu}

\author[0009-0006-7990-0547]{James Freeburn}
	\affiliation{University of North Carolina at Chapel Hill, 120 E. Cameron Ave., Chapel Hill, NC 27514, USA}
	\email{jamesfreeburn54@gmail.com}

\author[0000-0003-3270-7644]{Daniel Gruen}
	\affiliation{University Observatory, Faculty of Physics, Ludwig-Maximilians-Universität München, Scheinerstr. 1, 81679 Munich, Germany}
	\affiliation{Excellence Cluster ORIGINS, Boltzmannstr. 2, 85748 Garching, Germany}
	\email{daniel.gruen@lmu.de}

\author[0000-0002-9364-5419]{Xander J. Hall}
	\affiliation{McWilliams Center for Cosmology and Astrophysics, Department of Physics, Carnegie Mellon University, Pittsburgh, PA 15213, USA}
	\email{xjh@andrew.cmu.edu}

\author[0000-0002-6011-0530]{Antonella Palmese}
	\affiliation{McWilliams Center for Cosmology and Astrophysics, Department of Physics, Carnegie Mellon University, Pittsburgh, PA 15213, USA}
	\email{palmese@cmu.edu}

\author[0000-0002-4299-2517]{Tyler Parsotan}
	\affiliation{Astrophysics Science Division, NASA Goddard Space Flight Center, 8800 Greenbelt Rd, Greenbelt, MD 20771, USA}
	\email{tyler.parsotan@nasa.gov}

\author[0000-0003-0020-687X]{Samuele Ronchini}
	\affiliation{Department of Astronomy and Astrophysics, The Pennsylvania State University, 525 Davey Lab, University Park, PA 16802, USA}
    \affiliation{Gran Sasso Science Institute (GSSI), 67100, L'Aquila, Italy}
	\email{sjs8171@psu.edu}

\author[0000-0002-2810-8764]{Aaron Tohuvavohu}
	\affiliation{Cahill Center for Astronomy and Astrophysics, California Institute of Technology, Pasadena, CA 91125, USA}
	\email{aaron.tohu@gmail.com}

\author[0000-0002-0025-3601]{Maia A. Williams}
	\affiliation{Department of Physics and Astronomy, Northwestern University, Evanston, IL 60208, USA}
        \affiliation{Center for Interdisciplinary Exploration and Research in Astronomy (CIERA), Northwestern University, 1800 Sherman Avenue, Evanston, IL 60201, USA}
	\email{maiawilliams2030@u.northwestern.edu}

\begin{abstract}
GRB 250702B is an exceptional transient that produced multiple episodes of luminous gamma-ray radiation lasting for $>$\,$25$ ks, placing it among the class of ultra-long gamma-ray bursts (GRBs). However, unlike any known GRB, the \textit{Einstein Probe} detected soft X-ray emission up to 24 hours before the gamma-ray triggers. We present comprehensive X-ray observations of the transient's ``afterglow'' obtained with the \textit{Neil Gehrels Swift Observatory}, the \textit{Nuclear Spectroscopic Telescope Array}, and the \textit{Chandra X-ray Observatory} between $0.5$ to $65$ days (observer frame) after the initial high-energy trigger. The X-ray emission decays steeply as $\sim$\,$t^{-1.9}$, and shows short timescale X-ray variability ($\Delta T/T$\,$<$\,$0.03$) in both \textit{Swift} and \textit{NuSTAR}, consistent with flares superposed on an external shock continuum. Serendipitous  detections by the \textit{Swift} Burst Alert Telescope (BAT) out to $\sim$0.3 days and continued \textit{NuSTAR} variability to $\sim$2 days imply sustained central engine activity; including the early \textit{Einstein Probe} X-ray detections, the required engine duration is $\gtrsim$\,$3$ days. Afterglow modeling favors the combination of forward and reverse shock emission in a wind-like ($k$\,$\approx$\,$2$) environment. These properties, especially the long-lived engine and early soft X-ray emission, are difficult to reconcile with a collapsar origin, and GRB 250702B does not fit neatly with canonical ultra-long GRBs or relativistic tidal disruption events (TDEs). A ``hybrid'' scenario in which a star is disrupted by a stellar-mass black hole (a micro-TDE) provides a plausible explanation, although a relativistic TDE from an intermediate-mass black hole remains viable. 


 
\end{abstract}

\keywords{\uat{X-ray astronomy}{1810} --- \uat{X-ray transient sources}{1852} ---- \uat{Gamma-ray transient sources}{1853} --- \uat{Gamma-ray bursts}{629} --- \uat{Relativistic jets }{1390} --- \uat{Black holes} {162} --- \uat{High energy astrophysics}{739} --- \uat{Time domain astronomy}{2109}}


\section{Introduction} 
\label{sec:intro}


Gamma-ray bursts (GRBs) are among the most energetic explosions in the Universe, releasing $10^{48-52}$ erg of energy (collimation corrected) in the X-ray and gamma-ray regime ($1$\,$-$\,$10,000$ keV). They are commonly classified as short or long GRBs based on their prompt emission duration \citep{Kouveliotou1993}, and hardness ratio \citep{Nakar2007,Bromberg2013}. The prompt duration is typically linked to the lifetime of the central engine after the jet breaks out from the surrounding stellar material \citep[e.g.,][]{Bromberg2012,Bromberg2013}. A subset of GRBs, referred to as ultra-long GRBs, display exceptionally extended prompt gamma-ray emission lasting thousands to tens of thousands of seconds \citep[e.g.,][]{2005AstL...31..291T,Gendre2013,Levan2014,Ioka2016}.

An alternative origin for some ultra-long events is that they arise from relativistic jetted tidal disruption events (TDEs) in which a star is shredded by a massive black hole, launching a jet along our line of sight \citep{Bloom2011,Levan2011,Cenko2012,Pasham2015,Andreoni2022,Pasham2023}. This was the case for GRB 110328A/Sw J1644+57 \citep{Burrows2011,Bloom2011}, which was the first event discovered in this class. 
Unlike classical GRBs, the recognized sample of relativistic TDEs lacks highly variable prompt gamma-ray emission, and these events were instead discovered via longer-duration ``image'' triggers \citep{Cummings2011J1644,Sakamoto2011J1644,Burrows2011} or searches using multi-day stacked datasets \citep{Cenko2012,Brown2015}.


Differentiating ultra-long GRBs from relativistic TDEs remains a major observational challenge. The clearest 
discriminant is the identification of either \textit{i}) a supernova, which solidifies a GRB progenitor, such as in GRB 111209A/ SN 2011kl \citep{Griener2015}, or \textit{ii}) a rapid X-ray shutoff, which is likely indicative of a transition in the accretion mode and is taken as robust evidence for a TDE origin \citep[e.g.,][]{Pasham2015,Eftekhari2024}. Another key distinction lies in the transient's location within its host galaxy. Long GRBs often occur in star-forming regions offset from their host galaxy's center, whereas TDEs are expected to originate at the precise nuclei of their host galaxies, coincident with supermassive black holes (SMBHs). In addition, relativistic TDEs typically exhibit prolonged (weeks to months) and luminous X-ray emission, exceeding that of standard or ultra-long GRBs \citep{Andreoni2022,Pasham2023}. This is usually considered the signpost of their discovery, and has been used as an argument for the identification of other relativistic TDE candidates \citep[e.g., EP240408a;][]{OConnor2025,Zhang2025}. 

On July 2, 2025, a series of multiple high-energy bursts were detected by numerous satellites, including the \textit{Fermi Gamma-ray Space Telescope} \citep{Meegan2009}, \textit{Konus-Wind} \citep{Aptekar1995}, the \textit{Space Variable Objects Monitor} (\textit{SVOM}; \citealt{SVOM}), the \textit{Neil Gehrels Swift Observatory} \citep{Gehrels2004}, the \textit{Monitor of All-sky X-ray Image}  \citep[\textit{MAXI};][]{MAXI}, and the \textit{Einstein Probe} \citep[EP;][]{Yuan2025}. These multiple gamma-ray triggers (GRB 250702B/D/E; \citealt{Neights2025}) were later noticed to originate from the same sky location \citep{Elizagcn,JimmyGCN,EPgcn,maxigcn,konusgcn,svomgcn}, and the event was subsequently designated GRB 250702BDE/ EP250702a. We adopt the name GRB 250702B for consistency with standard GRB nomenclature.

Due to its on-sky location's close proximity to the Galactic plane ($b$\,$\sim$\,$5$ deg), its nature was initially ambiguous, with possible interpretations ranging from a new X-ray binary in outburst or a peculiar extragalactic transient. While debate over its classification continues \citep{Levan2025,Oganesyan2025,Neights2025,Carney2025,Gompertz2025,Beniamini2025,Granot2025}, the most plausible scenarios are that it represents either an ultra-long GRB or a relativistic TDE. The duration of the source's flaring activity, lasting for at least $>$\,$25$ ks (observer frame) at gamma-ray wavelengths \citep{konusgcn,Neights2025}, easily places it among the class of ultra-long gamma-ray bursts. However, the \textit{Einstein Probe} mission identified source activity at soft X-rays ($0.5$\,$-$\,$4$ keV) starting nearly a full 24 hours before the first gamma-ray trigger \citep{EPgcn,EP250702a-arxiv}, which is not easily reconciled with standard GRB models. 

Following the arcminute localization by the \textit{Einstein Probe} \citep{EPgcn,EP250702a-arxiv}, X-ray observations with the \textit{Swift} X-ray Telescope (XRT; \citealt{Burrows2005}) identified a rapidly fading X-ray source and provided an arcsecond localization \citep{kenneagcn}. Near-infrared observations with the Very Large Telescope revealed an extremely red, quickly decaying counterpart \citep{Levan2025}. Subsequent \textit{Hubble Space Telescope} (\textit{HST}) observations provided conclusive evidence that the transient lies on top of the light of an irregular host galaxy with a negligible probability ($<$\,$0.1\%$) of chance alignment \citep{Levan2025}, confirming the extragalactic nature of GRB 250702B. A spectroscopic redshift of the host galaxy of $z$\,$=$\,$1.036$ was obtained by the \textit{James Webb Space Telescope} \citep{Gompertz2025}. Notably, the source is spatially offset from the center of the galaxy \citep{Levan2025,Carney2025}, complicating a classical TDE interpretation. 

We present comprehensive X-ray observations of GRB 250702B with \textit{Swift}, \textit{NuSTAR}, and \textit{Chandra}, aimed at characterizing its temporal and spectral evolution. At first glance, the observed multi-wavelength behavior is consistent with two leading interpretations: \textit{i}) a relativistic jetted TDE, potentially originating from an offset intermediate-mass black hole (IMBH), or \textit{ii}) a peculiar ultra-long GRB featuring a soft X-ray ``precursor''. We explore these possibilities in detail, highlighting the conflicting lines of evidence that make this event an outlier that resists a straightforward classification within existing high-energy transient populations. 


Throughout the manuscript we adopt a standard $\Lambda$CDM cosmology \citep{Planck2020} with $H_0$\,$=$\,$67.4$ km s$^{-1}$ Mpc$^{-1}$, $\Omega_\textrm{m}$\,$=$\,$0.315$, and $\Omega_\Lambda$\,$=$\,$0.685$.

\section{Observations}
\label{sec:obs}

\subsection{Neil Gehrels Swift Observatory}

\subsubsection{Burst Alert Telescope}
\label{sec:BAT}

The \textit{Neil Gehrels Swift Observatory} \citep{Gehrels2004} Burst Alert Telescope \citep[BAT;][]{Barthelmy2005} is a large field of view, ($\sim$\,$2$ sr) coded mask imager, capable of imaging in the $14$\,$-$\,$195$ keV energy range. Additionally, BAT is sensitive to impulsive emission originating from outside its coded field of view in the $50$\,$-$\,$350$ keV energy range, though at a reduced sensitivity compared to inside the coded field of view.  

On July 2, the \textit{Fermi} Gamma-ray Burst Monitor \citep[GBM;][]{Meegan2009} sent out low-latency alerts for four bursts associated with GRB 250702B \citep{Elizagcn,Neights2025}, triggering the Gamma-ray Urgent Archiver for Novel Opportunities \citep[\texttt{GUANO};][]{Tohuvavohu2020} system for \textit{Swift} to save BAT time-tagged event (TTE) data around the trigger times of each alert that would otherwise not be saved. The Non-Imaging Transient Reconstruction and Temporal Search \citep[\texttt{NITRATES};][]{NITRATES} pipeline was used to analyze the \texttt{GUANO} TTE data, and significantly detected transient emission around the ``D'', ``C'', and ``E'' bursts, while the ``B'' burst was occulted by the Earth for \textit{Swift} (see Table \ref{tab:triggertimes}; \citealt{Elizagcn}). Since these three detections were outside of the coded field of view, they had large localization areas, but all included the common position of GRB 250702B. The combined localizations of the four GBM triggers and three \texttt{NITRATES} detections resulted in a 90\% credible region of 142 square degrees \citep{JimmyGCN}.

We note that while GRB 250702C was dissociated from GRB 250702B and classified as a short GRB originating from another part of the sky \citep{Elizagcn}, a flare from GRB 250702B was coincidentally observed at the same time. The localization of the simultaneous short GRB was completely behind the Earth for \textit{Swift}, thus the detection of the simultaneous flare from GRB 250702B by \texttt{NITRATES} was unaffected. This robustly confirms the source flaring activity at the time of GRB 250702C (MJD 60858.6177). See Table \ref{tab:guanofluxes} for the peak fluxes of the bursts observed by BAT. Due to the long duration of these bursts, the background estimation may be contaminated by the source flux causing the fluxes to be underestimated. 



BAT TTE data is only available around onboard triggers or when \texttt{GUANO} is triggered by an external alert. When there is no TTE data, sky images can still be created from the BAT survey product, detector plane histograms (DPHs), which are detector plane images binned into 80 channels and $\sim$\,$300$ s exposures. Using the \texttt{BatAnalysis} package \citep{batanalysis} to create and analyze sky images from the DPHs, we searched for emission from the position of GRB 250702B at times when it was inside the coded field of view of BAT. 

After the initial GBM trigger, there were five observations on July 2 where GRB 250702B was inside of the BAT field of view, but none of them coincided with a reported burst. In two of these five observations, emission was significantly detected in images over the full observations. The first detection was in an $856$ s exposure from 2025-07-02 16:30:11 to 2025-07-02 16:44:27, which is minutes after the GRB 250702E trigger. The second detection was in an $1058$ s exposure from 2025-07-02 19:36:20 to 2025-07-02T19:53:58, which is hours after the last GBM trigger (GRB 250702E). The other three observations resulted in upper limits, with the deepest upper limit from an observation between the two detections from 2025-07-02 17:52:13 to 2025-07-02 18:06:50. We also did not identify any significant emission from the position of GRB 250702B in any other observation within $\pm 4$ days from the GRB 250702D trigger time. The $5\sigma$ upper limits as well as the flux levels of the two detections derived from BAT survey mode data ($14$\,$-$\,$195$ keV) are tabulated in Table \ref{tab:batlimits}. We additionally confirmed there were no previous detections of GRB 250702B by BAT in the $\sim$\,$1$ month before the initial series of high-energy triggers. We tested multiple bin sizes between $1$\,$-$\,$7$ d, but did not identify any significant signals.




\subsubsection{X-ray Telescope}
\label{sec:XRT}


Following the arcminute X-ray localization from EP/WXT \citep{EPgcn}, we rapidly requested a Target of Opportunity (ToO) monitoring campaign (ObsID: 19906, Submitter: O'Connor; Obsid: 19928, GO program 1922200; PI: Pasham) with the \textit{Neil Gehrels Swift Observatory} \citep{Gehrels2004} X-ray Telescope (XRT; \citealt{Burrows2005}). A highest urgency ToO detected the X-ray source with the XRT starting at $0.5$ d after the initial \textit{Fermi} trigger \citep{kenneagcn}. The source was localized \citep{kenneagcn} to RA, DEC (J2000) = $18^{h}58^m 45^{s}.61$, $-07^\circ 52\arcmin 26.9\arcsec$ with an uncertainty of $2.0\arcsec$ at the 90\% confidence level (CL). The initial observations consisted of multiple $\sim$\,$3$ ks exposures over the first $\sim$\,$3$ days. Continued monitoring (PI: Pasham) extended to 45 d (2025-08-16) with $\sim$\,$1$ ks per day for a total of 45 ks exposure\footnote{\url{https://www.swift.psu.edu/operations/obsSchedule.php?source_name=&ra=284.6898&dec=-7.8746}} (Table \ref{tab: observationsXray}). All XRT data were obtained in in Photon Counting (PC) mode.

We retrieved the automated XRT lightcurve analysis\footnote{\url{https://www.swift.ac.uk/LSXPS/transients/9377}} from The Living Swift XRT Point Source Catalogue \citep[LSXPS;][]{LSXPS}. 
We applied a constant energy conversion factor (ECF) of $8.68\times10^{-11}$ erg cm$^{-2}$ cts$^{-1}$ to convert between count rate and unabsorbed flux using the best-fit spectral parameters from the initial XRT observation (see \S \ref{sec:spec} for details). We further retrieved time-sliced spectral files through the XRT Build Products tool\footnote{\url{https://www.swift.ac.uk/user_objects/index.php}}. We likewise used this tool to obtain the $0.3$\,$-$\,$2$ and $2$\,$-$\,$10$ keV count rate lightcurves and derive hardness ratios.

\subsection{NuSTAR}
\label{sec:nustar}

The \textit{Nuclear Spectroscopic Telescope Array} \citep[\textit{NuSTAR};][]{Harrison+2013} observes between $3$\,$-$\,$79$ keV using two co-aligned focal plane modules (FPM), referred to as FPMA and FPMB. 
Following the initial XRT detections, we triggered our approved \textit{NuSTAR} ToO program (GO program 11282; PI: Pasham) for two epochs of $\sim$\,$20$ ks (Table \ref{tab: observationsXray}) starting at $1.3$ and $5.4$ d after the \textit{Fermi} trigger. A third \textit{NuSTAR} observation with a $\sim$\,$40$ ks exposure occurred at $9.5$ d through a Director's Discretionary Time (DDT) request (PI: Kammoun; see \citealt{Oganesyan2025}). 

We used standard tasks within the \textit{NuSTAR} Data Analysis Software pipeline (\texttt{NuSTARDAS}; \texttt{CALDB 20250714}) within \texttt{HEASoft} \citep{2014ascl.soft08004N} to reprocess the data, and then extract lightcurves and spectra. In the first epoch, the X-ray counterpart is clearly detected with a mean count rate of $\sim$\,0.3 cts s$^{-1}$ per FPM \citep{nustargcn}. We extracted the source counts from a circular region with radius 45\arcsec{} from both FPMA and FPMB, and utilized a source free region with radius 120\arcsec{} to extract the background. The background region was placed on the same detector as the transient. After subtracting the background, the mean count rate is $\sim$\,0.25 cts s$^{-1}$ per FPM.

Due to the rapid decay of the X-ray lightcurve, the source is only very weakly detected in the second epoch with a mean count rate of $\sim$\,$0.03$ cts s$^{-1}$. The second observation is also contaminated by strong radiation belt activity (up to $\sim$\,15 cts s$^{-1}$), and we therefore re-processed the data  with strict South Atlantic Anomoly (SAA) filtering (\texttt{saacalc=3, saamode=STRICT, tentacle=yes}). We are left with an effective exposure of $18.2$ ks for FPMA and $17.8$ ks for FPMB. We extracted source counts using a circular region of radius 20\arcsec{}, and background counts from a region of radius 120\arcsec{}.

In addition to our two epochs of \textit{NuSTAR} observations, we analyzed a third epoch of public DDT data, which was previously presented in \citet{Oganesyan2025}. We applied strict SAA filtering cuts (as noted above), leaving us with an exposure of $31.7$ ks for FPMA and $32.8$ ks for FPMB. We identify a weak source in FPMB, and extract source counts from both FPM using a  circular region of radius 20\arcsec{}. Background counts were extracted from similar regions as described above. 



 
Source and background lightcurves were extracted in the $3$\,$-$\,$6$, $10$\,$-$\,$20$, and $3$\,$-$\,$79$ keV energy ranges in time bins of $150$ and $500$ s. Spectra were grouped to a minimum of 1 count per energy bin. Photon arrival times were barycenter corrected using the latest clock file (Clock Correction file v209 released on 2025-07-15).

\begin{figure*}
    \centering
\includegraphics[width=2.1\columnwidth]{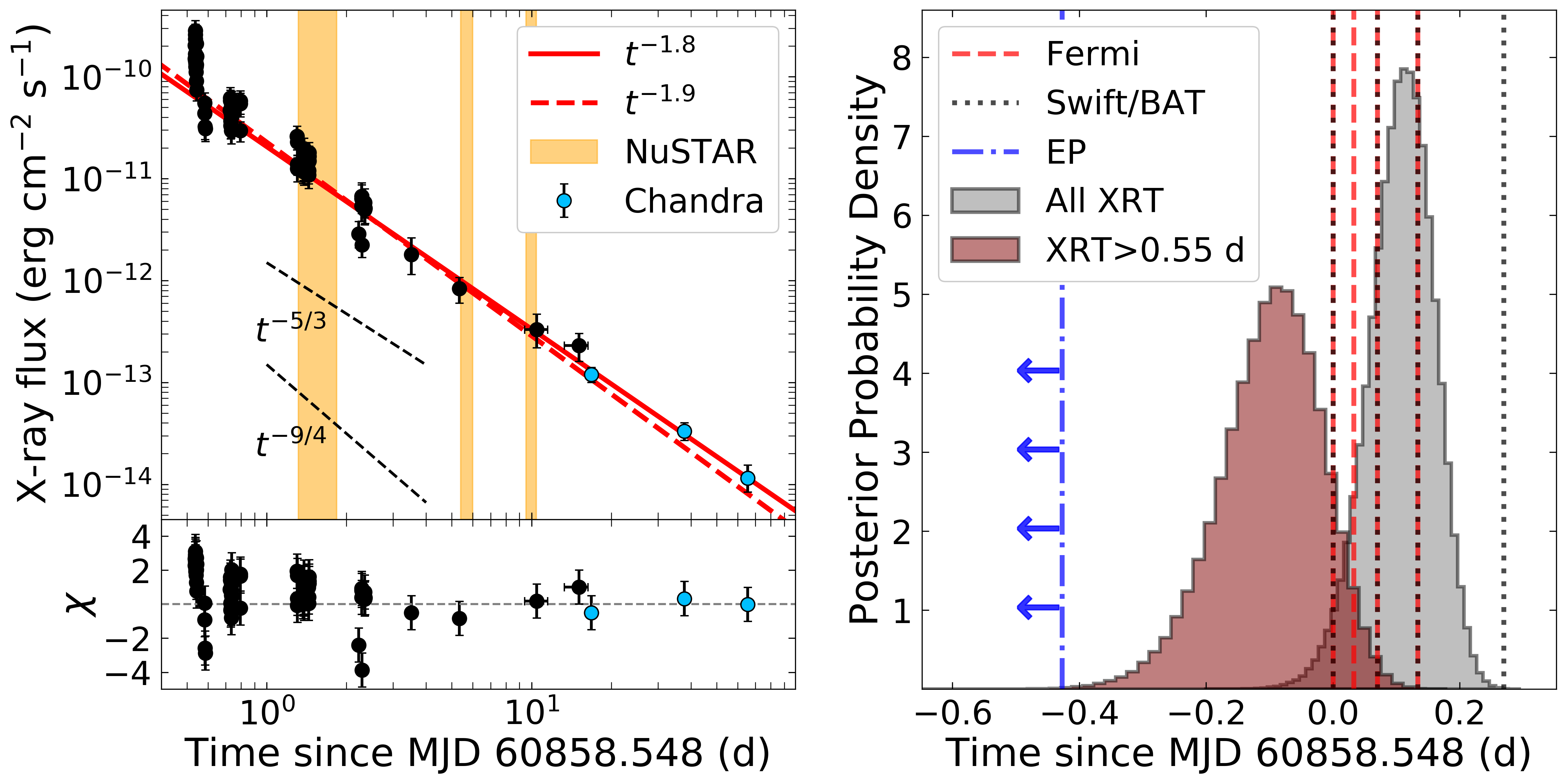}
    \caption{\textbf{Left:} X-ray lightcurve of GRB 250702B combining \textit{Swift} and \textit{Chandra} in the $0.3$\,$-$\,$10$ keV energy range. The start time $T_0$ is the GBM ``D'' burst. The red lines show the best-fit  decay from the initial trigger time of GRB 250702D (Table \ref{tab:triggertimes}). The dashed red line ($t^{-1.9}$) includes the first XRT orbit in the fit, whereas the solid red line excludes the first XRT orbit and provides a better fit to the late-time data ($t^{-1.8}$). The observation windows of our \textit{NuSTAR} data are shown as orange shaded regions. For reference we show the expected slopes for a relativistic TDE, corresponding to both complete ($t^{-5/3}$) and partial ($t^{-9/4}$) disruption. The bottom panel shows the $0.3$\,$-$\,$10$ keV lightcurve fit residuals relative to the solid red line. 
    \textbf{Right:} Histogram showing the posterior probability density of the best-fit $T_0$ time for the X-ray afterglow emission. The gray histogram shows the $T_0$ posterior obtained when fitting all XRT data, and the red histogram shows the $T_0$ posterior when excluding the first XRT orbit. 
    The trigger times reported by \textit{Fermi} (red) and EP (blue) are shown as vertical lines (Table \ref{tab:triggertimes}). The July 1 window in which EP reports an initial detection is to the left of the vertical blue line. We also show as vertical black lines the times of \textit{Swift}/BAT gamma-ray detections. All times are relative to the trigger of GRB 250702D at MJD 60858.548. 
    }
    \label{fig:temporalfit}
\end{figure*}

\begin{figure*}
    \centering
\includegraphics[width=2.0\columnwidth]{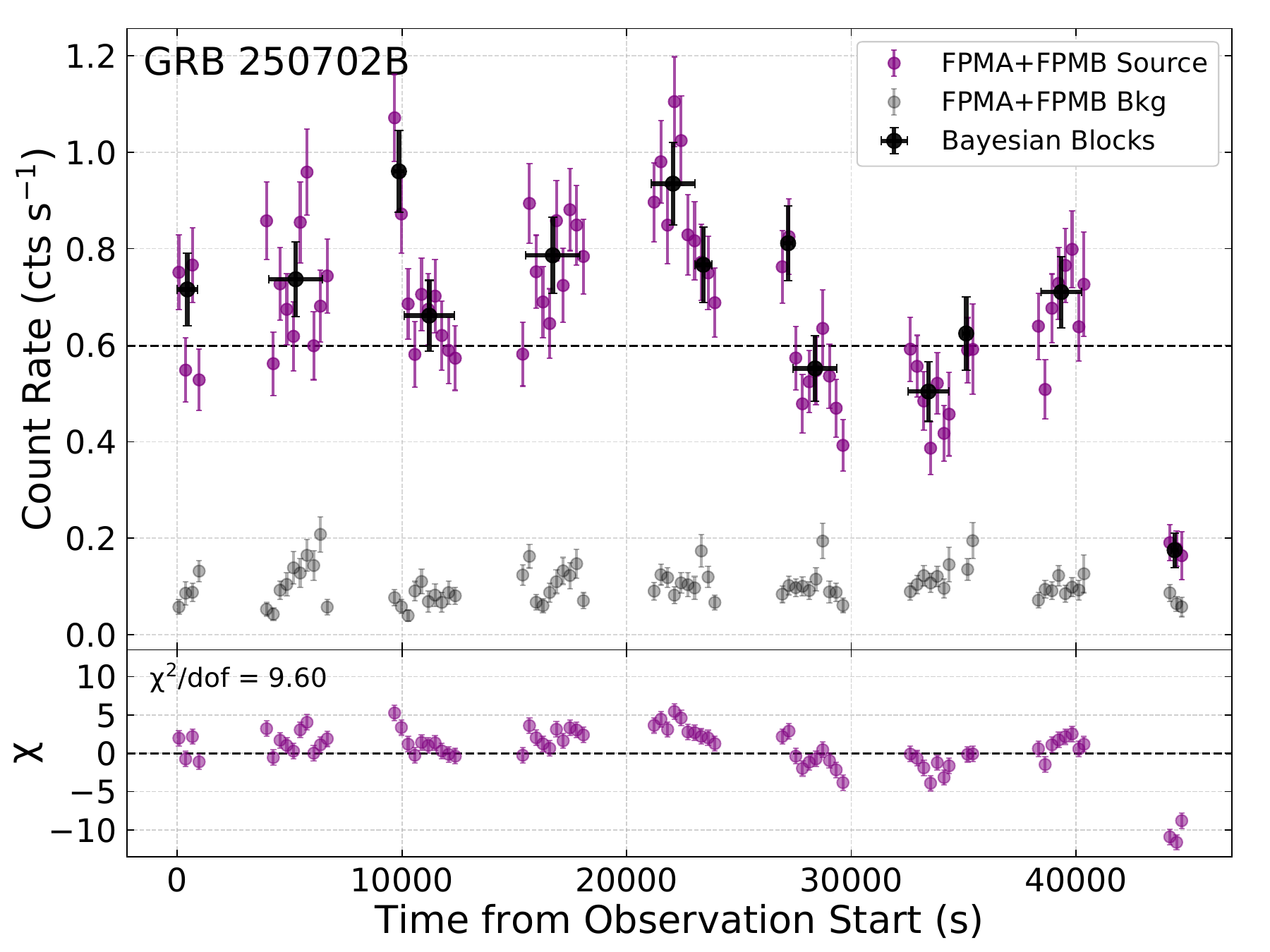}
    \caption{The $3$\,$-$\,$79$ keV \textit{NuSTAR} lightcurve (FPMA+FPMB) in 300 s bins (purple). The lightcurve is not background subtracted, but the background lightcurve (gray) is also shown for comparison. The dashed black line shows the mean count rate. The Bayesian Blocks results are shown as larger black circles. The bottom panel shows the residuals with respect to the mean count rate to underscore the source's short timescale variability.
    }
    \label{fig:lcrate}
\end{figure*}

\subsection{Chandra X-ray Observatory}
\label{sec:chandra}

We carried out a late-time X-ray observation with the \textit{Chandra X-ray Observatory} (CXO) through a DDT request (PI: O'Connor; ObsID: 31011) starting on 2025-08-09 at 05:49:28 UT (corresponding to $\sim$\,37.69 d after discovery) for $\sim$\,$27.7$ ks using ACIS-S \citep{chandragcn}. The target was centered at the aimpoint of the S3 chip. The \textit{Chandra} data were retrieved from the \textit{Chandra} Data Archive (CDA)\footnote{\url{https://cda.harvard.edu/chaser/}}. We re-processed the data using the \texttt{CIAO v4.17.0} data reduction package \citep{Ciao} with \texttt{CALDB v4.11.6}. At the subarcsec location of the near-infrared counterpart \citep{Levan2025}, we identify a clear X-ray source. We find 25 counts ($0.5$\,$-$\,$8$ keV) in a circular source region with radius $1.5\arcsec$ centered on the localization obtained with \texttt{wavdetect}. We extracted background counts from a nearby source free region with radius $30\arcsec$. Using the best fit X-ray spectrum (\S \ref{sec:spec}), we derive an unabsorbed $0.3$\,$-$\,$10$ keV flux of $(3.3^{+0.7}_{-0.6})\times10^{-14}$ erg cm$^{-2}$ s$^{-1}$. The \textit{Chandra} data confirms that there are no contaminating point sources within either the \textit{Swift} or \textit{NuSTAR} source extraction regions.

Additional \textit{Chandra} DDT observations were obtained and we re-analyze them here. We analyzed these observation in the same manner outlined above. The first (PI: Li; \citealt{EP250702a-arxiv}) began on 2025-07-18 at 19:24:33 UT, corresponding to 16.8 d after discovery, using ACIS-I for 14.9 ks. We find 40 counts in a $1.5\arcsec$ radius, and obtain an unabsorbed $0.3$\,$-$\,$10$ keV flux of $(1.2\pm0.2)\times10^{-13}$ erg cm$^{-2}$ s$^{-1}$ assuming out best fit X-ray spectrum (\S \ref{sec:spec}). The third DDT observation (PI: Eyles-Ferris; \citealt{chandragcn2,Eyles-Ferris2025}) started on 2025-09-05 at 23:26:41 UT, corresponding to 65.5 d after discovery, using ACIS-S for a total of $39.55$ ks. We find 12 counts in a $1.5\arcsec$ radius, and obtain an unabsorbed $0.3$\,$-$\,$10$ keV flux of $(1.15^{+0.40}_{-0.30})\times10^{-14}$ erg cm$^{-2}$ s$^{-1}$.

We caution that all three \textit{Chandra} DD observations (PIs: Li, O'Connor, and Eyles-Ferris) suffer from warm ACIS focal plane temperature's\footnote{\url{https://cxc.cfa.harvard.edu/ciao/why/acis_fptemp.html}} with extreme episodes of heating and cooling throughout their exposures. Warm focal plane observations suffer from gain issues (gain under-corrected) and we caution against strong interpretations of the spectral analysis (especially due to the low 30-40 counts regime) without the proper gain corrections which are not updated yet for this observation period (Chandra Helpdesk\footnote{\url{https://cxc.harvard.edu/help/}}, private communication). This gain correction has likely impacted the spectral analysis presented by \citet{Eyles-Ferris2025} and \citet{EP250702a-arxiv}.


\begin{figure}
    \centering
\includegraphics[width=\columnwidth]{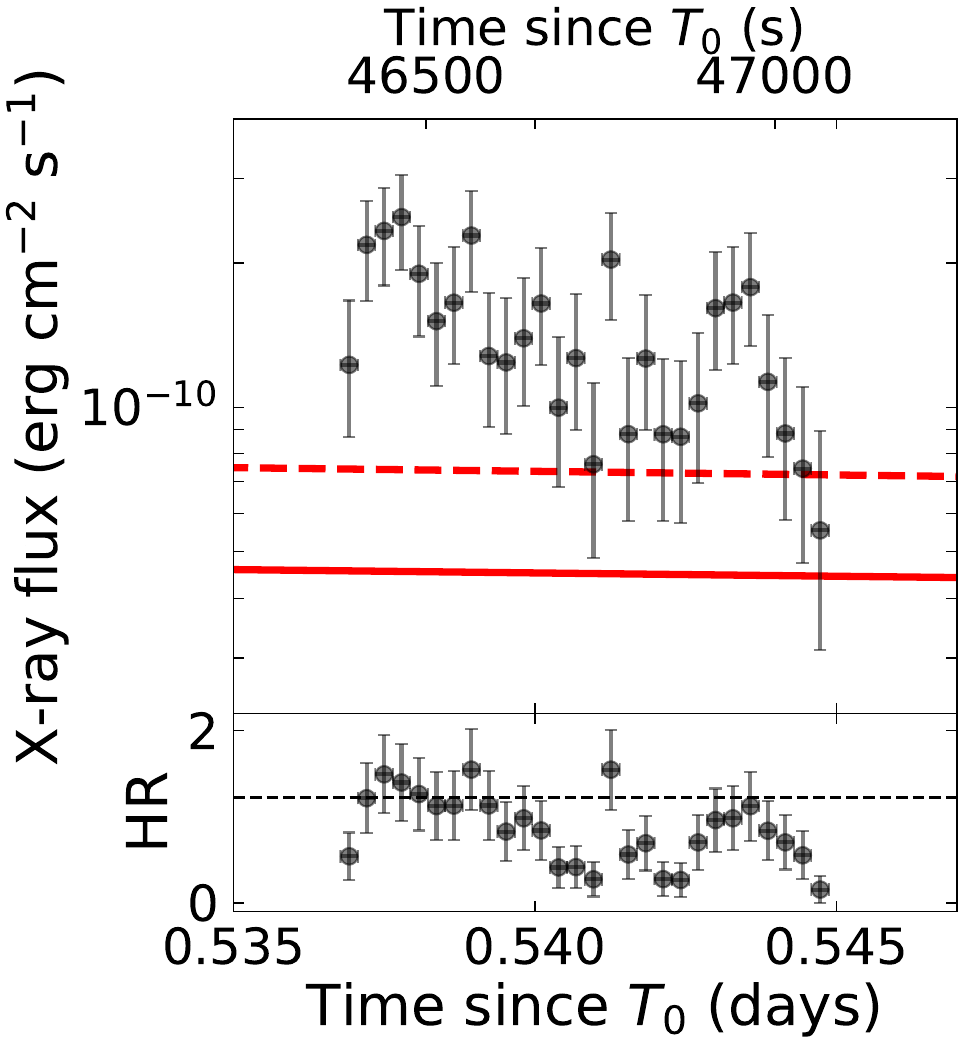}
    \caption{
    Zoom on the early \textit{Swift}/XRT lightcurve using 25 s bins. The start time $T_0$ is the GBM ``D'' burst. The red lines is the same as the temporal fit shown in Figure \ref{fig:temporalfit} (left panel). The hardness ratio (HR) between $2$\,$-$\,$10$ and $0.3$\,$-$\,$2$ keV is shown in the bottom panel. The dashed black line shows the mean hardness ratio of the full lightcurve.
    }
    \label{fig:XRTflaring}
\end{figure}

\begin{figure*}
    \centering
\includegraphics[width=2.0\columnwidth]{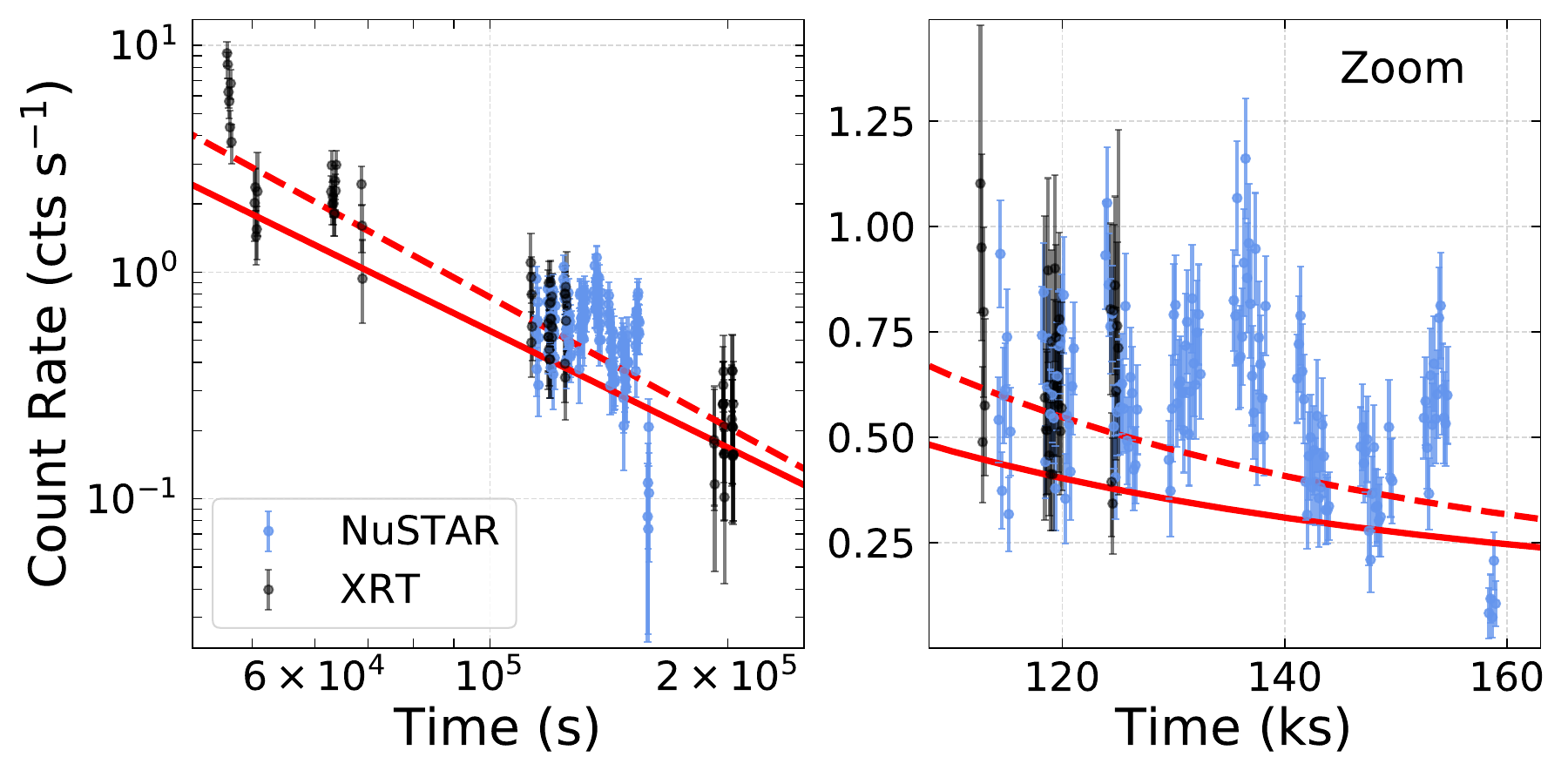}
    \caption{\textbf{Left:} X-ray lightcurve comparing \textit{NuSTAR} (FPMA+FPMB) data to the XRT lightcurve and best-fit temporal powerlaw as shown in Figure \ref{fig:temporalfit}. The start time $T_0$ is the GBM ``D'' burst. The solid red line is the best-fit ($t^{-1.7}$) when including only XRT at $>2.5$ d, and the red dashed line is the best-fit ($t^{-1.9}$) when including all XRT data. \textit{NuSTAR} data is in 150 s bins and XRT in 100 s bins. The XRT count rate and powerlaw fit are re-scaled to their expected \textit{NuSTAR} $3$\,$-$\,$79$ keV count rate. We emphasize that this scale factor is not arbitrary and is computed directly based on the best-fit spectral shape (\S \ref{sec:spec}).  
    \textbf{Right:} Same as the left panel but zoomed in on the \textit{NuSTAR} data in linear-linear space. Note that the temporal axis is in units of ks in this panel.  
    }
    \label{fig:xrt+nustar}
\end{figure*}

\section{Analysis and Results}
\label{sec:results}

\subsection{Analysis of the X-ray Lightcurve}

\subsubsection{Temporal Modeling}
\label{sec:temporalfit}

We modeled the $0.3$\,$-$\,$10$ keV X-ray lightcurve (\textit{Swift}/XRT and \textit{Chandra}) with a powerlaw of the form $t^{-\alpha}$ using the \texttt{emcee} package \citep{emcee}. As the exact onset time of the transient is unclear (Table \ref{tab:triggertimes}), we simplify this consideration by focusing on the start time of the afterglow $T_0$. This should not be confused with the onset time of the transient, which is set by the first detection by the \textit{Einstein Probe} \citep{EPgcn,EP250702a-arxiv}, despite this early detection preceding the peak of the X-ray and gamma-ray lightcurves. We use the combined \textit{Swift}/XRT and \textit{Chandra} X-ray lightcurve to determine the afterglow start time assuming it decays as a powerlaw. We performed multiple different fits to the data to explore the range of possible afterglow start times and temporal decay rates. We first allowed the afterglow start time $T_0$ to be a free parameter, and then performed an additional fit with $T_0$ fixed to the first high-energy trigger time (the discovery time) of GRB 250702D by \textit{Fermi} (see Table \ref{tab:triggertimes}).

We first modeled the data with a single powerlaw. The early XRT data obtained show significant variability on short timescales, providing scatter around the best fit powerlaw decay. We obtain a reduced chi-squared of $\chi^2$/dof\,$=$\,$2.36$ for a fixed $T_0$, and $\chi^2$/dof\,$=$\,$2.32$ for a free $T_0$. The poor chi-squared values are the result of the intrinsic variability at early times and can be observed in the bottom panel showing fit residuals in Figure \ref{fig:temporalfit} (left). 
In Figure \ref{fig:temporalfit} (left), we show the best fit powerlaw decay when allowing $T_0$ to be free. The best fit start time $T_0$ (MJD\,$=$\,$60858.657^{+0.054}_{-0.063}$), and the general posterior of $T_0$ values, agrees with a start time at the peak of gamma-ray activity \citep{Neights2025}. The full posteriors for both fits are shown as corner plots in Figure \ref{fig:temporalcorner}. The exact inferred powerlaw slope depends strongly on $T_0$. These two fits yield similarly steep slopes and agree at the $\sim$\,$1.5\sigma$ level, favoring an X-ray lightcurve decay of between $\approx$\,$t^{-1.7}$ to $t^{-1.9}$.

We performed these fits both with and without the inclusion of the initial XRT orbit at $<$\,$0.55$ days after discovery (see Figure \ref{fig:temporalcorner}). We demonstrate that the initial XRT observation shows flaring activity (\S \ref{sec:var}) and may bias the temporal analysis. Ignoring the initial flaring episode does not significantly change the inferred decay slope (see Figure \ref{fig:temporalcorner}), but does shift the inferred value of $T_0$ (Figure \ref{fig:temporalfit}; right panel) towards earlier times. This demonstrates that the exact decay slope and afterglow start time depend strongly on the assumptions taken for the origin of the X-ray lightcurve.

While it is standard in GRB afterglow modeling to take the initial trigger time as the afterglow start time $T_0$, in this case the start time is complicated by the early July 1 detection \citep{EPgcn,EP250702a-arxiv}. We performed comprehensive temporal fits to investigate the onset of the afterglow component. That the posterior of afterglow onset times agree with the later gamma-ray triggers, as opposed to the first detections by the \textit{Einstein Probe} is slightly unexpected, but clearly depends on the choice of powerlaw versus broken powerlaw (Figure \ref{fig:temporalcornerbkn}). If the true onset time occurred 24 hours earlier on July 1st then the inferred initial power-law decay is significantly steeper ($\sim$\,$t^{-3.5}$), and a break to a shallower slope ($\sim$\,$t^{-5/3}$) would be required to match the late-time X-ray data (see Figure \ref{fig:temporalcornerbkn}). This is plausible since the early emission is likely flaring, which then subsides to reveal the slowly fading afterglow. However, the simplest explanation is that either the onset of the relativistic jet's interaction with its surrounding environment occurs around the peak of gamma-ray activity or that the bulk of its energy is released at this phase, in either case leading to a single power-law afterglow decay. As the trigger of GRB 250702E has the highest fluence (i.e., the bulk of the energy release) of the \textit{Fermi} triggers \citep{Elizagcn,Neights2025}, it would make sense that this would be favored as the onset time of the jet's afterglow in this context.

\begin{figure*}
    \centering
\includegraphics[width=1.0\columnwidth]{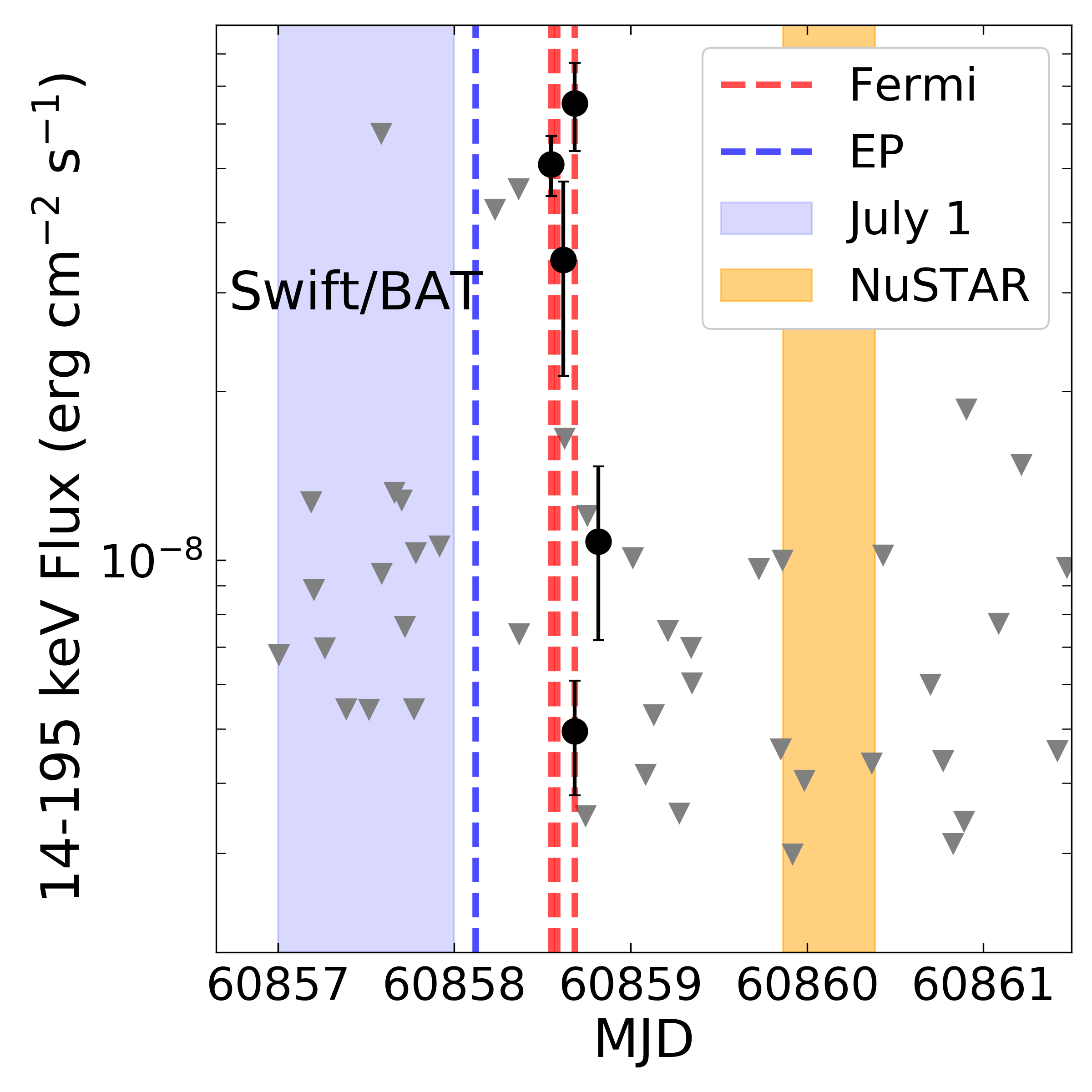}
\includegraphics[width=1.0\columnwidth]{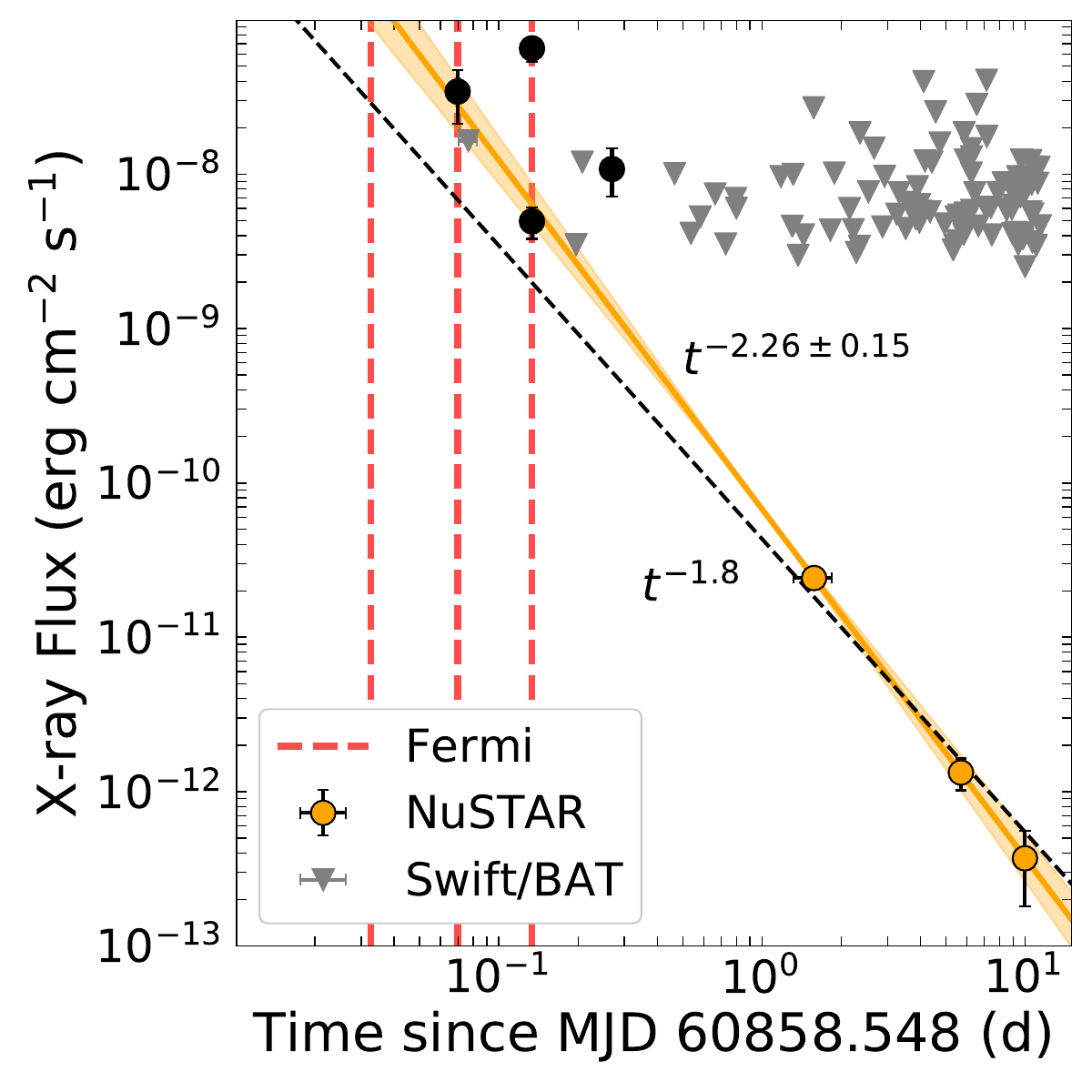}
    \caption{\textbf{Left:} \textit{Swift}/BAT survey mode ($14$\,$-$\,$195$ keV) on emission from GRB 250702B from the start of July 1, 2025 to the end of July 4, 2025. Black points show the \textit{Swift}/BAT detections ($14$\,$-$\,$195$ keV; Tables \ref{fig:batlimits} and \ref{tab:guanofluxes}) from both \texttt{GUANO} and survey data. Gray downward triangles represent $5\sigma$ upper limits from \textit{Swift}/BAT in the $14$\,$-$\,$195$ keV energy range. The trigger times reported by \textit{Fermi} (red) and EP (blue) are shown as vertical lines (Table \ref{tab:triggertimes}). The July 1 window in which EP reports an initial detection is shown as a blue shaded region. The time window of our initial \textit{NuSTAR} observation (starting 1.3 d after the initial \textit{Fermi} trigger) is shown as an orange shaded region. 
    \textbf{Right:} Best-fit temporal powerlaw to the \textit{NuSTAR} lightcurve in $3$\,$-$\,$79$ keV. The $1\sigma$ error on the fit is shown as a shaded orange region. \textit{Swift}/BAT upper limits are shown as gray triangles and detections as black circles. The start time $T_0$ is taken as the trigger time of GRB 250702D. The vertical red lines show the \textit{Fermi} trigger times of GRBs 250702B, 250702C, and 250702E (see Table \ref{tab:triggertimes}). 
    }
    \label{fig:batlimits}
\end{figure*}

Using the trigger of GRB 250702D as $T_0$, we similarly modeled the \textit{NuSTAR} $3$\,$-$\,$79$ keV flux (see \S \ref{sec:spec}) across the three epochs (with mid-times of 1.58, 5.70, 9.98 d; Table \ref{tab: observationsXray}) with a temporal powerlaw model. We derive a temporal index of $\alpha$\,$=$\,$2.26\pm0.15$. This is slightly steeper, but consistent at the $\sim$\,$2\sigma$ level, with the XRT fit using the same fixed start time (see Figure \ref{fig:temporalcorner}). We note that the \textit{NuSTAR} data is limited to three epochs and the third epoch suffers from large error bars, which suggests that the slope derived from the combined \textit{Swift} and \textit{Chandra} lightcurve is more accurate. The first epoch of \textit{NuSTAR} data also suffers from flaring (see \S \ref{sec:var}), which can also explain its higher flux that can lead to a steeper inferred slope. The shallower slope measured from \textit{Swift} and \textit{Chandra} is also more consistent with the early \textit{Swift}/BAT survey mode upper limits (see \S \ref{sec:BAT}; Table \ref{tab:batlimits}). This is shown in Figure \ref{fig:batlimits}.  While the \textit{NuSTAR} extrapolation goes through two BAT detections of the prompt emission, this depends strongly on the assumed temporal slope and we emphasize the BAT detections are not included in the fit.

\subsubsection{Short Timescale X-ray Variability}
\label{sec:var}

Our initial \textit{NuSTAR} observation starting at $T_0+1.3$ d and extending to 1.9 d after the initial \textit{Fermi trigger} (``D'' Burst), shows significant X-ray variability on $\sim$\,$1-2$ ks timescales. In Figure \ref{fig:lcrate}, we show the $3$\,$-$\,$79$ keV lightcurve (300 s bins; FPMA+FPMB) of GRB 250702B. The observed X-ray variations are seen in both FPMs independently (Figure \ref{fig:lcrateapp}). In particular, in the last good time interval (GTI) the count rate drops significantly (see the residuals in the bottom panel in Figure \ref{fig:lcrate}). The difference between the end of the second to last GTI and the start of last GTI is 3.7 ks (observer frame), which is taken to be a conservative upper bound to the minimum variability timescale at this phase. This implies $\Delta T/T$\,$<$\,$0.03$, which is difficult to produce by emission from an external shock and implies that there is late-time central engine activity extending out to $\sim$\,$1.6\times10^{5}$ s (observer frame). We confirmed that this variability is unrelated to the particle background during the observation by applying different combinations of SAA filtering and selecting different background regions for background lightcurve generation and subtraction. The background lightcurve is also shown for comparison in Figure \ref{fig:lcrate} for FPMA+FPMB, and in Figure \ref{fig:lcrateapp} for both FPMA and FPMB individually. 

In Appendix \ref{sec:grbvarappendix}, we describe our analysis of past \textit{NuSTAR} observations of GRB afterglows (e.g., GRB 130427A and 221009A), which do not display short term X-ray variability (see also Figure \ref{fig:grbvar}). We compute the reduced chi-squared with regard to the mean count rate in each observation, finding $\chi^2$/dof\,$\approx$\,$1.0$ for these other GRBs. This compares to $\chi^2$/dof\,$=$\,$2.8$ for GRB 250702B (Figure \ref{fig:lcrateapp}), demonstrating that the source is significantly more variable on short timescales. Even excluding the last GTI where a steep drop in count rate is clearly observed, we find $\chi^2$/dof\,$=$\,$1.9$. The reduced chi-square depends on the binning, as larger time bins reduce the errors and lead to larger values of the reduced chi-squared, as seen in Figure \ref{fig:lcrate} (300 s bins), where $\chi^2$/dof\,$=$\,$9.6$. We note that while Figure \ref{fig:lcrate} uses 300 s bins for visualization purposes, our comparison to GRB variability in Appendix \ref{sec:grbvarappendix} was consistently performed using 150 s bins for both GRB 250702B (Figure \ref{fig:lcrateapp}) and the other GRBs (Figure \ref{fig:grbvar}).

In addition, the existence of short time scale variability is independently supported by the early \textit{Swift}/XRT data where the count rate is high enough to have short time bins (Figure \ref{fig:XRTflaring}). In the initial XRT orbit we find significant flaring and changes in the hardness ratio (Figure \ref{fig:XRTflaring} and Appendix \ref{sec:HRev}). The correlated change in hardness ratio is typical of flaring activity, and in contrast with afterglow emission. The observed flares have a duration of $\sim$\,$150$ s at $\sim$\,$0.54$ d (observed time since ``D'' burst). This leads to a very short variability timescale of $\Delta T/T$\,$\approx$\,$0.003$, where $\Delta T$ is denoted as the difference ($T_\textrm{start}$\,$-$\,$T_\textrm{stop}$) between the start $T_\textrm{start}$ and stop $T_\textrm{stop}$ of the flare divided by the observed peak of flaring activity $T_\textrm{peak}$ \citep[e.g.,][]{Swenson2014}. We note that this is extremely fast even for X-ray flares observed in GRBs, which also generally have higher amplitudes \citep{Curran2008,Margutti2011,Bernardini2011,Swenson2014}.

In order to explore the nature of the variability observed by \textit{NuSTAR} (Figure \ref{fig:lcrate}), we cross-correlated the \textit{NuSTAR} data with simultaneous XRT data covering the first few orbits. This is displayed in Figure \ref{fig:xrt+nustar}, where the XRT count rate lightcurve has been re-scaled to the expected \textit{NuSTAR} $3$\,$-$\,$79$ keV count rate based on the observed joint XRT and \textit{NuSTAR} spectral fit. See \S \ref{sec:spec} for details for the spectral modeling and its results. We emphasize that this scale factor is not arbitrary and derived using the \texttt{PIMMS} software\footnote{\url{https://heasarc.gsfc.nasa.gov/cgi-bin/Tools/w3pimms/w3pimms.pl}}. This scale factor provides good agreement between the simultaneous XRT and \textit{NuSTAR} data. We then compared these data to the best-fit powerlaw decay to the soft X-ray data (Figure \ref{fig:temporalfit} and \S \ref{sec:temporalfit}), which was also re-scaled using the same factor. We caution that this scale factor depends on the spectral shape, which likely changes as a result of the observed X-ray variability (Figure \ref{fig:XRTflaring}), although we do not find strong evidence for this in the \textit{NuSTAR} data (Figure \ref{fig:HR} and Appendix \ref{sec:HRev}). Regardless, in Figure \ref{fig:xrt+nustar} the \textit{NuSTAR} data shows a clear excess above the extrapolation of the XRT data. We therefore interpret the \textit{NuSTAR} variability as late-time flaring activity with duration of $\sim$\,$1$\,$-$\,$2$ ks, requiring late-time central engine activity out to $\sim$\,$2$ d (observer frame). However, due to the variability we cannot conclude whether the power-law fit to the XRT data is setting the continuum level of an external shock afterglow (serving as a baseline for the flares), or whether it is accretion driven emission originating from an internal dissipation region within the relativistic jet (see \S \ref{sec:int-vs-ext}). We note that while the last \textit{NuSTAR} GTI could be interpreted as a sharp drop in flux, we do not have clear evidence for the short timescale sharp ``dips'' in flux (by a factor of $\sim$\,10) observed from Sw J1644+57  \citep{Burrows2011,Bloom2011,Levan2011}. We discuss both possible interpretations of the X-ray lightcurve further in \S \ref{sec:afterglow} and \ref{sec:xrayafterdisc}. 

In order to more robustly explore the variability we made use of a Bayesian Blocks approach \citep{1998ApJ...504..405S,Scargle2013}. We extracted NuSTAR light curves from FPMA and FPMB in the $3$\,$–$\,$30$ keV energy range using the standard {\tt nuproducts} task, which applies the usual instrumental corrections for exposure, aperture/PSF, and vignetting. After restricting the data to overlapping intervals of the two sets of GTIs (FPMA and FPMB), we combined the light curves from both detectors and rebinned them to 300 s in order to improve the signal-to-noise ratio. Uncertainties were propagated in quadrature. We then performed a Bayesian Blocks analysis on the combined light curve on a per-GTI basis, adopting a false-positive probability parameter of $p_0$\,$=$\,$0.05$. This choice corresponds to a 5\% chance that any detected change point is spurious, thereby balancing sensitivity to genuine variability against the risk of over-segmentation (see, e.g., \citealt{1998ApJ...504..405S} for more details on Bayesian Blocks). The Python implementation provided in {\tt astropy.stats.bayesian\_blocks}\footnote{\url{https://docs.astropy.org/en/stable/api/astropy.stats.bayesian\_blocks.html}} was used, with uncertainties incorporated through the {\tt sigma} input parameter. For comparison, we also tested the analysis without rebinning, using the combined barycenter-corrected FPMA and FPMB source event files directly with {\tt fitness=`events'}, which yielded consistent results in terms of block boundaries. The Bayesian Blocks result is shown in Figure \ref{fig:lcrate} and supports the existence of short timescale X-ray variability.

\subsubsection{Pulsation Search}
\label{sec:timing}

We carried out a pulsation search for periodic X-ray signals. 
We analyzed the barycenter corrected events files for both FPMA and FPMB, individually, as well as the combined event file, to search for coherent periodicity. The event files were filtered to contain only the source region using \texttt{xselect}. We performed a Lomb-Scargle periodogram \citep{Lomb1976,Scargle1982} analysis using a bin size of $\delta t$\,$=$\,$0.003$ s to search between frequencies of $10^{-4}$ to $300$ Hz. We also constructed an averaged power spectrum using the \texttt{AveragedPowerspectrum} class in \texttt{Stingray} \citep{Huppenkothen2019} using 1000 s bins and a bin size of $\delta t$\,$=$\,$0.003$ s. We tested multiple different combinations of bins, but the results were unchanged. We performed the same analysis on multiple energy bins, including $3$\,$-$\,$6$, $3$\,$-$\,$10$, $3$\,$-$\,$20$ keV, and $3$\,$-$\,$79$ keV. 
We do not identify any significant signal (occurring in both FPMA and FPMB, or in the combined FPMA+FPMB events list) in the power spectra that exceeds the trial corrected 99.7\% confidence level. 


\subsubsection{Average Power Spectrum}
\begin{figure}
    \centering
\includegraphics[width=\columnwidth]{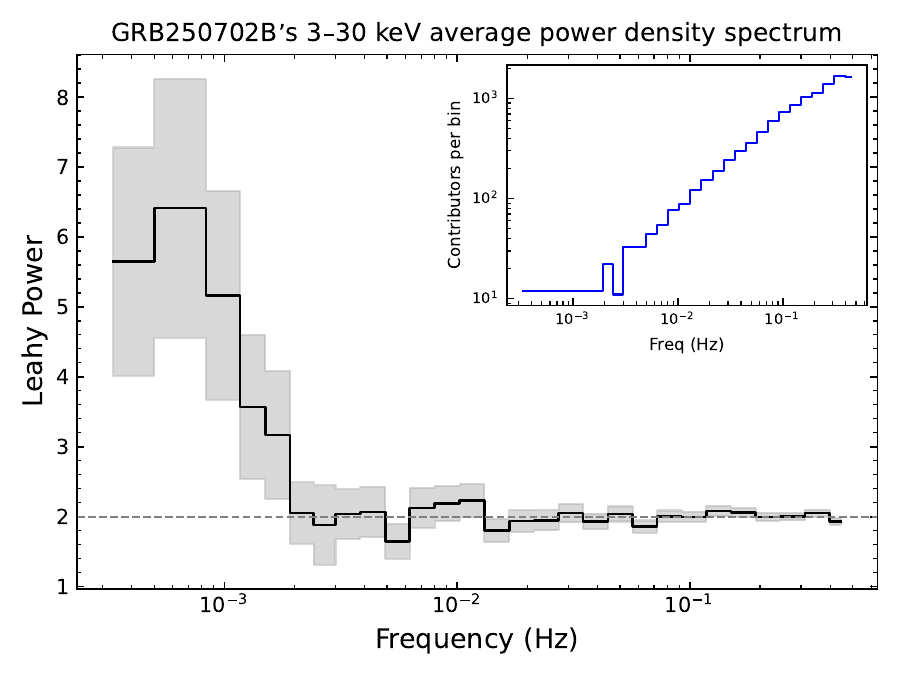}
    \caption{GRB250702B’s $3$\,$–$\,$30$~keV average power density spectrum from \textit{NuSTAR} FPMA+FPMB data. The black step line shows the mean Leahy-normalized power spectrum, the gray shading indicates $1\sigma$ uncertainties, and the dashed line marks the expected Poisson noise level of 2. The inset shows the number of contributing PSD values per frequency bin on log--log axes.
    }
    \label{fig:leahypsd}
\end{figure}

Motivated by claims of minutes-timescale X-ray quasi-periodic oscillations (QPOs) in some tidal disruption events \citep[e.g.,][]{2012Sci...337..949R, 2019Sci...363..531P} we also extracted an average power density spectrum focusing below 0.5 Hz as follows. We extracted \textit{NuSTAR}'s FPMA and FPMB's barycenter-corrected event lists in the 3--30~keV energy band and restricted the data to the intersection of their Good Time Intervals (GTIs), ensuring strictly simultaneous and gap-free coverage. Photon arrival times were binned into 1~s light curves, and power spectra were computed on a per-GTI basis using the Leahy normalization \citep{1983ApJ...266..160L}, for which purely Poissonian noise corresponds to a flat level of 2 \citep{Leahy1983}. To balance sensitivity to low frequencies with efficient use of exposure, we adopted a hybrid segmentation scheme: long GTIs were divided into equal segments of length $T_{\rm seg}=3000$~s, corresponding to a minimum sampled frequency of $1/T_{\rm seg}=3.3\times10^{-4}$~Hz, while shorter GTIs that could not accommodate such segments were retained in full as shorter stretches, contributing only above their respective $1/T_{\rm seg}$ frequency limits. Power spectra from all segments were computed independently and then mapped onto a common uniform Fourier frequency grid, padding with null values below each segment’s $1/T$ cutoff to avoid bias at low frequencies. The resulting power spectra were averaged in the Fourier domain, with each frequency bin weighted by the number of contributing segments. Finally, the combined spectrum was logarithmically rebinned into 30 bins across the accessible frequency range to provide a clearer representation of the broadband variability, and uncertainties in each bin were estimated as the mean power divided by the square root of the number of independent measurements contributing to that bin. The result is displayed in Figure \ref{fig:leahypsd}, which clearly shows a red noise contribution below $2\times10^{-3}$ s, supporting the prescence of short timescale X-ray variability out to 2 days after discovery. 

We note that a significantly shorter variability timescale of $\sim$\,$1$ s was found in the gamma-ray data from \textit{Fermi}/GBM \citep{Neights2025}. However, we caution that the low count rate of the \textit{NuSTAR} data cannot probe 1 s variability timescales and we cannot determine whether the lack of identified X-ray variability on shorter timescales during our observation is a physical property of the source.

\begin{figure}
    \centering
\includegraphics[width=1.0\columnwidth]{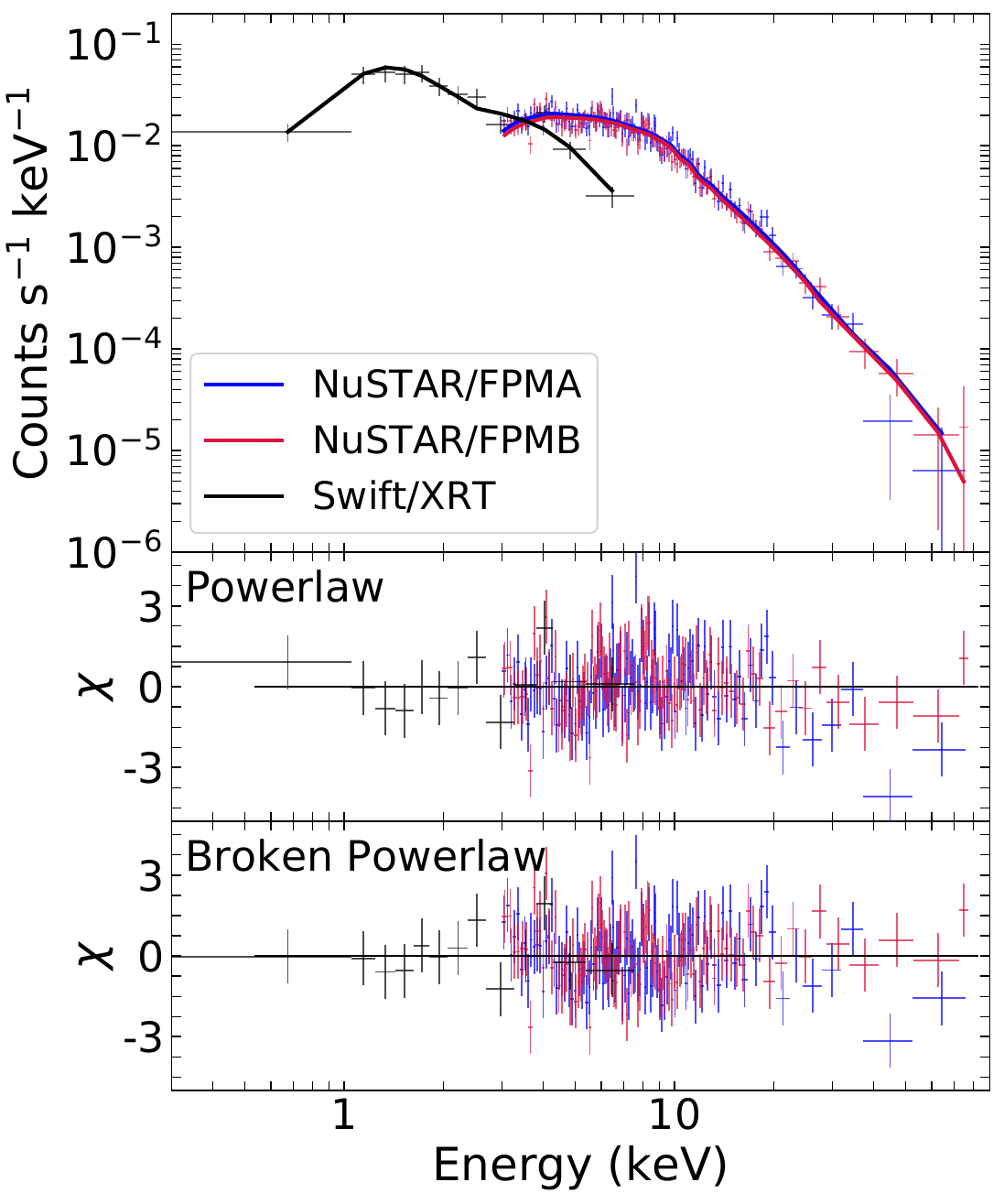}
    \caption{Joint spectral fit including \textit{Swift}/XRT (black) and \textit{NuSTAR} FPMA (blue) and FPMB (red) in the $0.3$\,$-$\,$30$ keV energy range. The middle panel displays the fit residuals for an absorbed powerlaw model, and the bottom panel shows the fit residuals for an absorbed broken powerlaw. The data has been rebinned within \texttt{XSPEC} for visualization purposes.
    }
    \label{fig:xrayspec}
\end{figure}

\subsection{Spectral Modeling}
\label{sec:spec}

\subsubsection{X-ray Spectra}

We obtained multi-epoch \textit{NuSTAR} observations with simultaneous \textit{Swift}/XRT coverage (Figure \ref{fig:xrt+nustar}), enabling spectral modeling at a mid-time of $1.58$ days post-trigger across the $0.3$\,$-$\,$79$ keV band. In subsequent epochs, the rapid fading of the X-ray counterpart leads to diminished spectral quality in both instruments, so we restrict our joint spectral analysis to the first epoch only. We additionally performed a fit to the final two epochs of \textit{NuSTAR} data by themselves.

We modeled all time-averaged X-ray spectra with \texttt{XSPEC v12.14.0} \citep{Arnaud1996} using the Cash statistic \citep[C-stat;][]{Cash1979} with the interstellar medium (ISM) abundance table set following \citet{Wilms2000} and photoelectric absorption cross-sections set following \citet{Verner1996}. All spectra were grouped to a minimum of 1 counts per bin. We performed a fit with an absorbed powerlaw model (\texttt{constant*tbabs*ztbabs*pow}) to the \textit{NuSTAR} (FPMA and FPMB) and simultaneous \textit{Swift}/XRT data. The normalization of FPMA was fixed to unity and we allowed the normalization of FPMB to vary to account for cross-calibration uncertainty. The constant normalization of FPMB varied by $\lesssim3\%$ compared to unity, likely due to a rip in the multi-layer insulation of FPMB \citep[][]{Madsen+2020}.

In all fits we have frozen the Galactic hydrogen column density to $N_\textrm{H,gal}$\,$=$\,$3.34\times 10^{21}$ cm$^{-2}$ \citep{Willingale2013}. The data were modeled between $0.3-10$ keV for Swift and $3-79$ keV for NuSTAR. Using an absorbed powerlaw model, we find an intrinsic hydrogen column density $N_\textrm{H,z}$\,$=$\,$(6.2\pm0.8)\times 10^{22}$ cm$^{-2}$ and X-ray photon index $\Gamma$\,$=$\,$1.82\pm0.03$ with Cstat/dof\,$=$\,$1277/1390$. The spectral fit and fit residuals are shown in Figure \ref{fig:xrayspec}. We do not find evidence for spectral features\footnote{We note the presence of residuals at $\sim$\,8 keV in FPMB, which were determined to be noise (Brian Grefenstette, private communication) and not a real feature.}, such as emission or absorption lines or reflection features. This further supports an extragalactic scenario for GRB 250702B as it would be peculiar for an X-ray binary in outburst.

We performed an additional fit to the first 24 hours of XRT data. The spectrum was extracted using the XRT Build Products tool. We similarly modeled the data using an absorbed powerlaw \texttt{tbabs*ztbabs*pow}. We find an intrinsic hydrogen column density of $N_\textrm{H}$\,$=$\,$(3.5^{+1.2}_{-1.1})\times 10^{22}$ cm$^{-2}$ with photon index $\Gamma$\,$=$\,$1.57\pm0.17$. We note that the the derived ECF (\S \ref{sec:XRT}) of $8.68\times10^{-11}$ erg cm$^{-2}$ cts$^{-1}$ differs by $\sim$\,$5\%$ compared to that derived by the automatic LSXPS tools ($9.07\times10^{-11}$ erg cm$^{-2}$ cts$^{-1}$) when not including an intrinsic absorber at $z$\,$=$\,$1.036$ \citep{Gompertz2025}.  
In any case, the derived photon index from \textit{NuSTAR} ($\Gamma$\,$=$\,$1.80\pm0.03$) is steeper than derived from the XRT data \citep[\S \ref{sec:XRT};][]{kenneagcn}. While this is only a $\sim$\,$1\sigma$ deviation, we note that the initial EP/FXT observation ($\sim$\,$0.5$ d), occurring near-simultaneously to the first orbit of XRT data, yielded a photon index of $\Gamma$\,$=$\,$1.57\pm0.07$ \citep{2025GCN.40917....1C}, which is a more significant deviation. However, it is necessary to note that these other spectra occur at earlier times (0.5 vs 1.6 d) and we can neither confirm or rule out an evolution of the photon index. 

In order to explore whether the steeper index can be explained by a spectral break within the \textit{NuSTAR} band, we further tested a broken powerlaw model (\texttt{bknpow}). The spectral break energy $E_\textrm{br}$ is measured in the observer frame. This yields $N_\textrm{H,z}$\,$=$\,$(5.0\pm0.8)\times 10^{22}$ cm$^{-2}$ and photon indices of $\Gamma_1$\,$=$\,$1.64\pm0.07$ and later photon index of $\Gamma_2$\,$=$\,$1.97\pm0.06$ with a break at $E_\textrm{br}$\,$=$\,$8.3^{+0.8}_{-0.5}$ keV. The fit has Cstat/dof\,$=$\,$1260/1388$. We then added the requirement that the deviation between the initial photon index $\Gamma_1$ and latter photon index $\Gamma_2$ is 0.5 as expected from synchrotron radiation with a cooling break in the X-ray band \citep[$\Delta\beta$\,$=$\,$0.5$;][]{Sari1998,Granot2002}. We find $N_\textrm{H,z}$\,$=$\,$(4.4\pm0.7)\times 10^{22}$ cm$^{-2}$, $\Gamma_1$\,$=$\,$1.55\pm0.04$, and $E_\textrm{br}$\,$=$\,$8.4\pm0.6$ keV for Cstat/dof\,$=$\,$1263/1389$. The break is identified even if we restrict the X-ray spectra to be fit at only $<$\,$30$ keV. We used the Akaike Information Criterion (AIC) to asses the preferred model between a powerlaw and broken powerlaw. The broken powerlaw model is favored with $\Delta$AIC\,$=$\,$-12$. Spectral breaks in the X-ray range are a property of synchrotron radiation \citep{Granot2002} and are found regularly in GRBs \citep[e.g.,][]{Filgas2011,Troja12}. An evolving spectral break was also identified in the relativistic TDE AT2022cmc \citep{Yao2024}. 

That said, the evidence is not overwhelming and a single powerlaw provides an adequate fit to the data (Figure \ref{fig:xrayspec}) without a very noticeable change in the residuals. The benefit of the broken powerlaw is that it solves the early (potential) discrepancy with the initial EP/FXT photon index, and is physically motivated. However, we cannot rule out variations in the photon index with time that could allow for a change of 1.6 to 1.8 in photon index between 0.5 and 1.6 days. 

We also modeled the second and third epochs (with mid-times of 5.70 and 9.98 d; Figure \ref{fig:batlimits}) of \textit{NuSTAR} data (Table \ref{tab: observationsXray}). In the second epoch we derive a powerlaw photon index of $\Gamma$\,$=$\,$1.68\pm0.19$ for fixed $N_\textrm{H,z}$\,$=$\,$5\times10^{22}$ cm$^{-2}$ with Cstat/dof\,$=$\,$126/164$. The column density was fixed as it is unconstrained by \textit{NuSTAR} alone. We find no evidence for a spectral break, and modeling with a broken powerlaw places the break energy far outside of the observed spectral range. We note that the spectrum has very low signal-to-noise at $>$\,$20$\,$-$\,$30$ keV. In the third epoch the source is significantly fainter and we restrict our spectral modeling to $<$\,$30$ keV. We derive $\Gamma$\,$=$\,$1.86\pm0.35$ with Cstat/dof\,$=$\,$93/95$.


\begin{figure}
    \centering
\includegraphics[width=1.0\columnwidth]{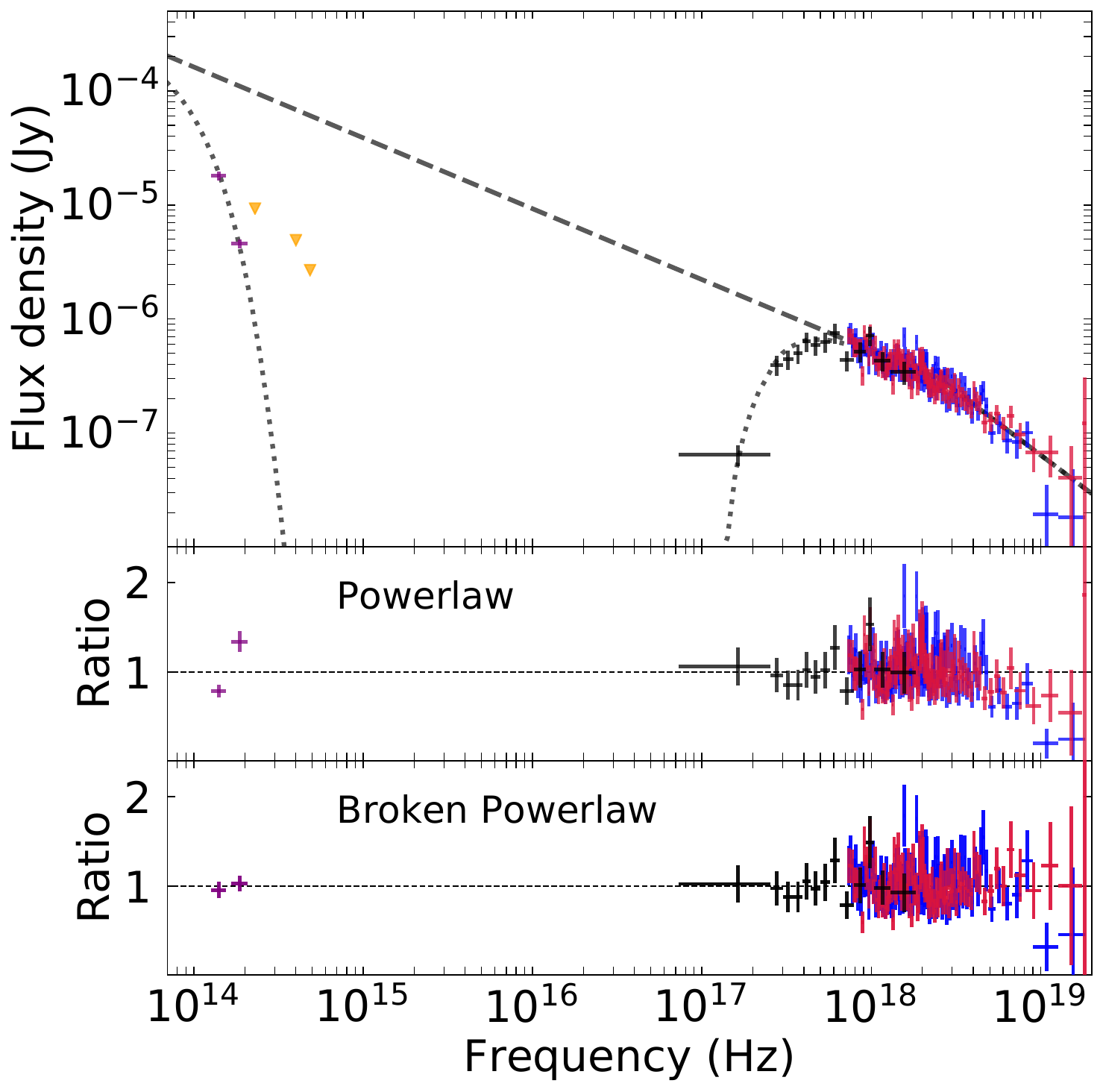}
    \caption{Observed X-ray to near-infrared spectral energy distribution with \textit{Swift}/XRT (black), \textit{NuSTAR} FPMA (blue) and FPMB (red), VLT HAWK-I \citep[$HK$ - purple;][]{Levan2025}, and optical upper limits \citep[$riJ$ - orange;][]{Carney2025}. The middle panel displays the fit residuals for an absorbed powerlaw model, and the bottom panel shows the fit residuals for an absorbed broken powerlaw. 
    }
    \label{fig:broadbandspec}
\end{figure}

Additionally, we modeled the spectra of the two \textit{Swift}/BAT survey detections (see \S \ref{sec:BAT} and Table \ref{tab:batlimits}) using the $\chi^2$ statistic, as the mask-weighted counts in the spectra follow Gaussian statistics. The spectra are binned into the eight, default survey analysis energy bins ranging from $14$\,$-$\,$195$ keV. Using a single powerlaw model, we find a photon index $\Gamma$\,$=$\,$1.44\pm0.38$ and a $14$\,$-$\,$195$ keV flux of $(4.95\pm1.14)\times 10^{-9}$ erg cm$^{-2}$ s$^{-1}$ for the first detection occurring shortly after the ``E'' burst (Table \ref{tab:triggertimes}),  and $\Gamma$\,$=$\,$1.77\pm0.58$ and a $14$\,$-$\,$195$ keV flux of $(1.08\pm0.39)\times 10^{-8}$ erg cm$^{-2}$ s$^{-1}$ for the second detection starting at $\sim$\,0.28 days post-trigger (measured from the ``D'' burst). We do not find any significant evidence for a high-energy cutoff to the powerlaw in either detection. As the BAT detection is not persistent and clearly fluctuates above earlier $5\sigma$ upper limits (Figure \ref{fig:batlimits}), these detections may be associated to late-time central engine activity (e.g., flares). If interpreted as continued prompt emission, it would further extend the duration of the prompt phase \citep[for a discussion, see][]{Neights2025}.

\subsubsection{X-ray to Near-infrared Spectral Modeling}
\label{sec:xoirSED}

\begin{figure*}
    \centering
    \includegraphics[width=1.0\columnwidth]{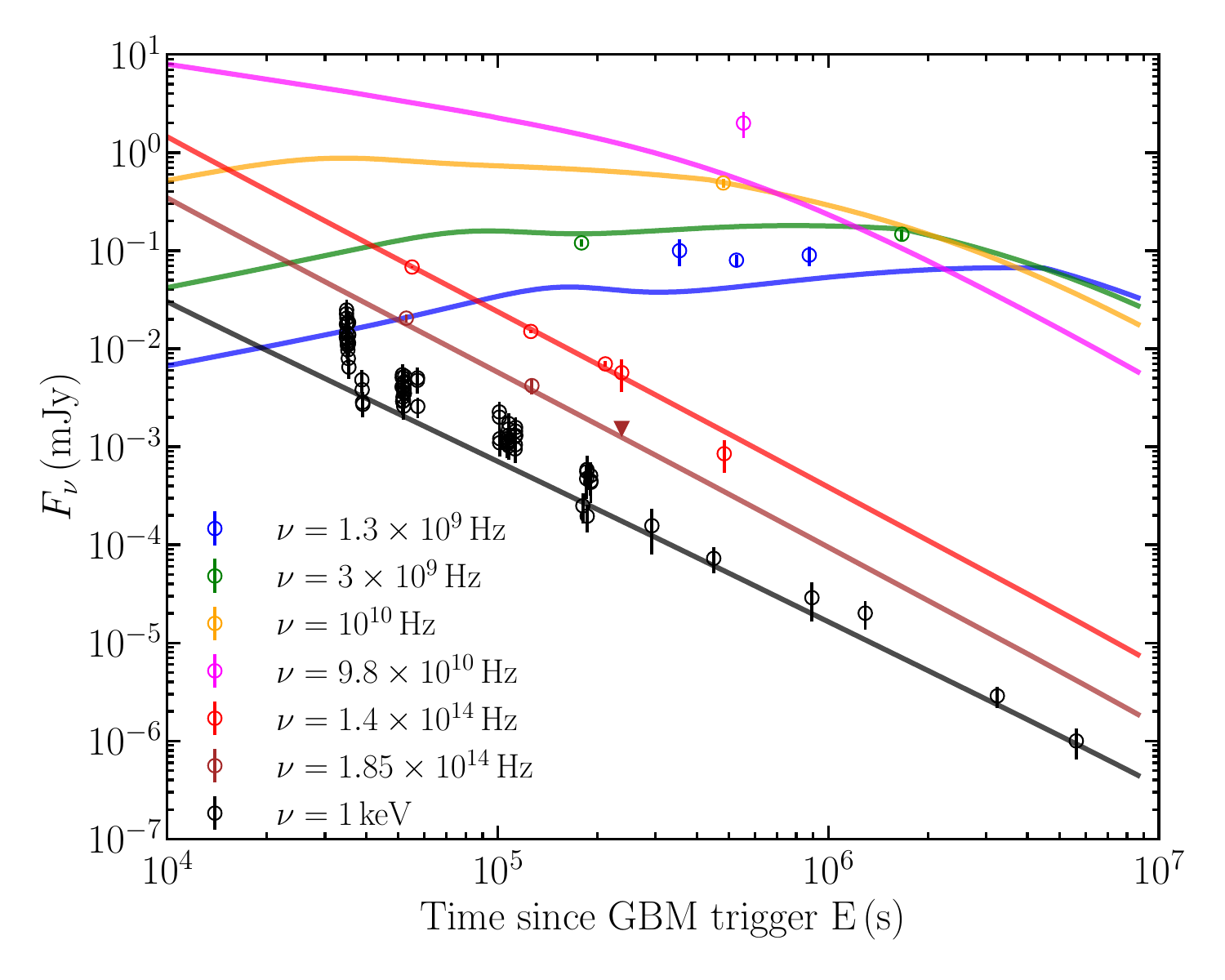}
    \includegraphics[width=1.0\columnwidth]{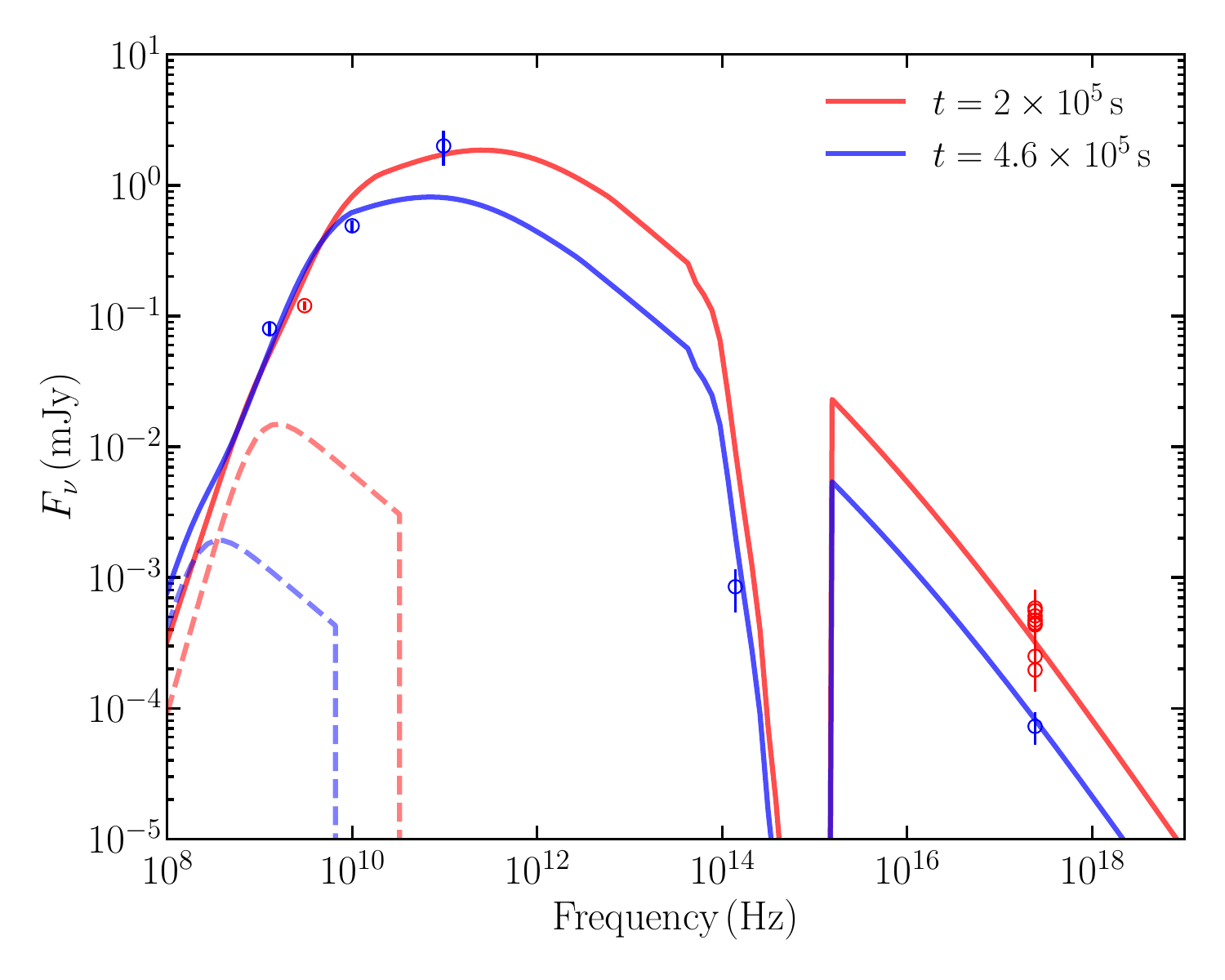}
    \caption{\textbf{Left:} Forward (FS) and reverse shock (RS) afterglow model fit to multi-wavelength observations. The start time $T_0$ is the GBM ``E'' burst. The MCMC parameter posterior distributions are shown in Figure\,\ref{fig:afterglowcorner} that only considers the last five X-ray detections for the X-ray lightcurve. The lightcurves are obtained for the following model parameters that keep the X-ray afterglow emission below the flaring X-ray emission at early times: $\theta_j=7.8\times10^{-3}$\,rad, $E_{\rm k,iso}=6.2\times10^{54}$\,erg, $\Gamma_0=306$, $n_0=0.6$ at $R_0=10^{18}$\,cm, $p=2.1$, $\epsilon_e=0.2$, $\epsilon_B=10^{-3}$, $k=2$. The same dynamical parameters are used for the RS emission with these shock microphysical parameters: $p_{\rm RS}=2.2$, $\varepsilon_{e,\rm RS}=0.1$, $\varepsilon_{B,\rm RS}=10^{-3}$, $t_{\rm grb}=10^3$\,s, and $g_{\rm RS}=2.0$. 
    The start time $T_0$ is taken as GRB\,250702E (Table \ref{tab:triggertimes}). 
    \textbf{Right:} Model spectrum comparison with observations at two different times, with RS spectrum shown with dashed lines.  
    }
    \label{fig:afterglow}
\end{figure*}

While the X-ray and near-infrared emission can have various explanations when considered between the progenitor classes of ultra-long GRBs and relativistic TDEs, here we evaluate their consistency in coming from the same emission component, namely an external forward shock (see also \S \ref{fig:afterglow}). 
To do this, we investigated the X-ray to near-infrared spectral energy distribution (SED) of GRB 250702B using near-infrared ($HK_s$ bands) detections reported by \citet{Levan2025} that were obtained near simultaneously with the mid-time of our initial \textit{NuSTAR} observation (1.58 versus 1.60 days). The source is extremely extinct due both to dust within our Galaxy ($A_\textrm{V}$\,$=$\,$0.847$ mag; \citealt{Schlafly2011} and additional intrinsic dust within its host \citep[$A_\textrm{V,z}$\,$>$\,$10$ mag;][]{Levan2025,Carney2025}, and only displayed detections in the near-infrared $H$ and $K_s$-bands despite deep limits at optical wavelengths \citep{Carney2025}. We modeled the X-ray to near-infrared spectrum with \texttt{XSPEC} using both an absorbed powerlaw model (\texttt{tbabs*ztbabs*zdust*pow}) and a broken powerlaw model (\texttt{tbabs*ztbabs*zdust*bknpow}). The near-infrared data were corrected for Galactic extinction prior to fitting \citep{Schlafly2011}. For intrinsic dust within the host galaxy, we utilized a Small Magellanic Cloud (SMC)
extinction law with $R_\textrm{V}$\,$=$\,$2.93$ \citep{Pei1992} and a redshift $z$\,$=$\,$1.036$ \citep{Gompertz2025}. For the broken powerlaw, we similarly required that $\Gamma_2-\Gamma_1$\,$=$\,$0.5$. The X-ray data were modeled using the Cash statistic and near-infrared data using chi-squared. The total fit statistic was minimized by the fit. 

For a single powerlaw between $10^{14}$ to $10^{19}$ Hz, we derive $N_\textrm{H,z}$\,$=$\,$(5.3\pm0.7)\times 10^{22}$ cm$^{-2}$, $\Gamma_\textrm{XIR}$\,$=$\,$1.72\pm0.04$, and reddening $E(B-V)_z$\,$=$\,$3.07\pm0.20$ mag with a total fit statistic of $1300$ for $1391$ dof. The reduced chi-squared is $\chi^2$/dof\,$=$\,$1.32$. For  $R_\textrm{V}$\,$=$\,$2.93$, the derived visual extinction is extremely large $A_\textrm{V,z}$\,$=$\,$R_\textrm{V}\times E(B-V)_z$\,$=$\,$9.0\pm0.6$ mag. The required visual extinction is a strong function of the dust law, assumed redshift, and near-infrared to X-ray spectral index. The lack of optical or near-infrared detections at frequencies higher than $H$-band is also a factor. 

A broken powerlaw fit yields $N_\textrm{H,z}$\,$=$\,$(3.0\pm0.6)\times 10^{22}$ cm$^{-2}$, $\Gamma_\textrm{XIR,1}$\,$=$\,$1.64\pm0.03$, $E_\textrm{br}$\,$=$\,$10.9^{+1.5}_{-1.1}$ keV, and reddening $E(B-V)_z$\,$=$\,$2.25\pm0.20$ mag with a total fit statistic of $1264$ for $1390$ dof. The reduced chi-squared is $\chi^2$/dof\,$=$\,$1.14$. The visual extinction is $A_\textrm{V,z}$\,$=$\,$6.6\pm0.6$ mag. This fit and residuals are shown in Figure \ref{fig:broadbandspec}. Based on the total fit statistic and the residuals, we favor the broken powerlaw fit as providing a better description of the X-ray to near-infrared data at $\sim$\,1.6 d. The derived spectral index is also consistent with the initial XRT and EP/FXT data. 

We caution that (while clearly significant) the exact dust contribution is uncertain and here depends on the assumption that both near-infrared and X-rays originate from the same emission component. However, this is unclear and not required by the data. It may simply be possible due to the additional free parameter of dust, which can modify the observed near-infrared spectral index to a broad range of intrinsic spectral indices. In addition, the X-rays are flaring at the time the spectra were acquired (which may require slightly more dust to match the higher flux level), but this is not necessarily going to impact the near-infrared emission, especially if that originates from an external shock.

\subsection{Broadband Afterglow Modeling}
\label{sec:afterglow}

The interaction between the local environment and the relativistic jets launched by both GRBs and jetted TDEs produce a broadband (radio to gamma-ray) synchrotron afterglow \citep{Sari1998,Granot2002}. 
We modeled the multi-wavelength lightcurves of GRB 250702B using the standard fireball model \citep[][]{Meszaros1997,Wijers1999,Granot2002} using the numerical methods of \citet{GG2018,Gill2023} to compute the forward and reverse shock (RS) emission from a non-spreading top-hat jet. 

The forward shock (FS) model depends on the isotropic-equivalent kinetic energy of the outflow $E_\textrm{kin}$ at the jet's core, initial bulk Lorentz factor $\Gamma_0$ at the jet's core, the jet's core half-opening angle $\theta_\textrm{j}$, surrounding density $n_{\rm ext}$, slope of the electron's powerlaw energy distribution $p$, and the microphysical parameters $\varepsilon_e$ and $\varepsilon_B$  that determine the partitioning of shock energy in electrons and magnetic fields, respectively. The RS depends on the additional parameters $p_\textrm{RS}$, $\varepsilon_{e, \textrm{RS}}$ and $\varepsilon_{B,\rm RS}$, and $g_{\rm RS}$ \citep[see][]{Gill2023}. To explore the nature of the surrounding environment, the powerlaw slope $k$ and normalization $n_0$ at $R_0=10^{18}$\,cm of the external density profile, $n_\textrm{ext}$\,$=$\,$n_0(R/R_0)^{-k}$, are left as free parameters. Another parameter that must be specified when modeling RS emission is the GRB duration, typically $T_{90}$, that determines the initial radial width of the ejecta, 
$\Delta_0 = cT_{90}/(1+z)$, to see if a thin-shell (Newtonian RS) or thick-shell (relativistic RS) solution is obtained \citep{Sari1995}. With multiple emission episodes in this GRB, that are also separated by long quiescent periods, the initial radial width of the ejecta is unclear. Therefore, we assume a thin-shell scenario in which the RS and FS afterglow lightcurve peak times do not exactly coincide and the FS afterglow peaks later. 
The FS+RS afterglow model therefore depends on 12 free parameters.

We supplemented our X-ray dataset with radio data compiled from GCN Circulars \citep{2025GCN.41059....1A,2025GCN.41147....1G,2025GCN.41145....1B,2025GCN.41061....1T,2025GCN.41053....1S,2025GCN.41054....1A}, the multi-wavelength data presented in \citet{Levan2025}, and additional optical/near-infrared (OIR) photometry from \citet{Carney2025}. The OIR photometry was corrected for Galactic extinction prior to fitting \citep{Schlafly2011}, and an additional intrinsic extinction component using an SMC dust law \citep{Gordon2024} was added. We note that the optical and near-infrared upper limits \citep{Carney2025,Levan2025} are included in the fit, but are not displayed in the figures for clarity. We used the \texttt{emcee} \citep{emcee} package to fit the broadband afterglow first with a FS only model. We then used this solution to add the RS afterglow by manually considering the shock microphysical parameters that yielded a close match to the observations. 

We set the start time $T_0$ of the afterglow as the GBM ``E'' burst (GRB 250702E occurs only $\sim$\,0.134 d after the initial trigger GRB 250702D; Table \ref{tab:triggertimes}), see \S \ref{sec:stellarmass} for further discussion of the likely start time. Additional fits were performed using the ``D'' burst as $T_0$, and are presented in Appendix \ref{fig:afterglowcorner-D}. They do not modify our overall conclusions and are broadly consistent with the fits using the ``E'' burst as $T_0$. We performed another fit using a $T_0$ that is 1 day prior to the ``D'' burst, but this does not provide suitable solutions to the multi-wavelength lightcurve. 

In our fits we have assumed that the near-infrared data originates from the external shock (although this is not guaranteed; \citealt{Levan2016}). We have further assumed that the initial X-ray data at $<$\,$2\times10^5$ s (relate to the ``E'' burst) is significantly contaminated by excess emission from late-time central engine activity. This is supported by the presence of short term variability in the \textit{NuSTAR} data out to $\sim$\,$2$ d (see \S \ref{sec:var} and Figures \ref{fig:lcrate} and \ref{fig:xrt+nustar}). We therefore have only included the final X-ray points at $>$\,$2\times10^5$ s in our modeling. It is worth pointing out that the lack of variability inferred from the late-time X-ray data is potentially due simply to the low count rates at that phase. In any case, we note that including all X-ray data in the fit does not significantly modify our afterglow solutions or the inferred parameters.

The best-fit afterglow model is shown in Figure \ref{fig:afterglow} and a corner plot is displayed in Figure \ref{fig:afterglowcorner}. Regardless of redshift the conclusion is that the afterglow requires a large (isotropic-equivalent) kinetic energy, an ultrarelativistic jet with a small jet opening angle propagating in a wind-like density profile. The high Lorentz factor is required such that the jet decelerates prior to our first X-ray observations, and the small jet opening angle allows for the afterglow to be in a steep decaying post-jet-break phase throughout all our multi-wavelength observations. We discuss these points further in \S \ref{sec:stellarmass}.

We find that the inferred visual extinction from the afterglow model is slightly smaller than derived using \texttt{XSPEC} (\S \ref{sec:spec}). The main conclusion that a significant dust column must lie along the line-of-sight is the same. We find that the difference is likely due to variations in the parametrization of the SMC dust law \citep{Pei1992,Gordon2024}, and due to the flatter spectral slope favored by the afterglow model which naturally requires a smaller dust correction. Our broadband spectral modeling (Figure \ref{fig:broadbandspec}) favors a larger value of $p$\,$=$\,$2.28\pm0.06$\footnote{This value is derived from spectral fits that including data that exhibits flaring episodes (Figure \ref{fig:xrt+nustar}) and therefore may differ from the spectral energy distribution of the external shock afterglow.} compared to the value derived from our afterglow modeling ($p$\,$=$\,$2.11\pm0.03$). The afterglow model prefers a smaller value of $p$ so as to provide a better match to the post-jet-break temporal decay slope (see discussion in \S \ref{sec:stellarmass}). The difference in the $p$ values is likely coming from the integration over the jet's surface of equal arrival times that is done in the afterglow model compared to the spectral fit which simply assumes line-of-sight emission. However, these two values of $p$ only deviate by $\sim$\,$2\sigma$ and neither is  inconsistent with the data.

We note that the afterglow model (Figure \ref{fig:afterglow}; right panel) places the cooling frequency $\nu_\textrm{c}$ in or near the X-ray band at a consistent frequency with our observations of a spectral break based on an analysis of the X-ray to near-infrared spectral energy distribution (Figures \ref{fig:xrayspec} and \ref{fig:broadbandspec}). This provides support to the presence of a break in the spectral energy distribution. As the jet model used here is non-spreading\footnote{The condition $\Gamma_0\theta_\textrm{j}$\,$\geq$\,$1$ is enforced to prevent the jet from spreading laterally in the coasting phase \citep{Rhoads1997,Rhoads1999}, which would break its approximation as a spherical outflow. We note that this pushes $\Gamma_0$ to high (but not unreasonable) values, see \S \ref{sec:stellarmass} for discussion. Even relaxing this requirement, we require Lorentz factors in the range $\Gamma_0$\,$\gtrsim$\,$20$\,$-$\,$50$.}, even after the jet-break the cooling frequency continues to evolve with the same dependence $\nu_\textrm{c}$\,$\propto$\,$t^{(3k-4)/2(4-k)}$, which is $\sim$\,$t^{0.3}$ to $t^{0.5}$ for our preferred environment. We see this in the right panel of Figure \ref{fig:afterglow}. Evolving the observed $\sim$\,$10$ keV break at 1.6 d (observer frame) back in time places it at the upper end of the XRT band ($\sim$\,$6$\,$-$\,$7$ keV) during the first orbit of XRT data. It would also evolve the cooling frequency to $\sim$\,$17$\,$-$\,$25$ keV by the time of our second \textit{NuSTAR} observation, likewise allowing it to be missed due to limited counts (low signal-to-noise) above those energies. If instead the jet is spreading laterally, then the cooling frequency would be frozen in place during the course of all our observations, and even more easily missed in the initial XRT spectra. An additional point is that the cooling frequency is likely smooth over at least an order of magnitude in frequency, such that minor changes in the frequency are not noticeable on the short timescales probed by XRT and \textit{NuSTAR}. 

We find that the radio data, except for the ALMA observations at $\nu=98$\,GHz \citep{2025GCN.41059....1A}, are well explained by our model and are dominated by the FS emission. This is in contrast with the modeling of \citet{Levan2025} who attribute all of the radio afterglow to RS emission. Continued radio monitoring is crucial to fully measure the evolution of the radio emission and accurately constrain the FS and RS contributions.

\section{The Peculiar High-Energy Properties of GRB 250702B}

GRB 250702B is a peculiar ultra-long X-ray and gamma-ray transient that does not neatly fit into any class of known high-energy transients (see \S \ref{sec:intro} for a summary). Its properties most closely resemble the classes of ultra-long gamma-ray bursts or relativistic jetted tidal disruption events. Here we focus on a few of its relevant high-energy properties in this context. In \S \ref{sec:interp}, we consider both of these possible progenitor scenarios and explain why its observed properties are not an exact match to either progenitor class. 

\begin{figure*}
    \centering
\includegraphics[width=2.0\columnwidth]{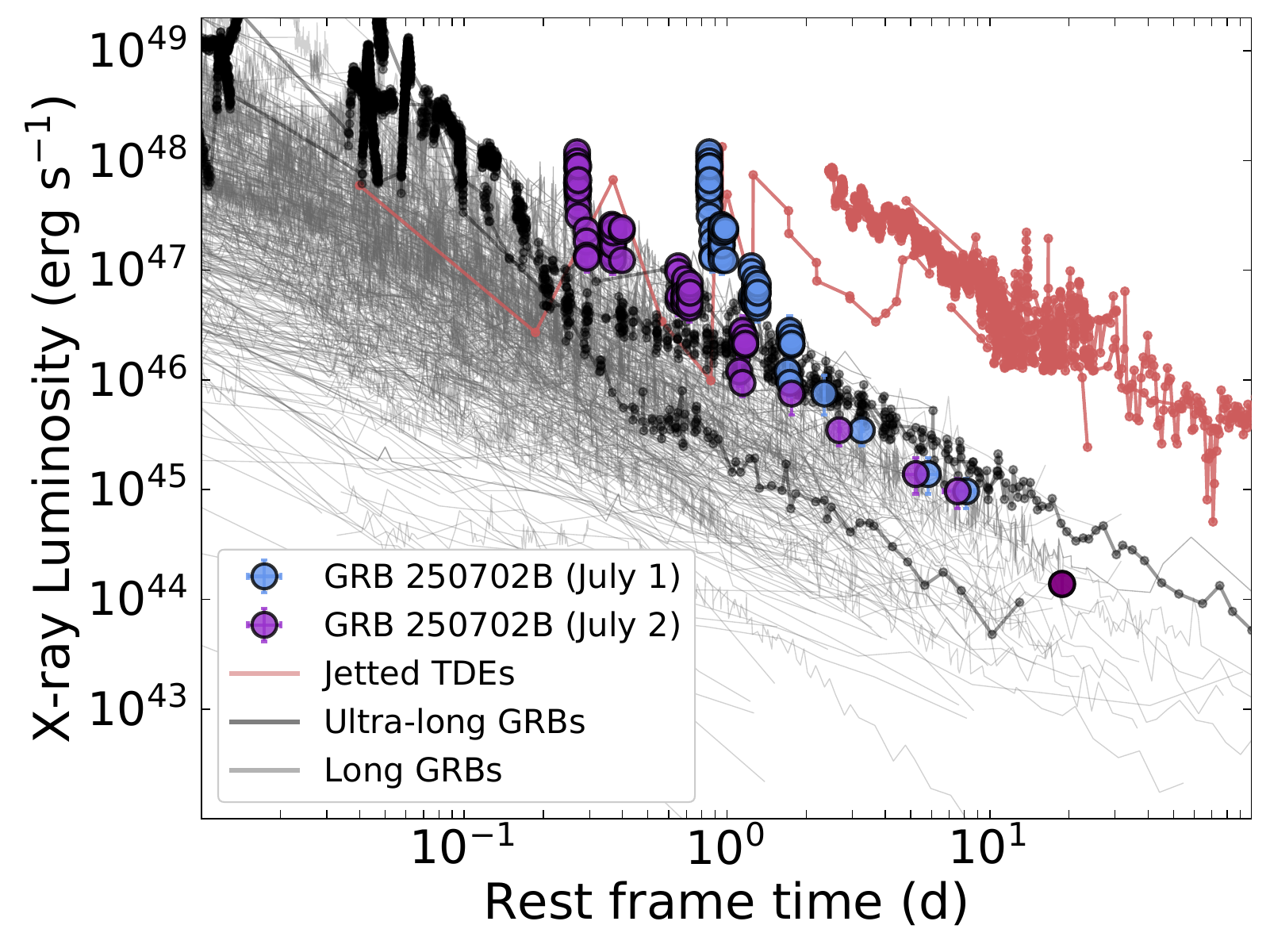}
    \caption{K-corrected rest frame ($0.3$\,$-$\,$10$ keV) X-ray lightcurves of \textit{Swift} long duration GRBs, ultra-long GRBs \citep[GRBs 111209A, 121027A, 130925A;][]{Levan2014,Evans2014-ultralong}, and relativistic jetted TDEs \citep{Bloom2011,Cenko2012,Brown2015,Pasham2023} compared to GRB 250702B for a few different $T_0$. For the July 2nd start time (purple), we have used the trigger time of GRB 250702D as $T_0$ (Table \ref{tab:triggertimes}). For the July 1 start time (blue), we use the beginning of the first EP detection window (Hui Sun, private communication). 
    }
    \label{fig:lclum}
\end{figure*}

\subsection{Ultra-long Prompt Emission}
\label{sec:promptdisc}

GRB 250702B displayed unusual prompt emission properties. For starters, it triggered a variety of high-energy monitors multiple times on July 2, 2025 with a gamma-ray duration in excess of $25$ ks \citep[see][for a detailed discussion]{Neights2025}. Additional high-energy detections were uncovered on July 1, 2025 by the \textit{Einstein Probe} and \textit{Fermi}, extending the prompt emission duration to longer than a day \citep{EPgcn,EP250702a-arxiv,Neights2025}. In fact, \textit{Swift}/BAT\footnote{As these \textit{Swift}/BAT detections occur prior to the start of soft X-ray observations with XRT, their interpretation as either additional prompt episodes or X-ray flares is unclear, though see \citet{Neights2025}.} (\S \ref{sec:BAT}; Figure \ref{fig:batlimits}) also detected the source in Survey mode on two additional instances that followed the \textit{Fermi} triggers, further pushing the prompt duration\footnote{We note that these \textit{Swift}/BAT detections are included in the 25 ks determination of the prompt emission duration by \citet{Neights2025}.} (and therefore central engine activity) to extreme values. 

As such, the exact start time of the event is unclear, though our analysis favors an afterglow onset timed with the July 2 triggers (Figure \ref{fig:temporalfit}; right panel). The delay between the July 1 and July 2 detections is unusual for GRBs, which have never before been found to have such long delay between the early ``precursor'' emission and the main burst episode. It is possible that this is a selection bias due to the lack of wide-field soft X-ray monitors like the \textit{Einstein Probe} in the past. It is therefore possible that some previously known ultra-long GRBs also show such sub-energetic precursor emission that went missed. However, the \textit{Einstein Probe} lightcurve shows a smooth rise to peak emission \citep{EP250702a-arxiv}, coinciding with the final \textit{Fermi} trigger \citep{Neights2025}, which excludes its interpretation as a ``precursor''. In any case, such underlying soft X-ray emission would easily be missed in previous ultra-long events. 


While these properties are atypical for GRBs, they have likewise never been observed from a relativistic TDE. While relativistic TDEs have been uncovered in binned gamma-ray data \citep{Cenko2012,Brown2015}, or 1000 s long image triggers \citep{Sakamoto2011J1644,Burrows2011}, they have never triggered high-energy monitors in the standard ``rate'' trigger method that detects typical GRBs, and they have definitely not done so multiple times in a 24 hour window. As near all-sky gamma-ray monitors have been operating for decades, multiple gamma-ray triggers would be more difficult to miss in the past than early soft X-ray emission that could only have been found by the \textit{Einstein Probe} which launched only in the last 2 years \citep{Yuan2025}. A detailed sub-threshold gamma-ray search in the day(s) preceding ultra-long GRBs is likely necessary to further investigate this possibility. 

An important clue can be obtained from the isotropic-equivalent gamma-ray energy $E_{\gamma,\textrm{iso}}$. The integrated energy obtained by \textit{Konus-Wind} \citep{konusgcn} is $E_{\gamma,\textrm{iso}}$\,$=$\,$(1.4^{+0.4}_{-0.2})\times10^{54}$ erg in the $1$\,$-$\,$10,000$ keV energy range \citep[see also][]{Neights2025}. This is at the high end of the isotropic-equivalent energies found in classical GRBs and ultra-long GRBs (see Figure \ref{fig:Egamma-Lx11}). This is also substantially higher than the energy released ($\approx$\,$10^{51}$ erg; $1$\,$-$\,$10,000$ keV) by Sw J1644+57 during its initial \textit{Swift}/BAT detection \citep[GRB 110328A;][]{Sakamoto2011J1644}. The individual energetics of the ``DBE'' events yield energies a factor of $10\times$ higher \citep{Neights2025}. However, Sw J1644+57 triggered BAT multiple times over the next two days before BAT triggers were disabled for the source \citep{Burrows2011}. We estimate the gamma-ray energy during this peak phase by adopting the average fluence reported by \citet{Burrows2011} over this 2 d window and adopting the spectral shape (peak energy $\sim$\,$70$ keV) reported by \citet{Levan2011}. This yields $E_{\gamma,\textrm{iso}}$\,$\approx$\,$(1.0\pm0.2)\times10^{53}$ erg in the $1$\,$-$\,$10,000$ keV. This value is derived during the peak phase ($\sim$\,$2$ d) of Sw J1644+57's emission (in both X-ray and gamma-rays), but Sw J1644+57 was detected by BAT both before and after this peak phase \citep{Burrows2011}. Therefore, our derived energy during this 2 d window is a lower limit to its full gamma-ray energy release. 
We show this lower limit as an orange shaded region in Figure \ref{fig:Egamma-Lx11} for comparison to the energetics of GRB 250702B and classical (short, long, and ultra-long) GRBs. 
Another notable difference to GRB 250702B is that the gamma-ray emission from Sw J1644+57 was substantially softer (peak energy $\sim$\,$70$ keV; \citealt{Levan2011}) than the $>$1 MeV emission observed from GRB 250702B \citep{Neights2025}.

In addition, the \textit{Fermi} data revealed a critical piece of evidence in the minimum variability time. \citet{Neights2025} identified a very short second duration ($\lesssim$\,$1$ s; observer frame) variability timescale, which is a factor of $100$\,$-$\,$1,000$ smaller than observed in relativistic TDEs \citep{Bloom2011,Pasham2023}. Relativistic TDEs, and accreting black holes in general, are limited by the Schwarzschild light crossing time $t_s$\,$=$\,$2R_\textrm{g}/c$, where $R_\textrm{g}$\,$=$\,$GM/c^2$ is the gravitational radius, which depends on the black hole mass. As larger masses lead to longer predicted minimum variability timescales, if a massive black hole is invoked it would require a small $\lesssim 5\times 10^{4} M_\odot$ intermediate-mass black hole. It is also important to note that the light crossing timescale argument is only a lower limit to the minimum variability timescale, and the black hole mass determined through this argument for past relativistic TDEs is larger than the latest analysis based on their jet shutoff timescales \citep{Eftekhari2024}.

However, we note that relativistic TDEs have been compared to a smaller version of a blazar \citep{Bloom2011}, which have been observed to have variability on timescales less than their light crossing times \citep{Giannios2009, BlazVarNature}. The shorter variability timescale observed from blazars has been explained by the ``jets in a jet'' model \citep{JetInJet}. Whether the same magnetic reconnection mechanism can explain the prompt emission of GRB 250702B is outside the scope of this paper, but we caution that this minijet scenario would mean that the short variability timescale does not directly constrain the black hole mass.

In any case, this seconds long variability is typical of that observed in long duration GRBs, and would therefore most obviously favor an interpretation that invokes a stellar-mass progenitor such as in GRBs. As an additional argument, in the collapsar progenitor channel, the central engine for long GRBs is likely a stellar-mass black hole \citep{Woosley1993}. Assuming a $\sim 5 M_\odot$ black hole, the predicted minimum variability timescale based on the Schwarzschild light crossing time is $5\times 10^{-5}$ s, which is significantly smaller (by orders of magnitude) than the minimum variability observed in GRBs \citep[see, e.g., $10^{-2}$\,$-$\,$100$ s;][]{Golkhou2014}. Following this argument, if applied to massive black holes, it would imply the minimum variability from light crossing is significantly smaller than the likely true intrinsic variability that can be observed. We therefore consider the second duration minimum variability of GRB 250702B  \citep{Neights2025} as strong evidence for a stellar-mass progenitor (\S \ref{sec:stellarmass}).

\subsection{X-ray Luminosity and Prompt-Afterglow Correlations}
\label{sec:xrayafterdisc}

The X-ray afterglow luminosity as a function of time is a key diagnostic between different classes of energetic transients. It has been used to initially identify between a GRB or TDE origin \citep[e.g,][]{Bloom2011,Burrows2011,Andreoni2022,Pasham2023}. In Figure \ref{fig:lclum}, we compare the rest frame X-ray lightcurve measured by \textit{Swift} and \textit{Chandra} to classical long GRBs\footnote{\url{https://www.swift.ac.uk/xrt_curves/}}, ultra-long GRBs, and relativistic jetted TDEs. The striking feature is the agreement with the observed luminosities and decay rates of \textit{Swift} long GRBs.

While we have taken the initial \textit{Fermi} trigger (``D'' burst) as the $T_0$ for this figure, it is possible that the start time is prior to this, despite our temporal modeling favoring otherwise (Figure \ref{fig:temporalfit}). However, regardless of the start time taken for the event, we are unable to reproduce the long-lived behavior observed in known relativistic TDEs. As shown in Figure \ref{fig:temporalcornerbkn}, an earlier $T_0$ requires a very steep initial decay of the X-rays that then returns to a GRB-like luminosity and decay. This additionally would strongly imply an extreme X-ray luminosity (higher than ultra-long GRBs or any observed GRB) at $>$\,$1$ d, before quickly returning to GRB afterglow-like levels (Figure \ref{fig:lclum}). While ultra-long GRBs show this rapid X-ray decay during the end of their prompt phases, they have all reached afterglow-like behavior a factor of 2 earlier in time in the rest frame (assuming a July 1 onset; Figure \ref{fig:lclum}). For these reasons, a much earlier $T_0$ is hard to believe, especially given the lack of detections in \textit{Swift}/BAT Survey mode data even extending to a month before discovery (\S \ref{sec:BAT}; Figure \ref{fig:batlimits}).

\begin{figure}
    \centering
\includegraphics[width=1.0\columnwidth]{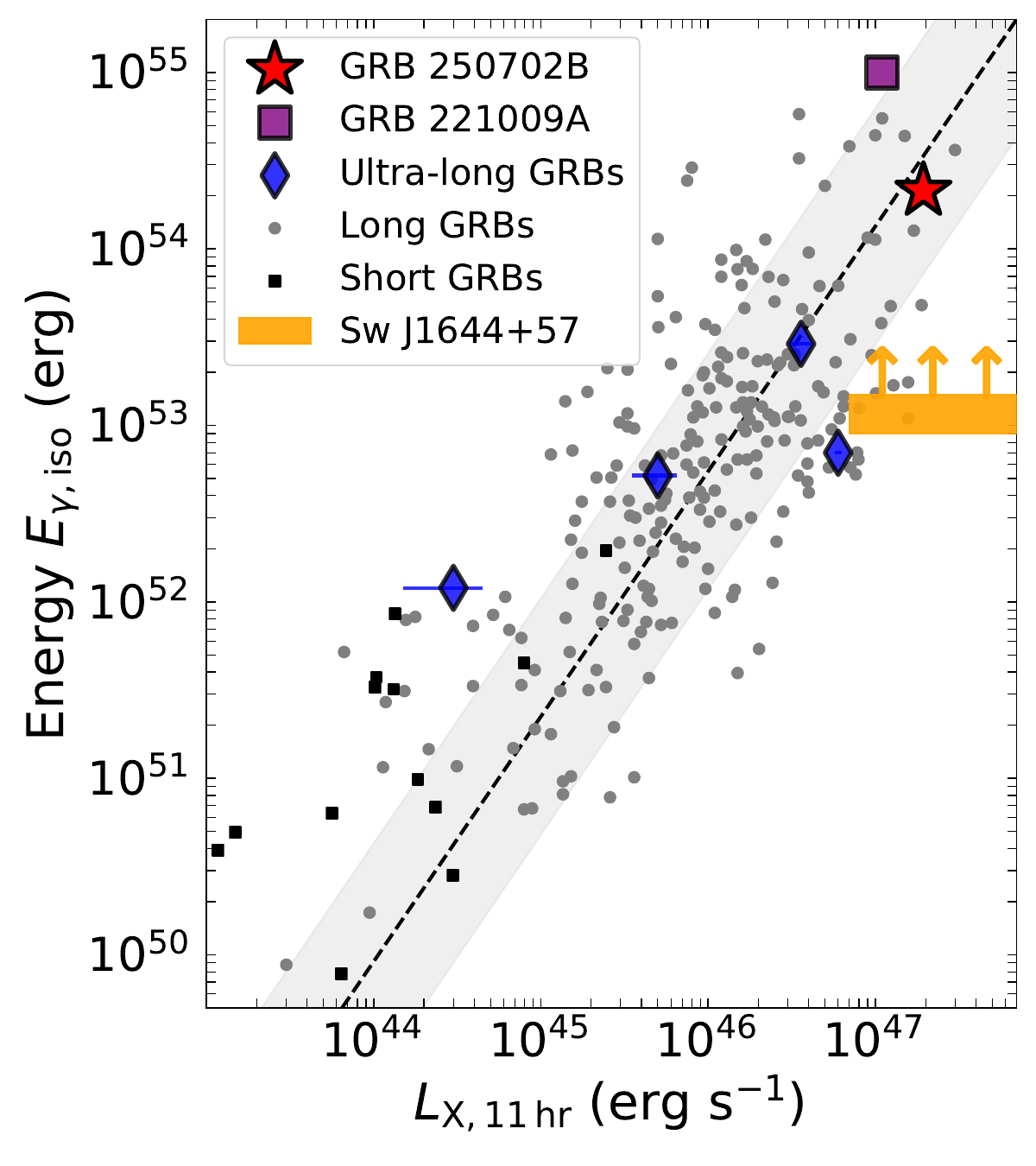}
    \caption{Rest frame X-ray luminosity ($0.3$\,$-$\,$10$ keV) at 11 hours (in the rest frame) from discovery versus gamma-ray energy ($1$\,$-$\,$10,000$ keV). GRB 250702B is shown by a red star. We measure 11 hours from the start of the ``D'' burst (discovery trigger). We show Sw J1644+57 as a shaded region (see text for details; \citealt{Burrows2011}). A sample of short (black squares) and long GRBs (gray circles) are compiled from \citet{Nysewander2009,Berger2014}. We show GRB 221009A as a purple square \citep{OConnor2023,Lesage2023,Burns2023}. A sample of ultra-long GRBs is shown for comparison \citep[GRBs 101225A; 111209A, 121027A, 130925A;][]{Levan2014,Evans2014-ultralong}. The long GRB correlation of $L_\textrm{X,11}$\,$\approx$\,$5.6\times10^{44} (E_{\gamma,\textrm{iso}}/10^{51} \,\textrm{erg})^{0.72}$ erg s$^{-1}$ is shown by a black dashed line \citep{Berger2014}. Reproduced from \citet{OConnor2025}. 
    }
    \label{fig:Egamma-Lx11}
\end{figure}

The range of X-ray afterglow temporal slopes (\S \ref{sec:temporalfit}; Figure \ref{fig:temporalcorner}) is consistent with a range of possibilities, such as fallback accretion for either complete stellar disruption ($t^{-5/3}$) or a post-jet-break afterglow (\S \ref{sec:afterglow}). The X-ray luminosity clearly falls in the typical range for GRB afterglows, and disfavors a classical relativistic TDE (though there are other reasons for this as well, see \S \ref{sec:massiveblackhole} for a discussion). While the exact nature of the X-ray emission (i.e., internal vs external) is uncertain (\S \ref{sec:int-vs-ext}), it is worth noting that this event follows the standard GRB correlations (Figure \ref{fig:Egamma-Lx11}) between the prompt gamma-ray energy and the X-ray luminosity at 11 hours, $E_{\gamma,\textrm{iso}}$-$L_\textrm{X,11}$ \citep{Nysewander2009,Berger2014,OConnor2020}. This correlation (within its scatter) holds for both classical GRBs and ultra-long GRBs (Figure \ref{fig:Egamma-Lx11}). 

In Figure \ref{fig:Egamma-Lx11}, we consider the X-ray luminosity at 11 hours (rest frame) from the discovery of each transient. Thus, for GRB 250702B we have adopted the ``D'' burst as $T_0$. However, the agreement of GRB 250702B with this correlation depends on the exact afterglow onset time (Figure \ref{fig:temporalfit}), as adopting an earlier start time (e.g., the initial EP detection) is likely to modify the X-ray luminosity (towards brighter values). In any case, we do find that utilizing the ``E'' burst as $T_0$ does not modify this conclusion.


Ultra-long GRBs also clearly follow this correlation within the observed scatter of classical GRBs. Each of the ultra-long GRBs (e.g., GRBs 101225A, 111209A, 121027A, and 130925A) has a gamma-ray duration of $\sim$\,$10^{4}$ s and displays non-afterglow-like soft X-rays behavior during this phase \citep{Levan2014}. Despite their long lived central engine activity, each of these sources returned to a more standard afterglow behavior with a shallow plateau-like decay by 11 hours (rest frame). That these long-lived, peculiar GRBs would also follow the $E_{\gamma,\textrm{iso}}$-$L_\textrm{X,11}$ correlation would circumstantially favor external afterglow emission from GRB 250702B at 11 hours after the initial \textit{Fermi} trigger. However, this interpretation is complicated by the observed X-ray variability at $\sim$\,$1$ d (rest frame), which is another behavior shared by ultra-long GRBs at $<$\,$10^{4}$ s \citep[see Figure 3 of][]{Levan2014}. Instead, it is possible that this correlation is simply a demonstration of the radiation efficiency of relativistic blastwaves (potentially those launched by stellar-mass black holes; \S \ref{sec:stellarmass}). We discuss likelihood of an external origin for the X-ray emission in \S \ref{sec:int-vs-ext}.

\subsection{An Internal or External Origin for the X-ray Lightcurve?}
\label{sec:int-vs-ext}

The observed short timescale X-ray variability by \textit{Swift} (Figure \ref{fig:XRTflaring}) and \textit{NuSTAR} (Figure \ref{fig:lcrate}) complicates the interpretation that the X-ray emission arises from an external shock. The short timescale ($\Delta T/T$\,$<$\,$0.03$) of these flares is not possible to reproduce with an external emission mechanism. Instead, this favors an internal origin for the X-ray lightcurve out to at least $\sim$\,$1$ d (rest frame), as also seen at early times ($<$\,$10^4$ s; rest frame) in ultra-long GRBs \citep{Levan2014}. 

However, unlike the current sample of ultra-long GRBs, the observed short timescale variability in GRB 250702B does not overlap with the (observed) duration of its prompt gamma-ray emission, which stops being detected prior to the first XRT observation \citep{Neights2025}. An additional puzzle piece is the clear agreement of the later time ($>$\,$1$ d; rest frame) X-ray lightcurve with GRB afterglows (Figure \ref{fig:lclum}). With these qualities in mind, the origin of the X-ray emission has two main possibilities:
\begin{itemize}
    \item A completely accretion driven relativistic outflow with an internal, non-thermal dissipation mechanism dominating the X-ray emission out to $>$\,$65$ days (observer frame), as observed in known relativistic TDEs,
    \item Or late-time central engine activity to produce short timescale variability on top of the non-thermal external shock continuum which dominates at $>$\,$2$ days (observer frame). 
\end{itemize}
The exact cause of the X-ray emission has strong implications for the precise nature of the progenitor system and its central engine. In particular, any central engine capable of reproducing the observed X-ray emission must remain active (for at least) between the initial EP detection \citep{EPgcn,EP250702a-arxiv} and the end of the first \textit{NuSTAR} observation. In fact, numerical simulations have shown that the soft X-ray emission is a better tracer of the jet activity duration than the gamma-ray duration \citep{Parsotan2024}. 
Cumulatively, the observations require an engine duration in excess of $\gtrsim$\,$3$ days in the observer frame, corresponding to $\gtrsim$\,$1.5$ days ($\gtrsim$\,$1.6\times10^{5}$ s) in the rest frame. 

The central engine is very likely due to accretion onto a black hole, and the main open questions are the mass of the black hole and its formation pathway  (see \S \ref{sec:interp} for further discussion). Relativistic TDEs such as Sw J1644+57 display accretion driven central engine activity, characteristic of sharp dips and flaring, out to hundreds of days \citep{Saxton2012,Mangano2016}, easily exceeding this requirement. In such a scenario, the X-ray emission arises from internal dissipation within the relativistic jet with dips in the X-ray flux likely linked to changes in the accretion rate of disrupted stellar material onto the black hole. For a small enough black hole mass (\S \ref{sec:massiveblackhole}), this scenario is easily capable of matching the observed flaring timescales, but may still be challenged to reproduce the short 1 s (observer frame) minimum variability of the prompt emission \citep{Neights2025}. 

Relativistic TDEs must also emit forward shock X-ray emission due to the interaction of their relativistic jet with its surrounding environment, which also gives rise to their luminous long-lived radio emission. However, the X-ray emission from the external forward shock is completely outshined (and therefore undetected) by the fallback accretion until late-times when the internal jet emission shuts off. If interpreting GRB 250702B as a relativistic TDE, the low X-ray luminosity of the fallback accretion, compared to known relativistic TDEs, must be explained. It is possible that this is due to a lower mass accretion onto the black hole or a lower radiation efficiency, but this is at odds with the significantly more energetic, flaring high-energy gamma-ray emission of GRB 250702B. In addition, the forward shock emission must be hidden at a lower flux level, which can cause challenges based on the required kinetic energy.  For example, in order for the X-rays generated by the external shock to lie an order of magnitude below the observed lightcurve requires a small kinetic energy ($<$\,$10^{52-53}$ erg) of the jet, which leads to extremely high gamma-ray efficiencies ($>90\%$) that we deem unrealistic \citep[e.g.,][]{Beniamini2016corr}.

In the second scenario for the X-ray emission, the external X-ray emission does not lie very far below the accretion driven excess that produces the flaring behavior (Figure \ref{fig:afterglow}). Our afterglow modeling shows that this scenario is easy to produce with reasonable kinetic energies (Figure \ref{fig:afterglowcorner}) with the forward shock emission beginning to dominate the observed lightcurve at $>$\,$2$ days. It is important to note that the X-ray variability observed for GRB 250702B (Figure \ref{fig:xrt+nustar}) is at a lower amplitude, and much less frequent, than observed in known relativistic TDEs (e.g., Sw J1644+57; \citealt{Burrows2011,Bloom2011}). The current sample of relativistic TDEs shows dramatic (up to $10\times$) dips in the flux on short timescales, which are not observed here. There is minor evidence for a dip at the end of the initial \textit{NuSTAR} observation (Figure \ref{fig:xrt+nustar}) and at $\sim$\,$2.2$ d (observer frame) by \textit{Swift} (Figure \ref{fig:temporalfit}), but the variability amplitude is significantly smaller than in Sw J1644+57 \citep{Bloom2011,Burrows2011} or AT2022cmc \citep{Andreoni2022,Pasham2023}. However, we point out that the jetted TDEs Sw J2058+0516 \citep{Cenko2012} and Sw J1112-8238 \citep{Brown2015} do not show as pronounced dips as in Sw J1644+57 \citep{Bloom2011,Burrows2011}. 
The cause of this in GRB 250702B could be due to the proximity of the external afterglow to the flux level of the accretion driven emission at $\sim$\,$2$ days (observer frame). 

While past GRB afterglow observations with \textit{NuSTAR} do not display this short timescale behavior (\S \ref{sec:grbvarappendix}; Figure \ref{fig:grbvar}), this is only a small sample of the GRB phenomena, which is notably diverse. Ultra-long GRBs (e.g., GRB 121027A) do display comparable X-ray variability and dips \citep{Levan2014} out to $\sim$\,$10^4$ s in the rest frame (Figure \ref{fig:lclum}), see, e.g., Figure 14 of \citet{Evans2014-ultralong}. This late-time X-ray variability is likely linked to ongoing central engine activity and ongoing prompt gamma-ray emission. However, the timescale required for GRB 250702B is significantly longer than the current ultra-long sample, and has the additional difficulty of producing the initial EP emission \citep{EP250702a-arxiv}. We further discuss the prospects for the progenitor of GRB 250702B in \S \ref{sec:interp}.

\subsection{Results of External Afterglow Modeling}
\label{sec:extshockdisc}

We have modeled the multi-wavelength lightcurve (\S \ref{sec:afterglow}) of GRB 250702B assuming that the near-infrared and radio emission are generated by the combination of forward and reverse shock emission (Figure \ref{fig:afterglow}), and that the X-ray lightcurve after $>$\,$2\times10^5$ s (relative to the ``E'' burst; observer frame) is dominated by the external shock. Here we outline our main results and how they impact the interpretation of GRB 250702B. 

The long-lasting prompt phase adds difficulty in determining the precise onset time of the afterglow. Corroborated by our temporal analysis of the X-ray lightcurve (\S \ref{sec:temporalfit}; Figure \ref{fig:temporalfit}), we performed our afterglow modeling relative to the GBM ``E'' burst (MJD 60858.682; Table \ref{tab:triggertimes}). We support interpreting this as the afterglow onset time by invoking a picture where each independent GBM trigger (or prompt episode) is due to an internal shell-shell collision and essentially marks the launch time of an ultra-relativistic shell. As each shell collides, they merge to form a larger, more energetic shell, which eventually catches up to, and then refreshes, the blast wave producing the forward shock \citep{SariMeszaros2000,KumarPanaitescu2000}. This refreshed shock then resets the inferred afterglow $T_0$ with the most energetic shell (the ``E'' burst) dominating the afterglow. The results of our afterglow modeling using the ``E'' burst as $T_0$ are shown in Figures \ref{fig:afterglow} and \ref{fig:afterglowcorner}. However, we note that utilizing the ``D'' burst as $T_0$ only marginally changes our results for the afterglow parameters, see Figures \ref{fig:afterglow-D} and \ref{fig:afterglowcorner-D}. 

As we do not observe the jet's deceleration (coasting) phase, our afterglow modeling provides a lower limit to the initial Lorentz factor of the jet required to decelerate prior to the start of the first multi-wavelength observations of the afterglow phase (Figure \ref{fig:afterglow}). 
The required Lorentz factor is strongly correlated with the jet's kinetic energy (Figure \ref{fig:afterglowcorner}), with larger energies requiring larger Lorentz factors as the (on-axis) deceleration time $t_\mathrm{dec}$ is given by \citep{Sari1999,Molinari2007} 
\begin{align}
\label{eqn:tdec}
    \frac{t_\mathrm{dec}}{1+z}=\left[
    \frac{3-k}{2^{5-s}\pi}
    \frac{E_\textrm{kin,iso}}{c^{5-k}\,A\, \Gamma_0^{8-2k} }\right]^{1/(3-k)},
\end{align}
where $\Gamma_0$ is the initial bulk Lorentz factor at the jet's core, $E_\mathrm{kin,iso}$ is the (isotropic-equivalent) kinetic energy, $A$\,$=$\,$m_\textrm{p}n_0 R_0^k$, $n_0$ is the density normalization at radius $R_0$\,$=$\,$10^{18}$ cm, $m_\textrm{p}$ is the proton mass, and $c$ is the speed of light. 

The modeling finds Lorentz factors in the range of $\Gamma_0$\,$\gtrsim$\,$100$ (Figure \ref{fig:afterglowcorner}), which can be taken as lower limits to the true Lorentz factor as higher values would allow for an earlier deceleration (i.e., consistent with observations). While we have required $\Gamma_0\theta_j\geq1$ in our modeling, to prevent lateral spreading during the coast phasing such that the spherical jet approximation can be applied, we note that relaxing this requirement still demands large Lorentz factors of $\Gamma_0$\,$>$\,$20-50$. However, if $\Gamma_0\theta_\textrm{j}<1$ then the inferred true prompt emisison fluence becomes larger by a factor of $(\Gamma_0\theta_\textrm{j})^{-2}$ as most of the $\Gamma_0^{-1}$ cone is actually not emitting, so the energy per solid angle is larger for the same fluence. This would increase the inferred gamma-ray energy, which is already quite large.

The derived range of Lorentz factors is consistent with inferences from the population of classical GRBs \citep{Ghirlanda2018}, which covers a similar range (for a wind environment). However, the range is significantly higher than expectations for  relativistic TDEs (\S \ref{sec:massiveblackhole}), which are generally thought to be less relativistic ($\Gamma_0$\,$\approx$\,$10$\,$-$\,$20$; e.g., \citealt[][]{Zauderer2011,Zauderer2013,Andreoni2022,Pasham2023}) than the jets launched by GRBs. The range of Lorentz factors we infer is consistent with the detection of $>$\,$1$ MeV photons during the prompt phase, which require $\Gamma_0$\,$>$\,$50$ \citep{Neights2025}.

Additionally, the steep decay of the X-ray and near-infrared data requires that the afterglow (\S \ref{sec:afterglow}) is in the post-jet-break phase prior to the start of the multi-wavelength observations (Figure \ref{fig:afterglow}). This is consistent with the afterglow modeling presented by \citet{Levan2025,Carney2025,Gompertz2025}. For a non-spreading jet \citep[e.g.,][]{Granot07}, the geometric steepening after the jet break is given by $\Delta\alpha$\,$=$\,$\frac{3-k}{4-k}$, which ranges between $\Delta\alpha$\,$=$\,$3/4$ in a uniform density environment ($k$\,$=$\,$0$) to $\Delta\alpha$\,$=$\,$1/2$ for a wind environment ($k$\,$=$\,$2$). Our afterglow modeling favors a steep external density profile $n_\textrm{ext}$\,$\propto$\,$R^{-k}$ where $k$\,$\simeq$\,$2$. In this case, the post-jet-break slope for a non-spreading jet is given by $\alpha_\textrm{j}$\,$=$\,$\frac{3}{4}(p-1)+\frac{k}{2(4-k)}+\frac{3-k}{4-k}$\,$=$\,$1.83\pm0.04$ (for $\nu_\textrm{m}<\nu<\nu_\textrm{c}$) accounting for the errors on $p$ and $k$ (Figure \ref{fig:afterglowcorner}). This is consistent with the range of temporal slopes inferred from the X-ray lightcurve (\S \ref{sec:temporalfit}). 
Including the addition of lateral jet spreading yields instead $\alpha_\textrm{j}$\,$=$\,$p$ \citep{Rhoads1999,SariPiranHalpern1999}, which would similarly prefer $p$\,$\approx$\,$2$ and favor a slightly earlier afterglow start time closer to the ``D'' burst (Figure \ref{fig:temporalcorner}). The overall conclusions based on the energy, density, Lorentz factor, and jet opening angle are unchanged (Figure \ref{fig:afterglowcorner-D}).

Our modeling prefers a high isotropic-equivalent blastwave kinetic energy $E_\textrm{kin,iso}$ in excess of $10^{54-55}$ erg, which is consistent with the large isotropic-equivalent gamma-ray energy $\sim$\,$2\times10^{54}$ erg \citep{Neights2025}. The true (beaming corrected) energy release can be derived by taking into account the jet's core half-opening angle $\theta_\textrm{j}$. The constraint on the jet's half-opening angle likewise scales strongly with the jet's kinetic energy as the jet break time $t_{\rm j}$ is given by
\citep{SariPiranHalpern1999,Rhoads1999,ChevalierLi2000,Frail2001}:  
\begin{align}
\label{eqn:tjeteqn}
    \frac{t_{\rm j}}{1+z}= \left[
    \frac{3-k}{2^{5-s}\pi} \frac{E_\textrm{kin,iso}\,\theta_\textrm{j}^{8-2k}}{c^{5-k}\,A }\right]^{1/(3-k)}.
\end{align} 
As our non-spreading jet solution requires an early jet-break, the range of the jet's core half-opening angles (Figure \ref{fig:afterglowcorner}) can therefore be taken as an upper limit to the jet's collimation. While smaller opening angles, down to $\theta_\textrm{j}$\,$\approx$\,$0.01$ rad are preferred by the fit, a rough upper limit of $\theta_\textrm{j}$\,$\lesssim$\,$0.02$ rad ($\lesssim$\,$1$ deg) can be obtained. This is at the extremely low end of jet opening angles observed for GRBs \citep[e.g.,][]{Frail2001,Cenko2010,Wang2018}, and also significantly smaller than inferred for relativistic TDEs \citep[$\theta_\textrm{j}$\,$\approx$\,$0.35$ rad (20 deg);][]{Beniamini2023TDE}.

In any case, the inferred collimation constraints can be used to compute upper limits to the true collimation-corrected energy release $E_\textrm{kin,beam}$\,$=$\,$(\theta_\textrm{j}^2/2)E_\textrm{kin,iso}$ for our assumed tophat jet model. The same formalism applies to the gamma-ray energy $E_{\gamma,\textrm{beam}}$. The true beaming corrected energy release is then $E_\textrm{true}$\,$=$\,$E_\textrm{kin,beam}$\,$+$\,$E_{\gamma,\textrm{beam}}$\,$=$ $(4.1^{+1.6}_{-1.2})\times10^{50}$\,erg, where we have $E_\textrm{kin,beam}$\,$=$\,$(3.5^{+1.4}_{-1.0})\times10^{50}$\,erg and $E_{\gamma,\textrm{beam}}$\,$=$\,$(5.8^{+8.8}_{-5.6})\times10^{49}$ erg. These values are consistent with the range of energies inferred for traditional GRBs \citep[e.g.,][]{Cenko2010}. We emphasize that due to the lack of a direct jet break measurement these should be treated as upper bounds on the energy release. However, it must be noted that the association of these values with upper bounds on the energy requires that we are observing a post-jet-break lightcurve. While our afterglow modeling favors this interpretation, if instead there has been no jet break out to 65 d (observer frame), as seen in a handful of GRBs \citep[e.g.,][]{OConnor2023}, it would instead imply a much wider jet ($>$\,$10$ deg) and significantly larger collimation-corrected energies ($>$\,$10^{53}$ erg) that are difficult to reconcile with a traditional collapsar progenitor channel \citep[see also][]{Carney2025}. 

We can likewise use the inferred kinetic energy and gamma-ray energy release to compute the gamma-ray efficiency $\eta_\gamma$\,$=$\,$E_{\gamma,\textrm{iso}}/(E_{\gamma,\textrm{iso}}+E_\textrm{kin,iso})$, where we have used the isotropic-equivalent definition for energies (though this does not change the result). We derive a gamma-ray efficiency of $\eta_\gamma$\,$=$\,$0.15^{+0.15}_{-0.08}$, which is consistent with values inferred for classical GRBs \citep{Nava2014,Beniamini2015}.

\begin{figure}
    \centering
\includegraphics[width=1.0\columnwidth]{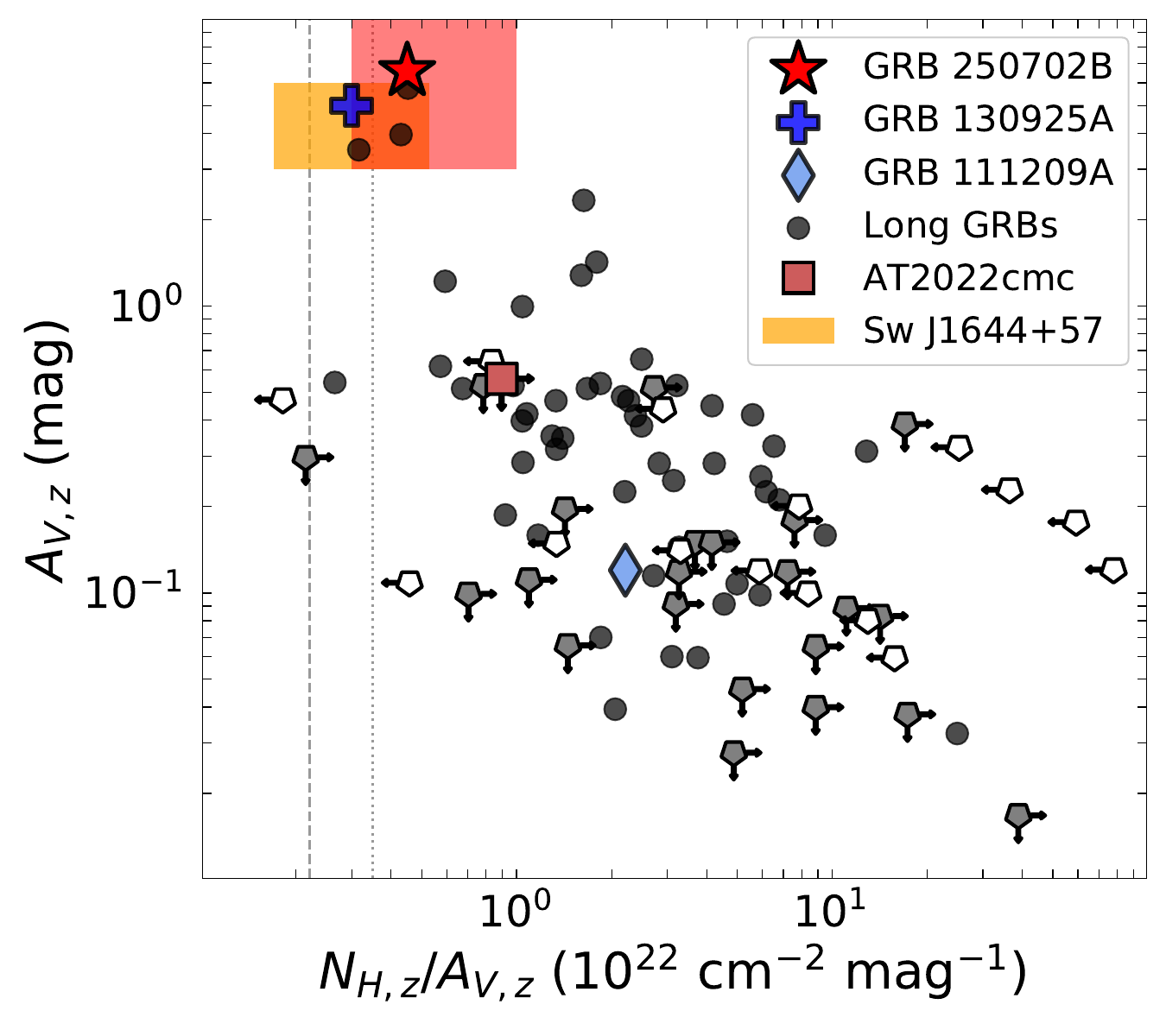}
    \caption{Intrinsic ratio of hydrogen column density to visual extinction versus visual extinction inferred from long GRB (black circles) afterglows \citep{Schady2010,Kruhler2011,Greiner2011,Perley2011,Zafar2011}. Sources with either upper limits on column density (white) or upper limits on extinction (gray) are shown as pentagons. GRB 250702B is shown as a red star based on our near-infrared to X-ray modeling. A more conservative range of the allowed values (even breaking the assumption that X-rays and near-infrared both originate from the external shock) is shown as a shaded red region. The ultra-long GRBs 111209A \citep{Stratta2013} and GRB 130925A \citep{Greiner2014} are also shown. We display as a shaded region the range for the relativistic TDEs Sw J1644 \citep{Levan2011,Bloom2011,Burrows2011,Levan2016} and as a triangle for AT2022cmc \citep[based on an upper limit on reddening; see][]{Yao2024}. The dashed and dotted line refer to the median values for the Milky Way \citep{Guver2009} and SMC \citep{Bouchet1985}, respectively. 
    The figure is reproduced from \citet{Kruhler2011}. 
    }
    \label{fig:gasdust}
\end{figure}

\subsection{Implications from the Gas-to-Dust Ratio}
\label{sec:gastodust}

The relation between hydrogen column density $N_\textrm{H}$ and visual extinction $A_\textrm{V}$ follows a well-known linear relationship in the Milky Way, e.g., $N_\textrm{H}/A_\textrm{V}$\,$=$\,$(2.21\pm0.09)\times10^{21}$ cm$^{-2}$ mag$^{-1}$ \citep[e.g.,][]{Predehl1995,Guver2009}. The exact slope of this relation can vary between galaxies and between different assumptions for the dust extinction law. Assuming the relation within our own Galaxy, we would obtain $A_\textrm{V}$\,$\approx$\,$13.6\pm2.8$ mag, which is higher than the inferred values (\S \ref{sec:xoirSED}).

Based on our modeling of the X-ray to near-infrared spectral energy distribution (Figure \ref{fig:broadbandspec}), we find a gas-to-dust ratio of $N_\textrm{H,z}/A_\textrm{V,z}$\,$=$\,$(4.5\pm1.0)\times10^{21}$ cm$^{-2}$ mag$^{-1}$ for GRB 250702B. This is approximately twice the ratio for the Milky Way \citep{Guver2009}. The inferred intrinsic dust extinction $A_\textrm{V,z}$ is strongly dependent on the assumed spectral shape between the X-ray and near-infrared bands (\S \ref{sec:xoirSED}), and the assumption they arise from the same emission mechanism (see \S \ref{sec:afterglow}). If instead the X-ray emission is dominated by an accretion driven outflow (\S \ref{sec:int-vs-ext}) and is unrelated to the near-infrared emission, under the assumption that the near-infrared emission arises from the external afterglow, we still require a significant dust contribution ($A_\textrm{V,z}$\,$\gtrsim$\,$3$) in our modeling.

The allowed range of values is displayed by the shaded red region in Figure \ref{fig:gasdust}. 
We compare this to the values observed for classical long GRBs, ultra-long GRBs, and relativistic TDEs in Figure \ref{fig:gasdust}. It is worth noting the ratio is comparable to that found in the relativistic TDE Sw J1644+57 \citep[$N_\textrm{H,z}/A_\textrm{V,z}$\,$\approx$\,$(2-5)\times10^{21}$ cm$^{-2}$ mag$^{-1}$;][]{Burrows2011,Bloom2011,Levan2011} and in the ultra-long GRB 130925A \citep[$N_\textrm{H,z}/A_\textrm{V,z}$\,$\approx$\,$3\times10^{21}$ cm$^{-2}$ mag$^{-1}$;][]{Evans2014,Greiner2014}.

The allowed range of $N_\textrm{H,z}/A_\textrm{V,z}$ ratios for GRB 250702B lies at the low end of the values inferred by \citet{Schady2010}, who had instead found (for a sample of 28 classical long GRBs) that the GRB derived $N_\textrm{H,z}/A_\textrm{V,z}$ ratio of $(3.3\pm2.8)\times10^{22}$ cm$^{-2}$ mag$^{-1}$ exceeded the values for the SMC, LMC, and Milky Way by at least a factor of 10 (see their Figure 5). 
\citet{Kruhler2011dust} analyzed a sample of the dustiest GRB afterglows ($A_\textrm{V,z}$\,$\approx$\,$3$\,$-$\,$5$ mag)
, and identified a possible inverse relationship between the intrinsic visual extinction $A_\textrm{V,z}$ and the intrinsic gas-to-dust ratio $N_\textrm{H,z}/A_\textrm{V,z}$ (see Figure \ref{fig:gasdust}). They found that the dustiest afterglows ($A_\textrm{V,z}$\,$>$\,$3$ mag) had smaller values of $N_\textrm{H,z}/A_\textrm{V,z}$ compared to the bulk sample. GRB 250702B agrees with this trend, and lies in the same location as the dustiest afterglows (see Figure \ref{fig:gasdust}; \citealt{Kruhler2011dust}). However, lower gas-to-dust ratios (order $10^{20}$ cm$^{-2}$ mag$^{-1}$) have been found for a handful of other ``dark'' bursts \citep{vanderhorst2009,Greiner2011}, such as GRBs 051022 \citep{Castro-Tirado2007} and 150309A \citep{Castro-Tirado2024}; see also \citet{Perley2009-hosts}. 

\citet{Kruhler2011dust} interprets the lower $N_\textrm{H,z}/A_\textrm{V,z}$ ratio as two independent absorbers with local ionized metals in the circumburst environment producing the inferred X-ray absorption $N_\textrm{H,z}$ and a physically separate dusty absorber in the host galaxy producing the high visual extinction $A_\textrm{V,z}$. This scenario is plausible for GRB 250702B and is consistent with the independent inference of a dusty galaxy based on the host spectral energy distribution modeling of \citet{Carney2025}. However, we cannot rule out that the local environment of GRB 250702B simply differs from classical GRBs and has SMC-like gas-to-dust properties \citep{Bouchet1985}. We do caution that $N_\textrm{H,z}/A_\textrm{V,z}$ ratios \citep[e.g.,][]{Guver2009} are calibrated in regions where $A_\textrm{V}$\,$<$\,$1$ mag and they may not apply in denser or dustier environments like observed for GRB 250702B. Regardless of the interpretation, we can conclude that all events with high $A_\textrm{V,z}$ fall in the same region of the parameter space, which is worth further investigation.
We note that there are existing selection effects against the discovery of the most highest extinguished GRBs, and that rapid, deep near-infrared imaging is required to identify other potential members of this class, which will aid in understanding the origin of these small gas-to-dust ratios.

\section{Interpretation of GRB 250702B}
\label{sec:interp}

The possible scenarios to explain GRB 250702B can be most simply broken down into two progenitor categories, those either involving \textit{i}) a massive black hole or \textit{ii}) a stellar-mass black hole central engine. The expected behavior, such as the minimum variability timescale or duration, is quite different between these different progenitor systems. Below we discuss these differences in more detail and highlight where each progenitor system has difficulties in reproducing the observed behavior.


\subsection{An Intermediate-mass Black Hole Progenitor}
\label{sec:massiveblackhole}


Our understanding of the relativistic jets launched by TDEs is limited due to the small number of existing candidates \citep{Bloom2011,Levan2011,Zauderer2011,Burrows2011,Cenko2012,Brown2015,Pasham2015,Andreoni2022,Pasham2023,Yao2024,OConnor2025}. These sources are thought to be produced when a main sequence (MS) star is shredded by a massive black hole ($M_\textrm{BH}$\,$\approx$\,$10^{5-8} M_\odot$), launching a narrowly collimated relativistic jet along our sightline\footnote{The inference of a highly collimated jet is required by the extremely super-Eddington accretion rates. A collimated jet drastically decreases the required energy budget compared to the inferred isotropic-equivalent energy.}. They have been found to emit extremely luminous, long-lived X-ray and radio emission. Their X-ray emission is highly variable with rapid short term ($\sim$ hours) variability. 

\begin{figure*}
    \centering
\includegraphics[width=2\columnwidth]{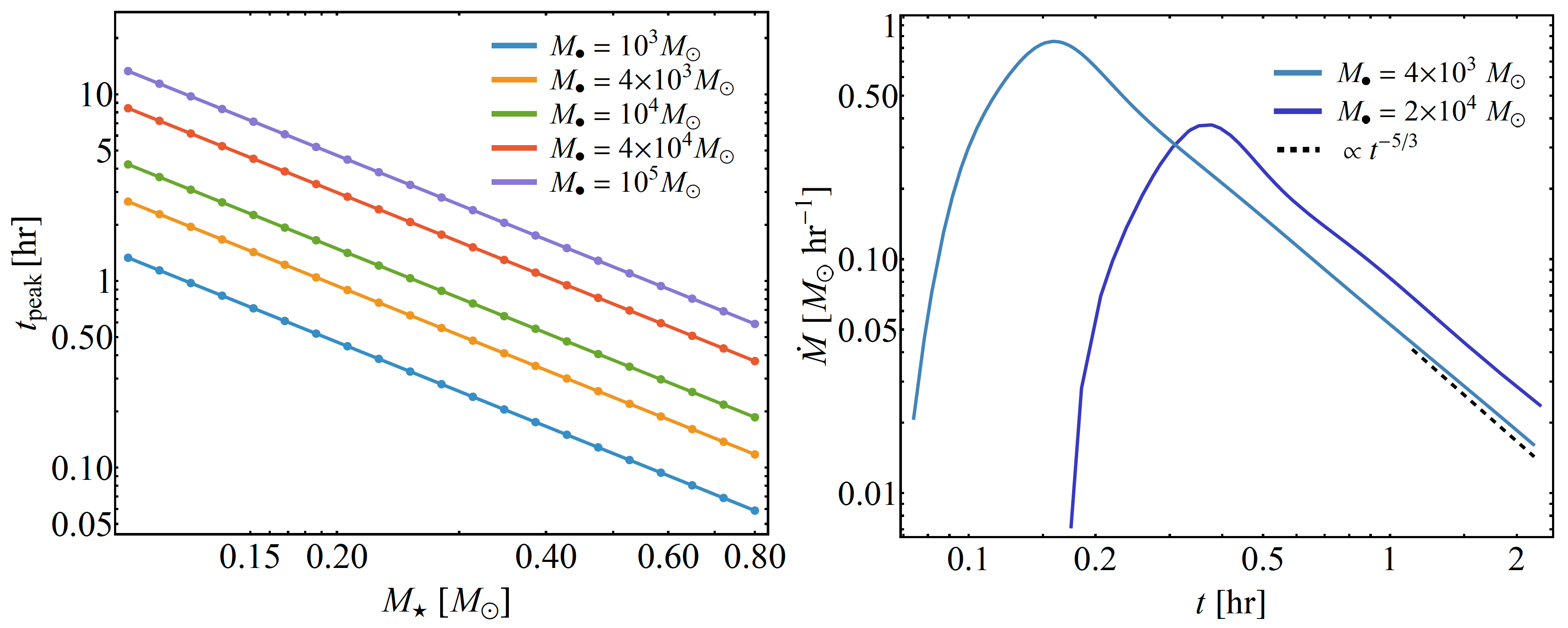}
    \caption{\textbf{Left:} The peak timescale (source frame) for accretion onto a black hole from a tidally disrupted white dwarf, as a function of the mass of the white dwarf. The different colors represent different black hole masses, as shown in the legend. \textbf{Right:} Fallback rate onto the black hole from a hydrodynamical simulation of a $0.6 M_\odot$ white dwarf modeled as a $5/3-$polytrope, being disrupted by a $4\times 10^3 M_\odot$ (light blue curve) and $2\times 10^4 M_\odot$ (dark blue curve) black hole. The black dashed line shows a $t^{-5/3}$ scaling, which is the expected late time scaling of the fallback rate if the white dwarf is completely tidally destroyed. Times are in the source frame.
    }
    \label{fig:wd_tpeak}
\end{figure*}

Relativistic TDEs are found at the center of their host galaxies\footnote{AT2022cmc is an exception as no host galaxy has been detected to date \citep{Andreoni2022}.}, coincident with the location of a massive black hole ($\gtrsim$\,$10^5 M_\odot$). These locations are possibly very dusty, as seen in Sw J1644+57 \citep{Levan2011,Levan2016}. This is consistent with the large dust column present in GRB 250702B. However, GRB 250702B is clearly offset by $\sim$\,$0.7\arcsec$ from the center of its host galaxy \citep{Levan2025,Carney2025}, corresponding to $\sim$\,$5.8\;$kpc at $z$\,$=$\,$1$ \citep{Gompertz2025}. The offset may be consistent with an intermediate-mass black hole scenario, and is not unprecedented in TDEs \citep[those without indications of relativistic outflows;][]{Jin2025-EP240222a,Yao2025}. However, the environment of an offset IMBH is expected to be significantly less dense, and likely less dusty, when compared to the environment surrounding an SMBH. In particular, a wind-like environment ($n_\textrm{ext}$\,$\propto$\,$R^{-2}$), as favored by our afterglow modeling (\S \ref{sec:afterglow}), is not expected around a massive black hole. We suggest this favors instead a stellar-mass black hole progenitor as discussed further in \S \ref{sec:stellarmass}. However, this is not far off from a Bondi-like profile (i.e., $r^{-3/2}$; see \citealt{Granot2025} for further discussion), which could be expected for the nucleus of a (dwarf) galaxy undergoing a gas-rich merger (as possibly favored by the host galaxy morphology; \citealt{Carney2025}).


Associating the observed prompt emission duration (\S \ref{sec:promptdisc}) with accretion onto the black hole following the tidal disruption of a star, the peak must have occurred on a timescale of at least $\sim$\,$1$ d (observer frame), allowing for the initial detection by the \textit{Einstein Probe} $\sim$\,$24$ hr before the \textit{Fermi} triggers (Table \ref{tab:triggertimes}). If the disrupted star was on the main sequence and completely destroyed, the time to peak emission (from initial pericenter) is $t_{\rm peak}$\,$\approx$\,$25 M_{\bullet, 6}^{1/2}\textrm{ d }$ \citep{coughlin22, bandopadhyay24, fancher25}, where $M_{\bullet,6}$ is the mass of the disrupting black hole in units of $10^6 M_{\odot}$. After the peak time of the fallback accretion rate (also referred to as the fallback time), the X-ray lightcurve will decay (e.g., $t^{-5/3}$) as observed in past relativistic TDEs (Figure \ref{fig:lclum}). Here, we observe the source to be fading from the initial XRT observation ($\sim$\,$0.5$ d after the initial triggers). Therefore, we consider a conservative bound on the peak time of fallback accretion to be $t_{\rm peak}$\,$\lesssim$\,$\big(\frac{1}{1+z}\big)1.5$ d.
Equating this peak timescale to $\lesssim$\,$\big(\frac{1}{1+z}\big)1.5$ d then constrains the mass of the disrupting black hole to $\lesssim$\,$ 9\times 10^{2} M_{\odot}$, which is firmly in the IMBH regime and could, e.g., be present in a low-mass satellite of the visually distorted massive parent galaxy \citep[see][]{Levan2025,Carney2025}. Allowing for an earlier disruption of the star (though disfavored by the existing data) up to a day before even the initial detections \citep{EPgcn,EP250702a-arxiv,Neights2025}, would extend this requirement on the black hole mass to $\lesssim$\,$2.5\times 10^{3} M_{\odot}$. The required black hole masses are consistent with the requirement ($\lesssim$\,$5\times10^{4} M_{\odot}$) from the Schwarzchild light crossing time (\S \ref{sec:promptdisc}). The partial disruption of the star would only lengthen this timescale, and while placing it on a tightly bound orbit could reduce the time to peak for a higher-mass black hole, it would require an additional mechanism, e.g., Hills capture \citep{Hills1988} or  disruption by a black hole binary, to generate such an orbit.

An alternative scenario is that the disrupted star was a white dwarf (WD), in which case the time to peak could have been significantly shorter (for the same black hole mass) as compared to the disruption of a main sequence star. Figure \ref{fig:wd_tpeak} (left panel) shows the prediction for the peak timescale, $t_{\rm peak}$, as a function of white dwarf mass, $M_{\star}$, that arises from equating the maximum gravitational field in the white dwarf interior to the tidal field of the black hole (i.e., by employing the model described in \citealt{coughlin22}), assuming that the white dwarf has a $5/3$-polytropic density profile and adopting the polytropic relationship $R_{\star} = 0.011\left(M_{\star}/(0.6M_{\odot})\right)^{-1/3} R_\odot$ \citep{Nauenberg1972}. We see that the peak timescale follows a power law scaling with the mass of the white dwarf ($t_{\rm peak} \propto M_{\star}^{-3/2}$), and for disruptions caused by a $4\times 10^3\,M_\odot$ black hole, ranges from $\sim0.1-2.7$ hr (source frame) for white dwarf masses ranging from $0.1-0.8 M_\odot$. Higher mass black holes yield longer peak timescales, up to $\sim 10$ hr for a low-mass white dwarf being disrupted by a $10^5 M_\odot$ black hole. These timescales are far too short to re-produce the long-lasting high-energy emission observed from GRB 250702B, especially as larger black holes also become strongly disfavored by the short minimum variability timescale \citep[see][]{Neights2025}.


The times to peak fallback shown in Figure \ref{fig:wd_tpeak} (left panel), as well as the scaling with the white dwarf mass, are in good agreement with the return time of the most bound debris that comes from the standard ``frozen-in'' model (as originally proposed in \citet{lacy82}; see also \citealt{rees88, lodato09, Stone2013}). Specifically, this approximation predicts (using the previously referenced mass-radius relationship)
\begin{equation}
\begin{split}
    T_{\rm fb} & = \frac{\pi R_{\star}^{3/2}}{\sqrt{2GM_{\star}}}\left(\frac{M_{\bullet}}{M_{\star}}\right)^{1/2} \\
    &\simeq 0.12\left(\frac{M_{\bullet}}{4\times 10^{3}M_{\odot}}\right)^{1/2}\left(\frac{M_{\star}}{0.6 M_{\odot}}\right)^{-3/2}\textrm{ hr}. \label{Tfbfr}
\end{split}
\end{equation}
These dependencies on black hole mass and stellar-mass are both present in Figure~\ref{fig:wd_tpeak} (left panel). 

To further substantiate this $\sim$ hour-long timescale, we performed hydrodynamical simulations of the disruption of a $0.6 M_\odot$ white dwarf modeled as a $\gamma = 5/3$ polytrope by a $4\times 10^3 M_\odot$ black hole and a $2\times 10^4 M_\odot$ black hole, using the smoothed particle hydrodynamics code {\sc phantom}~\citep{price18}. The radial density profile was mapped onto a three-dimensional particle distribution in {\sc phantom}, and relaxed for $\sim 10$ sound crossing times across the stellar radius. 
We used $10^5$ particles to model the white dwarf and an adiabatic equation of state with an adiabatic index $\Gamma=5/3$. Its center of mass was placed on a parabolic orbit with a pericenter distance $r_{\rm p} = r_{\rm t}$ (where $r_{\rm t} = R_\star (M_\bullet/M_\star)^{1/3}$ is its tidal radius), at an initial distance of $5 r_{\rm t}$ from the black hole. Figure \ref{fig:wd_tpeak} (right panel) shows the mass fallback rate onto the black hole. For the $4\times 10^3 M_\odot$ ($2\times 10^4 M_\odot$) black hole the fallback rate peaks at a time of $\sim 0.16$ hr ($0.38$ hr), which is roughly consistent with the expectations of the standard ``frozen-in'' model (Equation \ref{Tfbfr}), and agrees very well with the predictions from \citet{coughlin22} (shown in the right panel of Figure \ref{fig:wd_tpeak}). 

We note that the hydrodynamic simulation shown in Figure \ref{fig:wd_tpeak} (right panel) provides the fallback rate, which is not necessarily equal to the accretion rate, which is ultimately responsible for liberating the observed energy (by assumption here). In particular, there will be a substantial difference between the two if the viscous timescale is substantially longer than the timescale over which the material is supplied to the black hole, i.e., the fallback time \citep{rees88}. For standard $\alpha$-disc models and typical values of $\alpha$, the viscous time is generally short compared to the fallback time, and the timescale over which the viscous delays becomes significant is $\sim$ years \citep{cannizzo90, lodato11}. Additionally, recent works (e.g., \citealt{mockler19,nicholl22}) have found that the early-time luminosities of observed TDEs can be reproduced well with fallback models that impose zero viscous delay (though late-time observations of TDEs exhibit sustained levels of accretion that are consistent with a viscously spreading disc; \citealt{vanvelzen19}).

Based on Figure \ref{fig:wd_tpeak}, we  see that a white dwarf disruption would produce a timescale too short to be consistent with the observed, day-long duration associated with the prompt emission. Specifically, noting that the fallback rate peaks on a timescale of $\sim 0.16$ hr for a $4\times 10^{3} M_{\odot}$ black hole and using the $\propto M_{\bullet}^{1/2}$ scaling implied by Equation \eqref{Tfbfr} (and shown in the right panel of Figure \ref{fig:wd_tpeak}), a black hole mass of $M_{\bullet} \simeq 4\times 10^{7} M_{\odot}$ would be required to yield a peak timescale of $\sim$\,$\big(\frac{1}{1+z}\big)1.5$ days. This is well in excess of the Hills mass for the disruption of a white dwarf, which is comparable to $\sim a \,few \times 10^{5} M_{\odot}$ if the black hole is spinning and the orbital angular momentum of the star is aligned with the spin axis of the black hole. While the fallback rates shown in Figure~\ref{fig:wd_tpeak} (right panel) are obtained from the complete disruption of a white dwarf, a partial disruption could yield a longer peak timescale by a factor of at most a few\footnote{Figure 2 of \citet{garain24} shows fallback rates for the disruption of a $1M_\odot$ white dwarf by a $10^3M_\odot$ black hole for a range of orbital pericenter distances, wherein the peak timescale shows very little variation over the entire range.}. A longer peak timescale could also arise from the white dwarf being on an unbound (eccentricity $>1$) orbit about the black hole, but, as shown in Figure 4 of \cite{cufari22}, $t_{\rm peak}$ changes by less than a factor of $\sim 10$ over a wide range of orbital eccentricities, implying that the black hole mass required to generate a fallback rate having a peak timescale of at least a day would still exceed the Hills mass for the disruption of a white dwarf. A white dwarf does not produce a long lasting enough central engine as required by the prompt phase unless a tightly bound orbit (in which essentially all debris is accreted on the orbital timescale) combined with (rare) repeated partial disruptions are considered \citep{Levan2025}. Such scenarios likely require a (likely rare) Hills capture mechanism \citep{Hills1988}. Based on these arguments we strongly disfavor the disruption of a white dwarf by an intermediate-mass black hole as the progenitor of GRB 250702B. 

We therefore return to the consideration of a main sequence star disrupted by an intermediate-mass black hole. A characteristic feature of the known relativistic TDEs is a sharp cessation of their jet activity and a corresponding significant drop in luminosity, which can be interpreted as the transition from super-Eddington accretion to sub-Eddington accretion onto the black hole  \citep[e.g.,][]{Zauderer2013,Pasham2015,Eftekhari2024}. The observation of a steep drop in X-ray luminosity would be a smoking gun for a TDE progenitor, as it is not possible in an external shock scenario. Thus, long-term X-ray monitoring is required to search for this possibility. The shutoff time can be related to the mass of the black hole. Using the late-time \textit{Chandra} observations, we can confirm that the jet has not shut off prior to 65 days (observer frame). Given the time to peak in the fallback rate $t_{\rm peak}$ and the peak of the fallback rate itself as $\dot{M}_{\rm peak}$, one can estimate the subsequent and time-dependent accretion luminosity as $L = \eta c^2 \dot{M}_{\rm peak}\left(t/t_{\rm peak}\right)^{-5/3}$, from which the time to reach the Eddington limit, $t_{\rm Edd}$, is
\begin{equation}
    t_{\rm Edd} = t_{\rm peak}\left(\frac{\eta\kappa\dot{M}_{\rm peak}c}{4\pi G M_{\bullet}}\right)^{3/5}.
\end{equation}
Further letting $\dot{M}_{\rm peak} \simeq M_{\star}/(4 t_{\rm peak})$ and $t_{\rm peak} \simeq 25 M_{\bullet,6}^{1/2}\textrm{ d}$, this becomes (after setting $\eta = 0.1$ and $\kappa = 0.34$ cm$^2$ g$^{-1}$)
\begin{equation}
    t_{\rm Edd} \simeq 500 M_{\bullet,6}^{-2/5}M_{\star,\odot}^{3/5}\textrm{ d},
\end{equation}
where $M_{\star,\odot}$ is the mass of the star in solar masses. For a black hole mass of $\sim$\,$10^{3} M_{\odot}$, which is required for a peak timescale of $\lesssim$\,$\big(\frac{1}{1+z}\big)1.5$ days, this would predict a (source frame) time to reach the Eddington limit of $t_{\rm Edd} \simeq 20$ yr, i.e., well in excess of the observed 65 d lower limit (Figures \ref{fig:temporalfit} and \ref{fig:lclum}). On the other hand, the disruption of a lower-mass star (which is more likely if disrupted stars are drawn from a standard stellar population), a partial TDE, and a lower radiative efficiency -- which is likely, given the very high Eddington fraction and corresponding optical depths -- would all serve to reduce this timescale. It is worth noting that the time to Eddington level accretion for a WD-IMBH merger is likewise long (as the inital peak accretion rate is so highly super-Eddington), with typical (source frame) timescales of $\sim$\,$1$ yr for the fiducial parameters considered above. Late-time X-ray observations are strongly encouraged to constrain or measure this timescale, as this can be seen as ``smoking gun'' evidence for a TDE scenario. 


We note that a possible issue with the scenario in which a main sequence star is disrupted by a lower-mass black hole is that the relativistic advance of periapsis angle is $3\pi GM_{\bullet}/(c^2 r_{\rm p})$ to lowest order in $1/r_{\rm p}$, which amounts to $\sim 0.3^{\circ}$ for a solar-type star and $r_{\rm p} = r_{\rm t}$. Such a small periapsis advance could lead to the inefficient circularization of the debris, but this could be alleviated if the pericenter distance of the star is well within the tidal radius. In the Newtonian approximation and when the disrupted star is in the pinhole scattering regime (necessary for high-$\beta$ disruptions), the cumulative distribution function for $\beta$ is $F_{\beta}(\beta) = 1/\beta$, such that $\sim 2.5\%$ of stars 
reach $\beta \ge 40$ and have periapsis advance angles $\gtrsim 10^{\circ}$. This would therefore be a relatively rare event, but given that no other such gamma-ray bursts have yet been observed, we do not consider this an inconsistency.

Thus, while we can strongly disfavor a WD-IMBH merger, we cannot exclude a MS-IMBH merger \citep[for further discussion, see][]{Granot2025}. The MS-IMBH merger is consistent with many of the observed timescales, but still has difficulties as  \textit{i}) the multiple gamma-ray triggers are unlike any observed in past relativistic TDEs, \textit{ii}) the X-ray lightcurve is consistent with the luminosities of known GRBs and not known relativistic TDEs (Figures \ref{fig:afterglow} and \ref{fig:lclum}), and \textit{iii}) the event follows standard GRB prompt-afterglow correlations (Figure \ref{fig:Egamma-Lx11}). Additionally, the observed rate of GRBs, even ultra-long GRBs (though this depends on the exact duration requirement for this definition; see \citealt{Levan2014}), exceeds the known number of relativistic TDEs.

\subsection{A Stellar-mass Black Hole Progenitor}
\label{sec:stellarmass}

The observed properties of GRB 250702B are most similar to those found in the (notably small) sample of ultra-long GRBs. The main characteristics that make this event appear an outlier are the discovery of un-triggered high-energy emission up to a day prior to its first on-board gamma-ray triggers \citep{EPgcn,EP250702a-arxiv,Neights2025} and the continued short timescale flaring out to 2 days (\S \ref{sec:int-vs-ext}). Besides these factors, all other properties appear typical of GRBs. These include: the X-ray luminosity and rapid lightcurve decay, the triggered gamma-ray emission, the isotropic-equivalent gamma-ray energy release, the gamma-ray efficiency, the prompt emission's minimum variability timescale, the consistency of the multi-wavelength lightcurve with external afterglow emission in a dense wind environment, the dusty host galaxy, and the offset location within its host. In fact, each of these properties are instead atypical for a massive black hole progenitor (\S \ref{sec:massiveblackhole}).

However, the discovery of the early soft X-ray emission \citep{EP250702a-arxiv} and the extremely long ($>$\,$25$ ks; \citealt{Neights2025}) gamma-ray emission are difficult to reconcile in a traditional collapsar scenario, even considering a peculiar progenitor such as the collapse of a blue supergiant \citep{Meszaros2001BSG,Stratta2013,Nakauchi2013,Ioka2016}. 
It is worth noting that the \textit{Einstein Probe} is revealing that soft X-ray emission can extend for $10$\,$-$\,$100$ times the gamma-ray emission for what otherwise would appear a typical GRB at gamma-ray frequencies \citep[see, e.g.,][]{Levan2024,Yin2024,Liu2024,Yin2025ep250404a}. If this same phenomena can be extended to ultra-long GRBs, it could potentially explain the longer duration of the prompt emission. Previously existing selection biases, based on the lack of wide-field soft X-ray telescopes, may have played a role in limiting the discovery of soft precursor emission. It is possible that the initial soft X-ray emission detected by the \textit{Einstein Probe} \citep{EPgcn,EP250702a-arxiv} is due to the formation of a cocoon \citep[as in, e.g.,][]{Leung2025} as the jet (or multiple shells launched by the central engine) attempts to breakout of the extended stellar radius of a blue supergiant. As subsequent shells launched within the jet do not have as great a difficulty burrowing out of the stellar material, we have a clearer picture of their prompt gamma-rays (likely produced through internal shocks). However, the requirements for the duration of the central engine are still extreme even in such a scenario, and lead us to conclude that a collapsar (even from a blue supergiant) is an unlikely progenitor for GRB 250702B.

Based on all of the considerations presented in this manuscript, we find that while GRB 250702B displays similar properties of both classical and ultra-long GRBs (e.g., Figure \ref{fig:Egamma-Lx11}), with the main exception being the early pre-trigger X-ray emission \citep{EP250702a-arxiv}, it may be explained by a ``hybrid'' progenitor system involving a stellar-mass black hole, but with a TDE-like central engine. This is also favored by the consistency of the surrounding environment with the profile of a stellar wind. The main possibilities for this hybrid system are the merger of a helium star with a stellar-mass black hole \citep{Fryer1998,Woosley2012,Neights2025} or a micro-TDE by a stellar-mass black hole \citep{Perets2016,Beniamini2025}. These scenarios have been shown to be compatible with day long central engine activity as required by this event and may more naturally explain the early X-ray emission, 1 s minimum variability timescale of the prompt emission, long-lived central engine ($\gtrsim$\,$3$ days), and short timescale X-ray flaring produced by an accretion driven relativistic outflow. These progenitor scenarios can also potentially explain the current sample of ultra-long GRBs. A complete discussion of the micro-TDE scenario is provided by \cite{Beniamini2025}, and the helium merger scenario is discussed by \citet{Neights2025}. 


\section{Conclusions}
\label{sec:conclusions}

We present a comprehensive X-ray analysis of the ultra-long transient GRB 250702B using \textit{Swift}, \textit{NuSTAR}, and \textit{Chandra}. These observations trace the X-ray counterpart between $0.5$ to $65$ days from discovery and reveal a rapidly fading source at $0.3$\,$-$\,$79$ keV. By modeling the near-infrared to X-ray spectral energy distribution, we uncover a potential spectral break that is likely due to the synchrotron cooling frequency. The multi-wavelength lightcurve is found to be well described by the standard fireball model (forward and reverse shock emission) in a wind-like environment. 
Superposed on this external shock continuum, \textit{Swift} and \textit{NuSTAR} detect short timescale X-ray flares with $\Delta T/T$\,$<$\,$0.03$, implying engine-dominated X-ray emission persisting to at least $\sim$\,$2$ days that is most naturally attributable to accretion powered variability.

We demonstrate that the overall characteristics of GRB 250702B do not cleanly match either known relativistic TDEs or ultra-long GRBs. With the exception of the initial soft X-ray detection a full day before the initial gamma-ray triggers, the properties have a closer match to those seen by ultra-long GRBs, but require an even longer lived central engine ($\gtrsim$\,$3$ days). In particular, the early soft X-ray detections \citep{EPgcn,EP250702a-arxiv} are not easily explained in the standard collapsar model \citep{Woosley1993}. As such, we favor a hybrid stellar-mass black hole progenitor, such as a micro-TDE \citep{Beniamini2025}, though see also \citet{Neights2025} and \citet{Granot2025} for alternative explanations. 
Additional observations of similar transients, hopefully discoverable via the \textit{Einstein Probe} \citep{Yuan2025}, will aid in uncovering the true nature of GRB 250702B.

\begin{acknowledgments}
The authors thank the anonymous referee for their helpful comments that improved the clarity of the manuscript. 
The authors acknowledge Karl Forster and the \textit{NuSTAR} SOC for their rapid scheduling of our observations, as well as Pat Slane and the CXO staff for approval and scheduling of our \textit{Chandra} DDT request. B. O. acknowledges Brian Grefenstette, Oli Roberts, Eliza Neights, and Eric Burns for useful discussions. B. O. thanks Ben Gompertz, Andrew Levan, and Antonio Martin-Carrillo for sharing the redshift derived by \textit{JWST}.

B. O. is supported by the McWilliams Postdoctoral Fellowship in the McWilliams Center for Cosmology and Astrophysics at Carnegie Mellon University. J. D. acknowledges support from NASA under award numbers NAS5-0136, 80NSSC25K7567, and 80NSSC25K7788. D. R. P acknowledges support from NASA in the form of a NuSTAR guest observer program. E. R. C. acknowledges support from NASA through the Astrophysics Theory Program, grant 80NSSC24K0897. A. B. acknowledges support from NASA through the FINESST program, grant 80NSSC24K1548. M. M. is supported by an appointment to the NASA Postdoctoral Program at the NASA Goddard Space Flight Center, administered by Oak Ridge Associated Universities under contract with NASA. J. H. acknowledges support from NASA under award number 80GSFC21M0002. P. B. work was funded by a grant (no. 2020747) from the United States-Israel Binational Science Foundation (BSF), Jerusalem, Israel and by a grant (no. 1649/23) from the Israel Science Foundation. M. B. is supported by a Student Grant from the Wübben Stiftung Wissenschaft. Funded in part by the Deutsche Forschungsgemeinschaft (DFG, German Research Foundation) under Germany's Excellence Strategy – EXC-2094 – 390783311. 

This work used resources on the Vera Cluster at the Pittsburgh Supercomputing Center (PSC). Vera is a dedicated cluster for the McWilliams Center for Cosmology and Astrophysics at Carnegie Mellon University. We thank the PSC staff for their support of the Vera Cluster.

This work made use of data from the NuSTAR mission, a project led by the California Institute of Technology, managed by the Jet Propulsion Laboratory, and funded by the National Aeronautics and Space Administration. This research has made use of the NuSTAR Data Analysis Software (NuSTARDAS) jointly developed by the ASI Space Science Data Center (SSDC, Italy) and the California Institute of Technology (Caltech, USA). The scientific results reported in this article are based on observations made by the Chandra X-ray Observatory. This research has made use of software provided by the Chandra X-ray Center (CXC) in the application package CIAO. This work made use of data supplied by the UK \textit{Swift} Science Data Centre at the University of Leicester. This research has made use of the XRT Data Analysis Software (XRTDAS) developed under the responsibility of the ASI Science Data Center (ASDC), Italy. This research has made use of data and/or software provided by the High Energy Astrophysics Science Archive Research Center (HEASARC), which is a service of the Astrophysics Science Division at NASA/GSFC.

\end{acknowledgments}





%
\facilities{\textit{Neil Gehrels Swift Observatory}, \textit{Nuclear Spectroscopic Telescope Array}, \textit{Chandra X-ray Observatory}
}

\software{\texttt{Astropy} \citep{2018AJ....156..123A,2022ApJ...935..167A}, \texttt{SciPy} \citep{scipy}, \texttt{emcee} \citep{emcee}, \texttt{corner} \citep{corner},  \texttt{stingray} \citep{Huppenkothen2019}, \texttt{XSPEC} \citep{Arnaud1996}, \texttt{GUANO} \citep{Tohuvavohu2020}, \texttt{NITRATES} \citep{NITRATES}, \texttt{BatAnalysis} \citep{batanalysis},  \texttt{CIAO} \citep{Ciao}, \texttt{HEASoft} \citep{2014ascl.soft08004N}
}


\appendix

\section{Log of Observations}

We present the log of X-ray observations of EP250702a analyzed in this work in Table \ref{tab: observationsXray}. The flux values derived from \textit{Swift}/BAT Survey Mode data are presented in Table \ref{tab:batlimits}. In Table \ref{tab:guanofluxes}, we present the prompt detections of GRB 250702D, GRB 250702D, and GRB 250702E (Table \ref{tab:triggertimes}) by \textit{Swift}/BAT using the \texttt{GUANO} software (\S \ref{sec:BAT}).

\begin{table*}[ht]
\centering
\caption{Log of X-ray observations of EP250702a obtained with \textit{NuSTAR} (PI: Pasham) and \textit{Chandra} (PI: Li, ObsID: 31003; PI: O'Connor, ObsID: 31011; PI: Eyles-Ferris, ObsID: 31468). The fluxes are reported in $3$\,$-$\,$79$ keV for \textit{NuSTAR} and $0.3$\,$-$\,$10$ keV for \textit{Chandra}. The days from discovery, using the trigger time of GRB 250702D, is provided in the $\delta T$ column. 
}
\label{tab: observationsXray}
\begin{tabular}{lcccccc}
\hline\hline
\textbf{Start Time (UT)} & \textbf{$\mathbf{\delta T}$ (d)}  & \textbf{Telescope} & \textbf{Instrument} &  \textbf{Exposure (ks)} & \textbf{ObsID} & \textbf{Flux (erg cm$\mathbf{^{-2}}$ s$\mathbf{^{-1}}$)} \\
\hline
\hline
\multicolumn{7}{c}{\textbf{NuSTAR}} \\
\hline
 2025-07-03 20:46:11 & 1.32 & \textit{NuSTAR} & FPMA/B & 21.7  & 81102349002 & $(2.42\pm0.07)\times10^{-11}$\\[0.5mm]
 2025-07-07 22:56:12 & 5.41 & \textit{NuSTAR} & FPMA/B & 24.2  & 81102349004 & $(1.33\pm0.31)\times10^{-12}$ \\[0.5mm]
 2025-07-12 01:41:05 & 9.52 & \textit{NuSTAR} & FPMA/B & 38.8  & 91101324002 & $(3.7\pm1.9)\times10^{-13}$ \\[0.5mm]
\hline\hline
\multicolumn{7}{c}{\textbf{Chandra}} \\
\hline
2025-07-18 19:24:33 & 16.81 & \textit{Chandra} & ACIS-I &14.9& 31003 &$(1.2\pm0.2)\times10^{-13}$ \\[0.5mm]
2025-08-09 05:49:28	 & 37.69 & \textit{Chandra} & ACIS-S & 27.7 & 31011 &$(3.3^{+0.7}_{-0.6})\times10^{-14}$ \\[0.5mm]
2025-09-05 23:26:41	 & 65.52 & \textit{Chandra} & ACIS-S & 39.6 & 31468	&$(1.15^{+0.40}_{-0.30})\times10^{-14}$ \\[0.5mm]
\hline\hline
\end{tabular}
\end{table*}

\section{Log of Trigger Times}

Here we provide a compilation of the high-energy trigger times of GRB 250702B in Table \ref{tab:triggertimes}.

\begin{table*}[ht]
\centering
\caption{A compilation of the high-energy trigger times for this event, which we refer to as GRB 250702B. We list only the initial telescope to report the detection, but note that other high-energy monitors detected each of the associated gamma-ray burst triggers. For further discussion see \citet{Neights2025}. 
}
\label{tab:triggertimes}
\begin{tabular}{lcccc}
\hline\hline
\textbf{Telescope} & \textbf{Trigger Name} & \textbf{Time (UT)} & \textbf{MJD} & \textbf{Reference} \\
\hline
\textit{Einstein Probe} & EP250702a\tablenotemark{a} &  2025-07-01 01:40:26 & 60857.0697 & \citet{EP250702a-arxiv} \\
\textit{Fermi} & GRB 250702D\tablenotemark{b} & 2025-07-02 13:09:02  & 	60858.5479 & \citet{Elizagcn}\\
\textit{Fermi} & GRB 250702B\tablenotemark{c} & 2025-07-02 13:56:05 & 60858.5806 & \citet{Elizagcn}\\
\textit{Fermi} & GRB 250702C\tablenotemark{d} & 2025-07-02 14:49:31  & 60858.6177 & This work; \citet{Neights2025}\\
\textit{Fermi} & GRB 250702E & 2025-07-02 16:21:33 & 60858.6816  & \citet{Elizagcn}\\
\hline
\end{tabular}
\tablenotetext{a}{This is the start of the initial detection window for the July 1 detection by EP \citep{EP250702a-arxiv}}
\tablenotetext{b}{This burst was initially classified as ``Below horizon'', but later determined to be a real GRB based on a ground analysis \citep{Dburst}. This led to a deviation in the naming scheme with ``D'' occurring before ``B''.}
\tablenotetext{c}{This is the first reported trigger of the event and therefore taken as the name in this work \citep{Elizagcn}.}
\tablenotetext{d}{While the association was retracted by \textit{Fermi}, the transient has notable emission detected by \textit{Swift}/BAT at this time (\S \ref{sec:BAT}) that is robustly associated to this transient. A further discussion of its association is reported by \citet{Neights2025}.}
\end{table*}

\section{Temporal Fit Results}

\begin{figure*}
    \centering
\includegraphics[width=2.0\columnwidth]{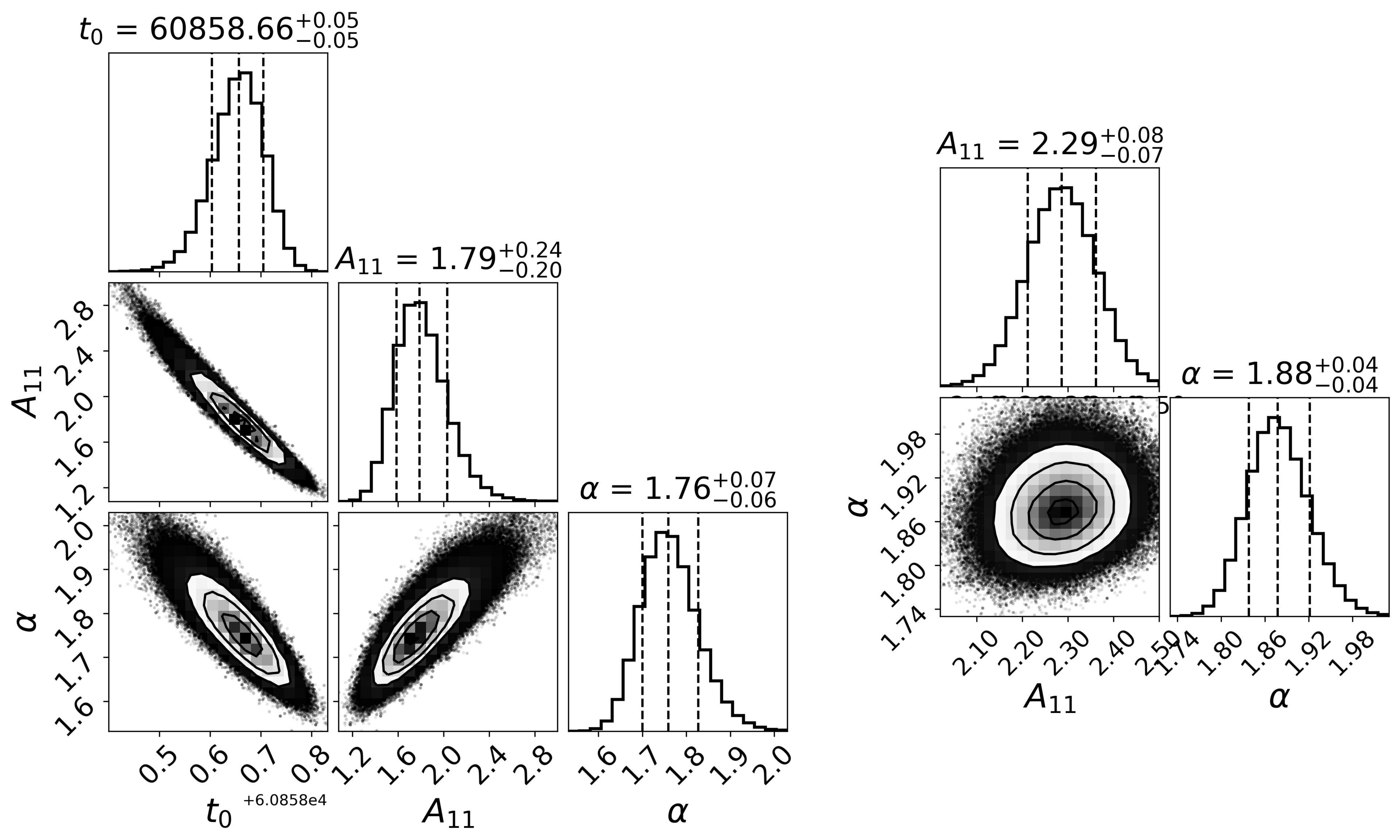}
\includegraphics[width=2.0\columnwidth]{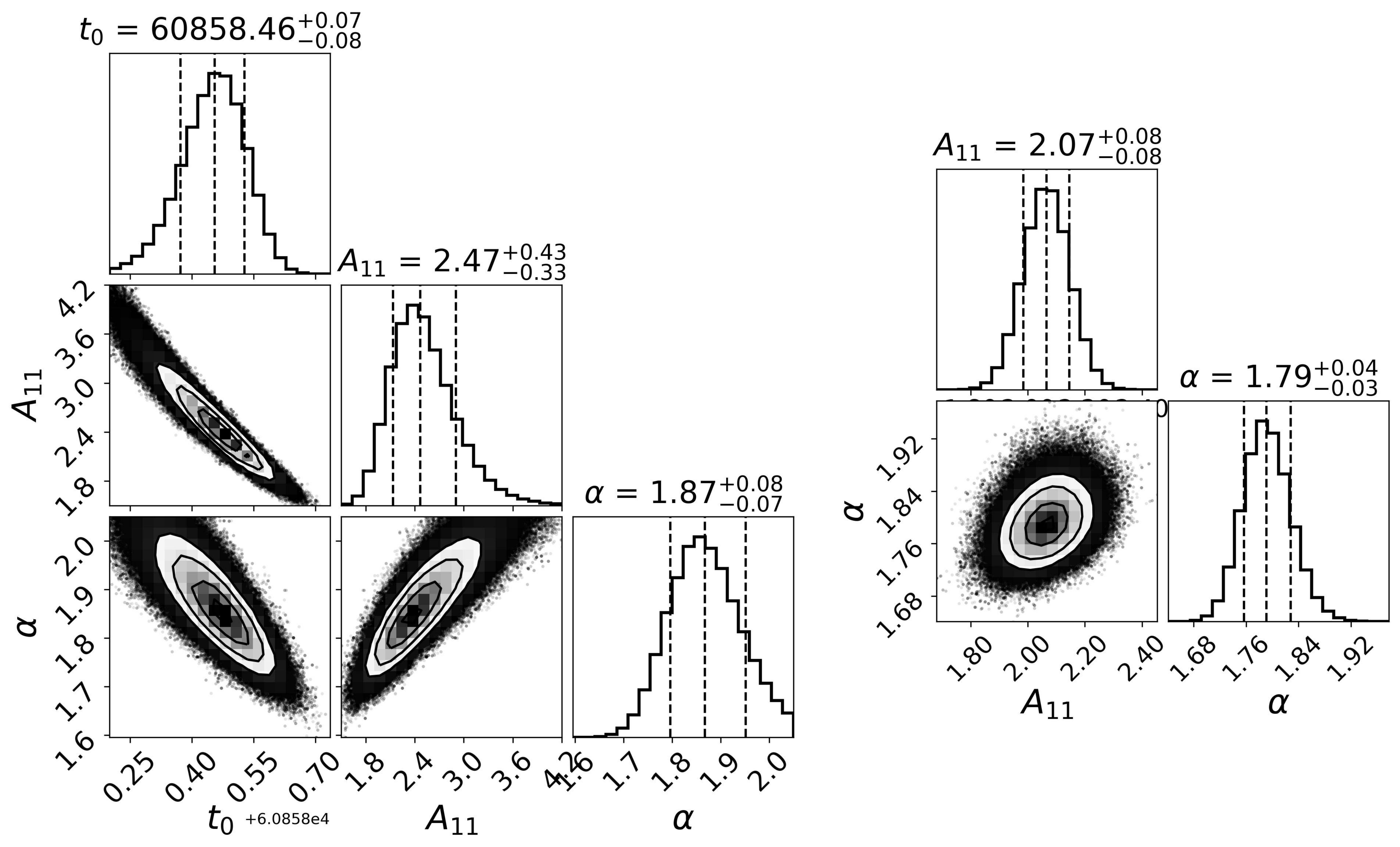}
    \caption{\textbf{Top Left:} Results of our temporal fit with the afterglow start time free. The fit includes all XRT and \textit{Chandra} data. \textbf{Top Right:} Results of our temporal fit with the start time $T_0$ fixed to GRB 250702D's trigger time. 
    \textbf{Bottom Left:} Same as top left, but only including XRT data after 0.55 days. \textbf{Bottom Right:} Same as bottom right, but only including XRT data after 0.55 days. The posterior distributions on $T_0$ are also shown in Figure \ref{fig:temporalfit}. 
    }
    \label{fig:temporalcorner}
\end{figure*}

\begin{figure*}
    \centering
\includegraphics[width=2.0\columnwidth]{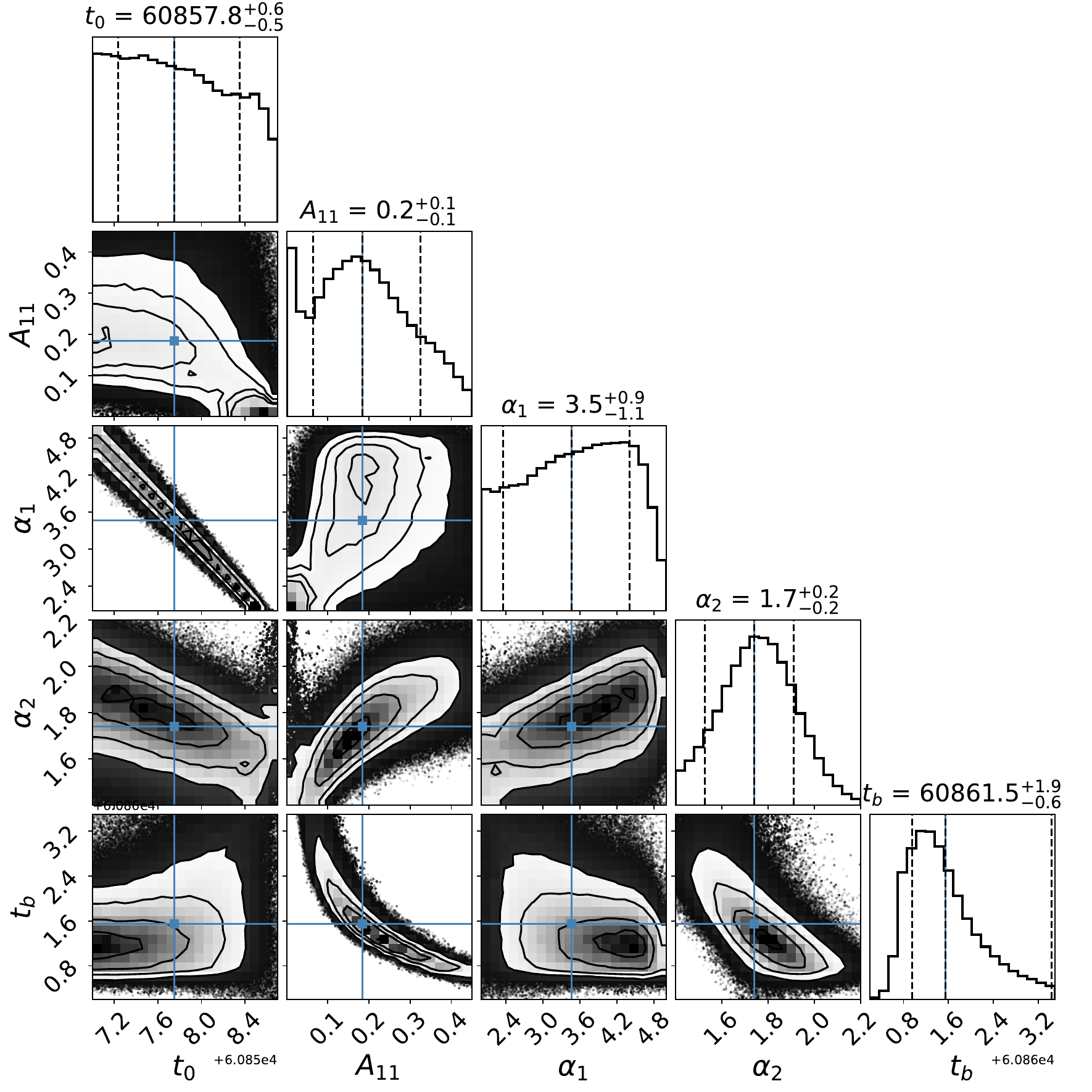}
    \caption{Results of our temporal fit with a broken powerlaw with the onset time $T_0$ as a free parameter.
    }
    \label{fig:temporalcornerbkn}
\end{figure*}

In Figure \ref{fig:temporalcorner}, we display the corner plot \citep{corner} showing the results of our powerlaw fit $F$\,$=$\,$A\, t^{-\alpha}$ to the X-ray lightcurve. The normalization $A$ is given in units of $10^{-11}$ erg cm$^{-2}$ s$^{-1}$, corresponding to the X-ray flux at 1 d from the given $T_0$. The left panel shows the results when the $T_0$ is left free, and the right panel shows the fit when $T_0$ is set to the trigger time of GRB 250702D. We performed an additional fit with a broken powerlaw with free $T_0$. We required that $T_0$ is between MJD 60857 (July 1) and MJD 60859. The results are shown in Figure \ref{fig:temporalcornerbkn}. The best-fit does not provide an improved reduced chi-squared ($\chi^2$/dof\,$=$\,$2.35$) compared to the single powerlaw fits. If we allow $T_0$ to push further backwards, i.e., before July 1, the results are similar, but the initial slope $\alpha_1$ becomes significantly steeper ($t^{-5}$), and the second slope $\alpha_2$ pushes closer to $t^{-2}$. 

\section{Comparison to Variability in Other GRB Afterglows with NuSTAR}
\label{sec:grbvarappendix}

\begin{figure*}
    \centering
\includegraphics[width=2.0\columnwidth]{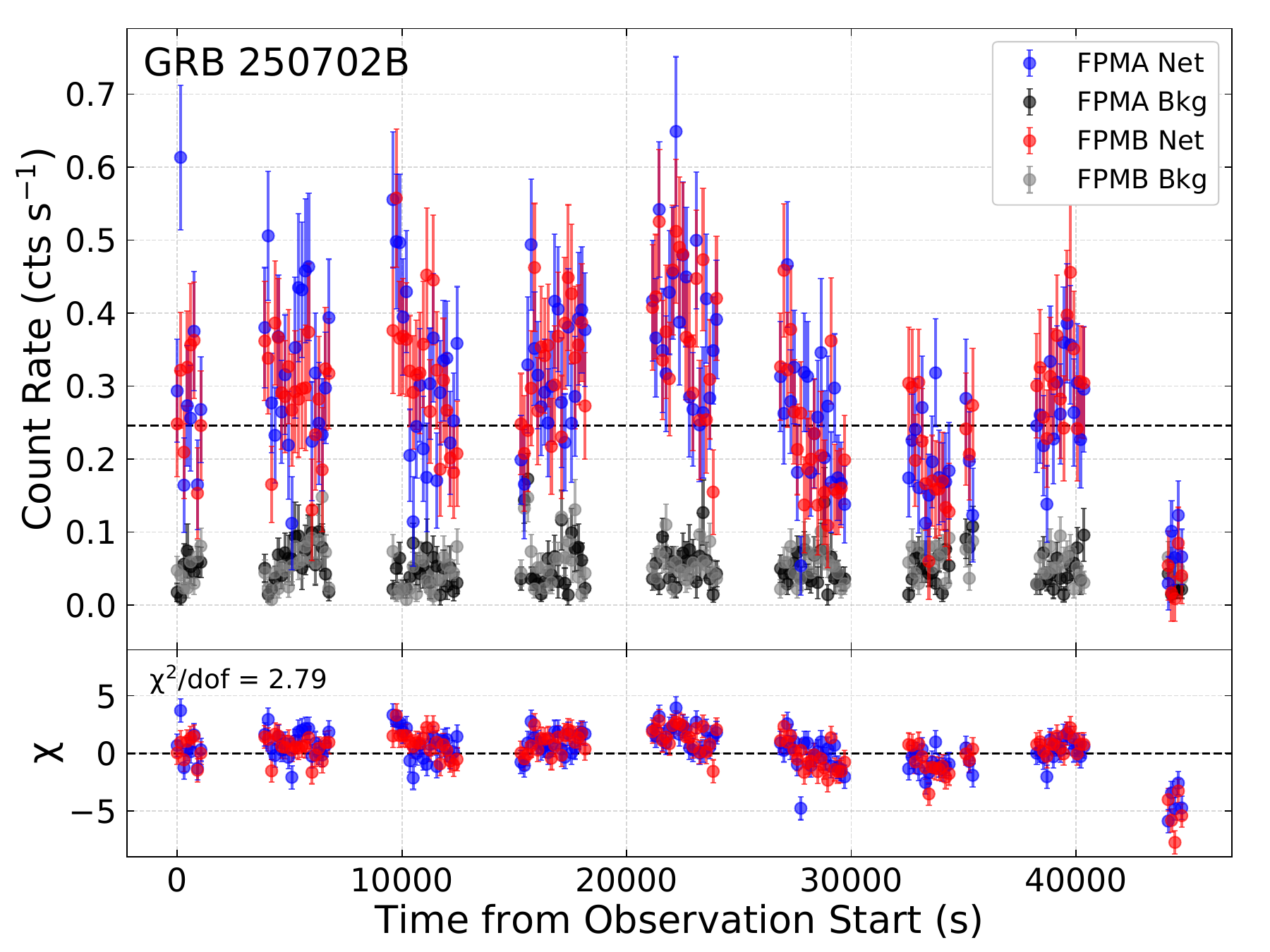}
    \caption{The background subtracted $3$\,$-$\,$79$ keV \textit{NuSTAR} lightcurve for both FPMA and FPMB individually in 150 s bins. The background lightcurve for both FPMA and FPMB is also shown for comparison. The dashed black line shows the mean (background subtracted) count rate in the observation ($\sim$\,$0.25$ cts s$^{-1}$). The bottom panel shows the residuals (for both FPMA and FPMB) with respect to the mean count rate to underscore the source's short timescale variability. 
    }
    \label{fig:lcrateapp}
\end{figure*}

\begin{figure*}
    \centering
\includegraphics[width=1.0\columnwidth]{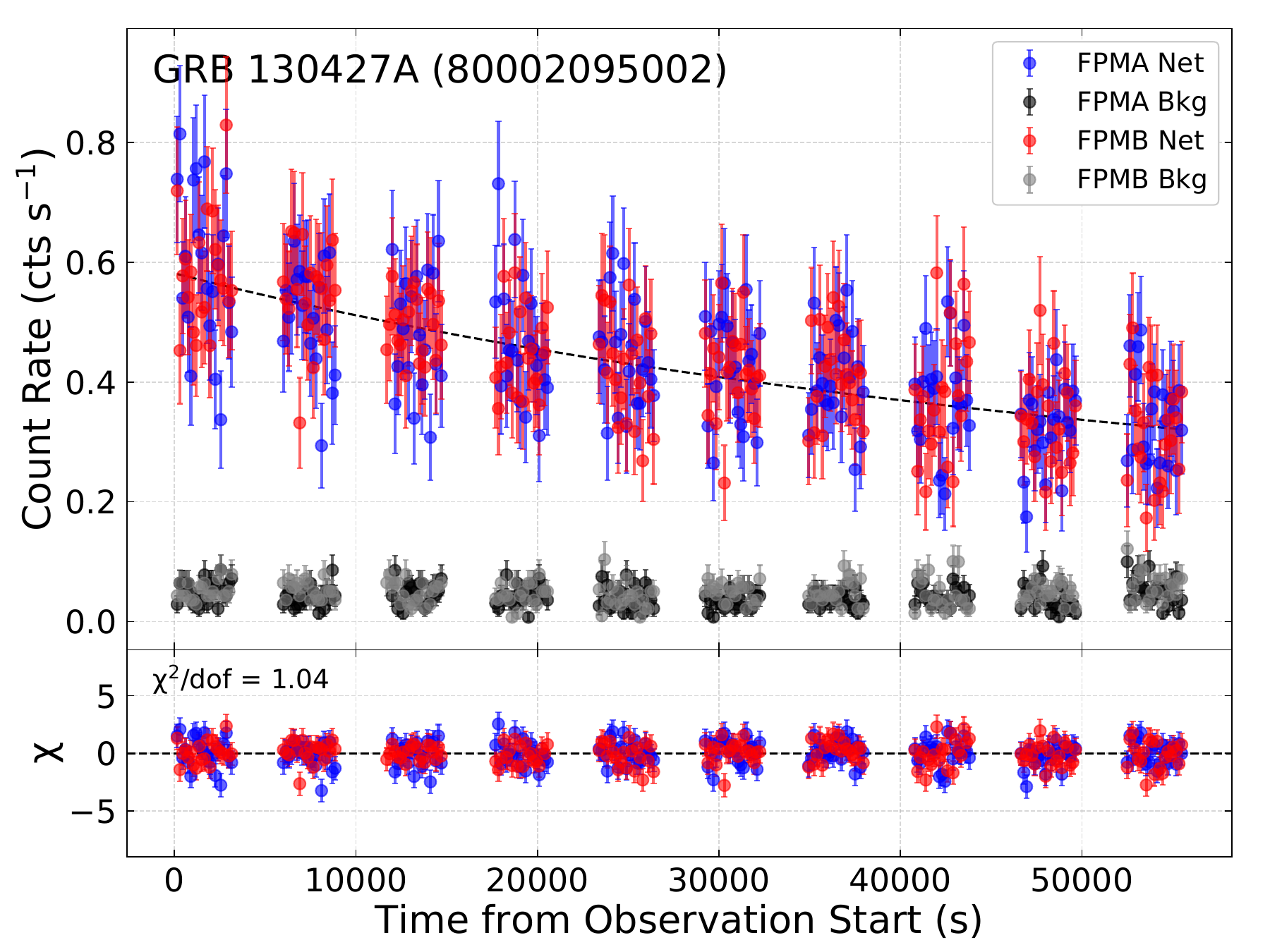}
\includegraphics[width=1.0\columnwidth]{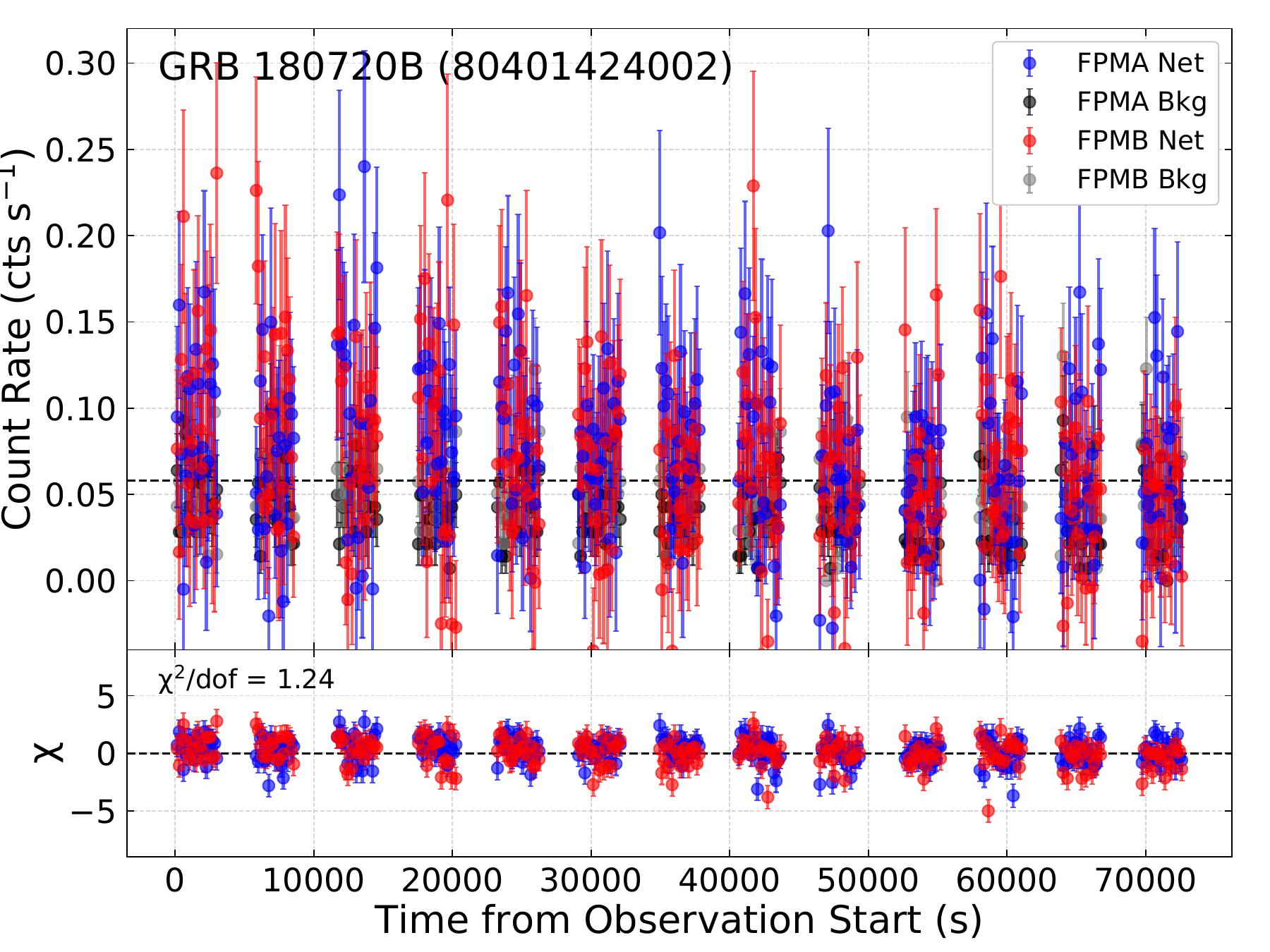}
\includegraphics[width=1.0\columnwidth]{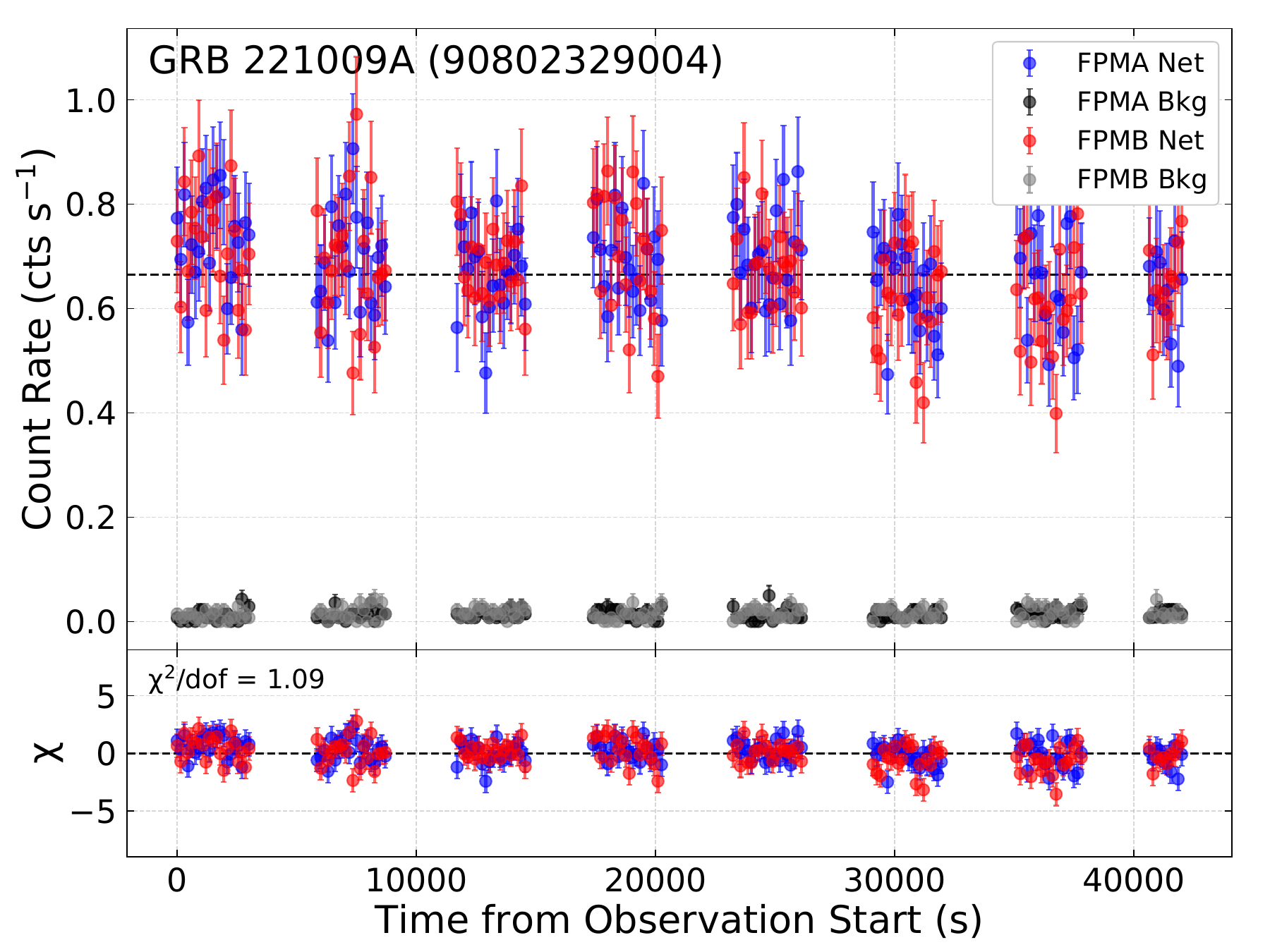}
\includegraphics[width=1.0\columnwidth]{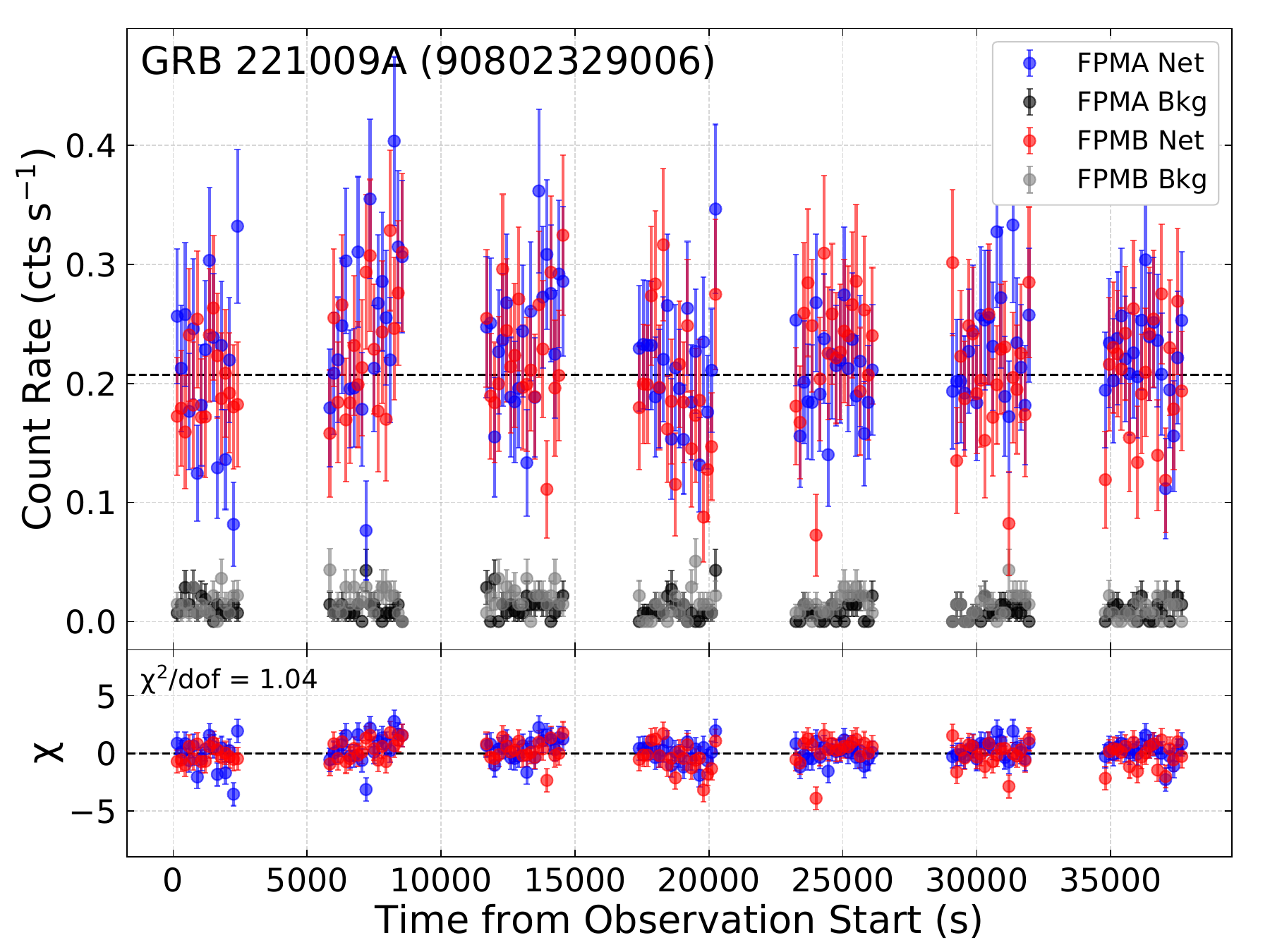}
    \caption{Same as Figure \ref{fig:lcrateapp} but for multiple \textit{NuSTAR} observations of GRB afterglows (see Appendix \ref{sec:grbvarappendix} for details). 
    }
    \label{fig:grbvar}
\end{figure*}

Since its launch in 2012, \textit{NuSTAR} has observed a handful of GRB afterglows, including GRBs 130427A, 130925A, 150201A, 170817A, 180720B, 190114C, 190829A, and 221009A. We analyzed \textit{NuSTAR} observations of these other GRBs to search for source variability ($3$\,$-$\,$79$ keV) in the afterglow phase for comparison to GRB 250702B (Figures \ref{fig:lcrate} and \ref{fig:lcrateapp}). We analyzed GRB 130427A (ObsIDs: 80002095002, 80002095004), GRB 130925A (ObsID: 80002096002, 80002096004), GRB 180720B (ObsID: 80401424002), GRB 190114C (ObsID: 90501602002), and GRB 221009A (ObsID: 90802329002, 90802329004, 90802329006). We find that other \textit{NuSTAR} observations of GRBs have lower count rates or resulted in non-detections (e.g., GRB 170817A). The data were reduced using the same procedures outlined in \S \ref{sec:nustar} with strict cuts on the particle background. In some cases (e.g., GRB 130925A) the source is observed at a lower count rate where the limits on source variability are less meaningful (e.g., GRB 180720B in Figure \ref{fig:grbvar}). In Figure \ref{fig:grbvar}, we present a few representative lightcurves for comparison. To assess variability, we computed the reduced chi-squared with regard to the weighted mean count rate per FPM. We find that these GRBs do not show evidence for variability in the $3$\,$-$\,$79$ keV energy range. For example, GRB 221009A shows much less variability at a similar count rate (e.g., ObsID 90802329004 and 90802329006 occurring at 6 and 11 days; \citealt{OConnor2023,Laskar2023}). 

\section{Hardness Ratio Evolution Analysis}
\label{sec:HRev}

\begin{figure*}
    \centering
\includegraphics[width=2.0\columnwidth]{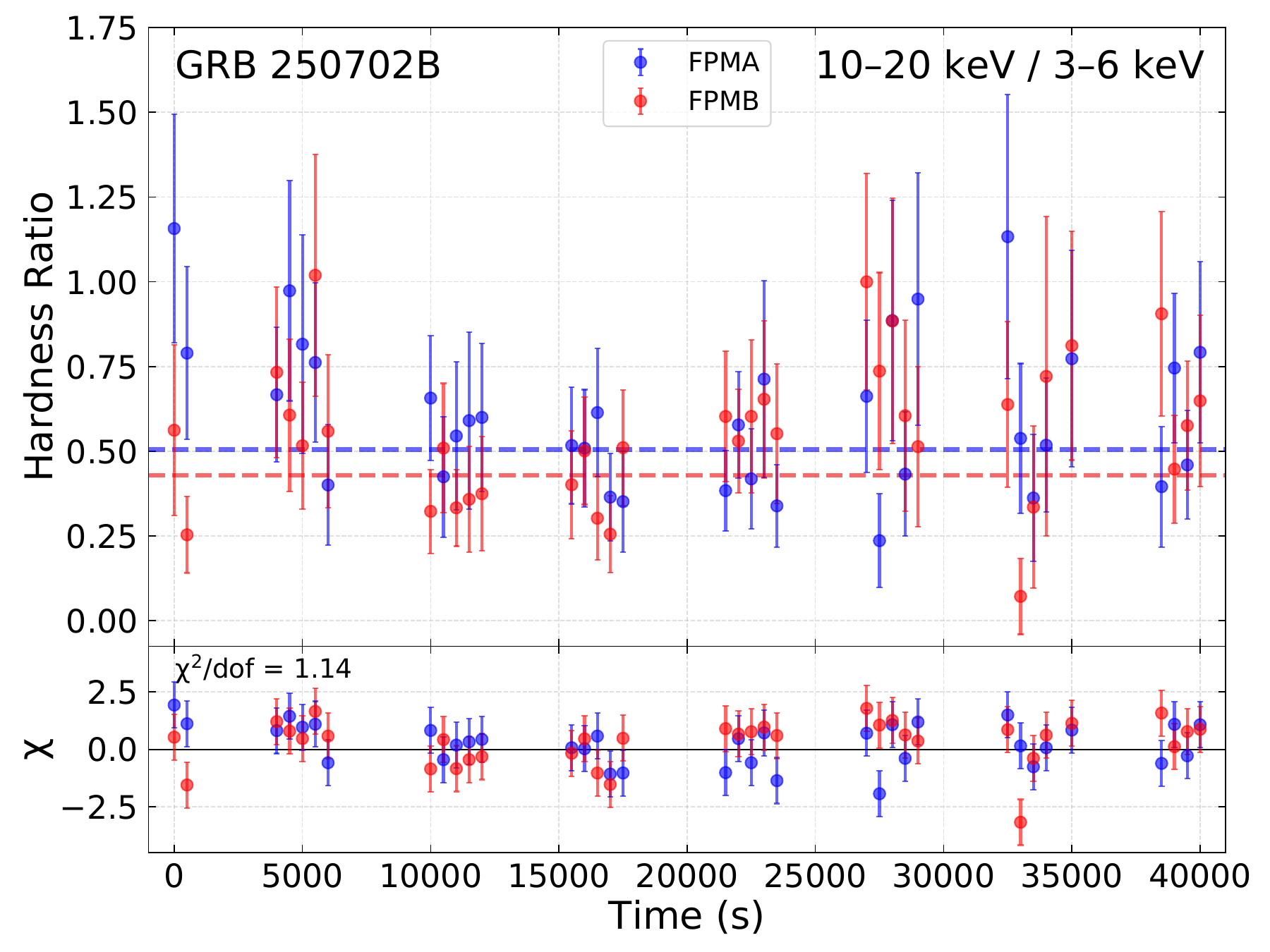}
    \caption{Hardness ratio (background subtracted) computed between $10$\,$-$\,$20$ and  $3$\,$-$\,$6$ keV in 500 s bins for both FPMA and FPMB. The weighted mean hardness ratio for both FPM are shown as horizontal lines. The bottom panel shows the residuals from the weighted mean hardness ratio.  
    }
    \label{fig:HR}
\end{figure*}

\begin{figure*}
    \centering
\includegraphics[width=2.0\columnwidth]{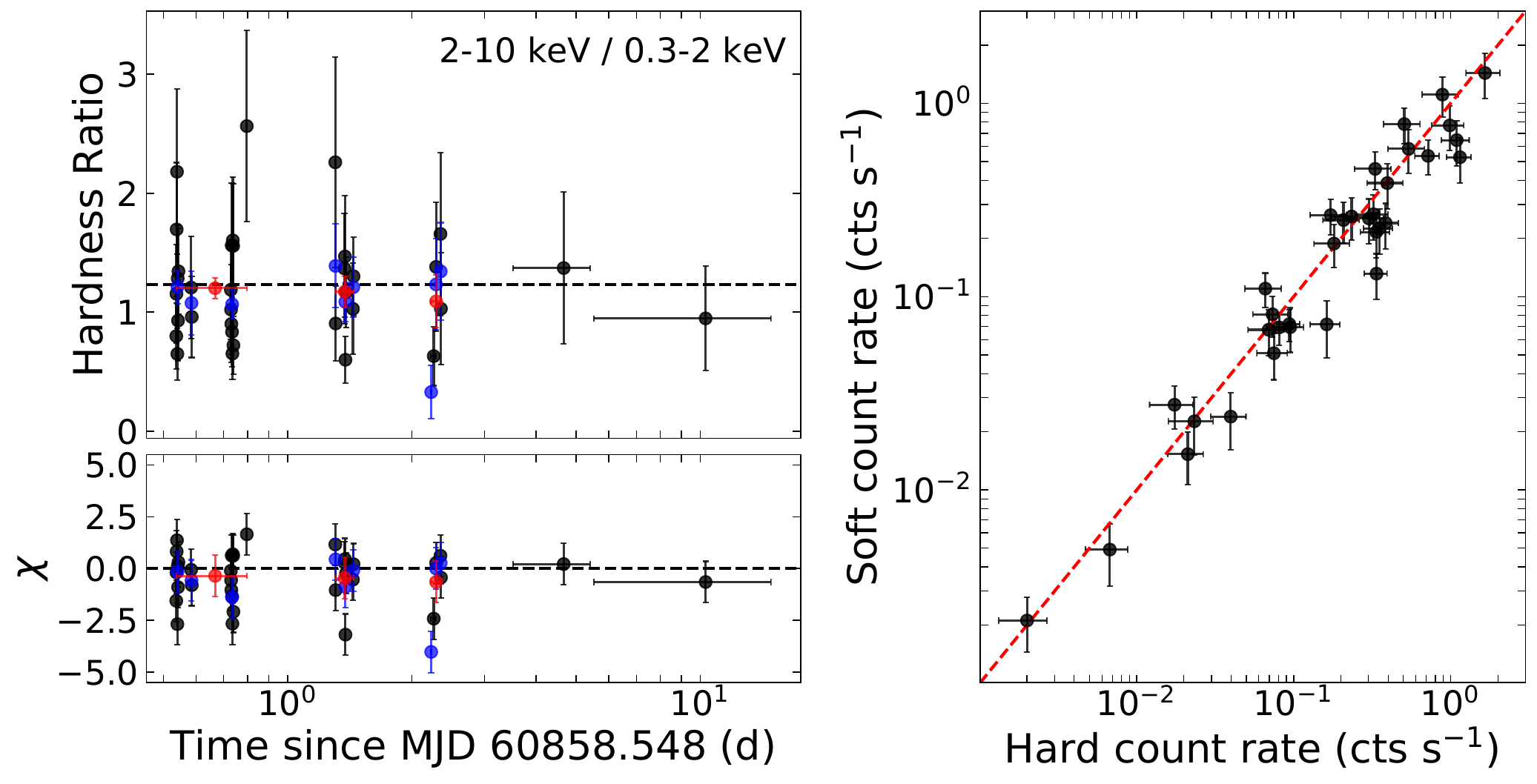}
    \caption{\textbf{Left:} \textit{Swift}/XRT hardness ratio computed between $2$\,$-$\,$10$ and $0.3$\,$-$\,$2$ keV. The mean hardness ratio is shown as a horizontal line. The bottom panel shows the residuals from the mean hardness ratio. We also show the HR on a per observation (red) and per snapshot (blue) basis for the first few days before mutliple observations must be combined to achieve a detection. \textbf{Right:} Hard ($2$\,$-$\,$10$ keV) versus soft ($0.3$\,$-$\,$2$ keV) count rates for the \textit{Swift}/XRT lightcurve. The dashed line shows the expectation for a one-to-one hardness ratio ratio. In both panels the data is binned to a minimum of 20 counts per bin.
    }
    \label{fig:HRxrt}
\end{figure*}

In Figure \ref{fig:HR} we show that the hardness ratio (HR), defined as the ratio of the counts in the hard band to soft band, is constant over the course of our first \textit{NuSTAR} observation. This is likely due to the low count rate in our observation as a spectral evolution would be expected given the source's lightcurve variability. We computed the hardness ratio between $10$\,$-$\,$20$ and $3$\,$-$\,$6$ keV in 500 s bins for both FPMA and FPMB. 
We also investigated the HR of the \textit{Swift}/XRT observations in the $0.3$\,$-$\,$2$ and $2$\,$-$\,$10$ keV energy bands. This is shown in Figure \ref{fig:HRxrt}, though see also Figure \ref{fig:XRTflaring}. The HR is does not show any significant long term deviations, though short term deviations were identified especially in the first orbit of data shown in Figure \ref{fig:XRTflaring}. 

\section{Afterglow Modeling Results}

\begin{figure*}
    \centering
\includegraphics[width=2.0\columnwidth]{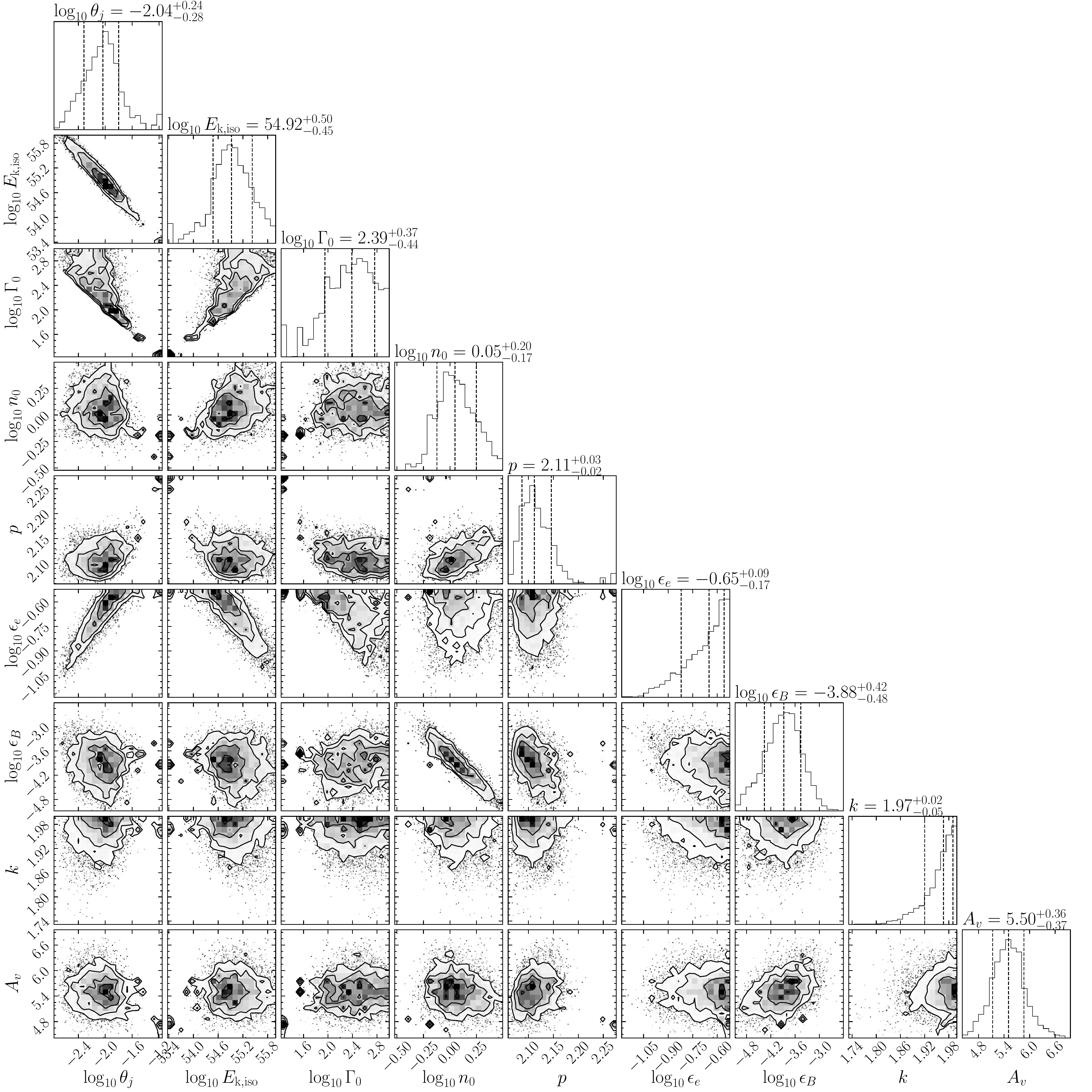}
    \caption{Corner plot showing the posterior parameter distributions of our forward shock afterglow modeling, as detailed in \S \ref{sec:afterglow}. We have used the GBM ``E'' burst as $T_0$. The condition $\Gamma_0\theta_j\geq1$ is enforced to prevent the jet from spreading laterally in the coasting phase.
    }
    \label{fig:afterglowcorner}
\end{figure*}

\begin{figure*}
    \centering
\includegraphics[width=2.0\columnwidth]{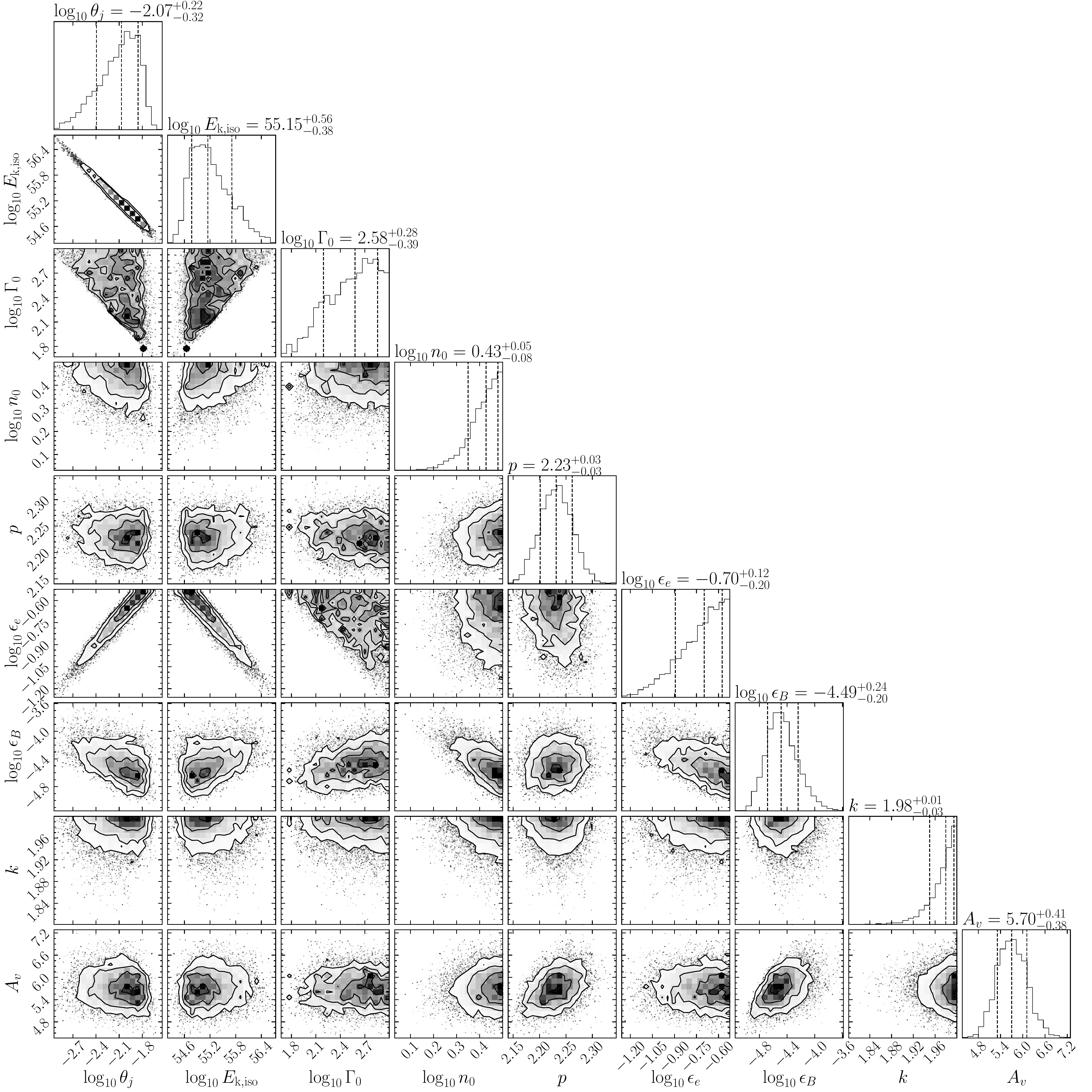}
    \caption{Same as Fig.\,\ref{fig:afterglowcorner}, but now using GBM ``D'' burst as $T_0$. 
    }
    \label{fig:afterglowcorner-D}
\end{figure*}

\begin{figure*}
    \centering
    \includegraphics[width=2.0\columnwidth]{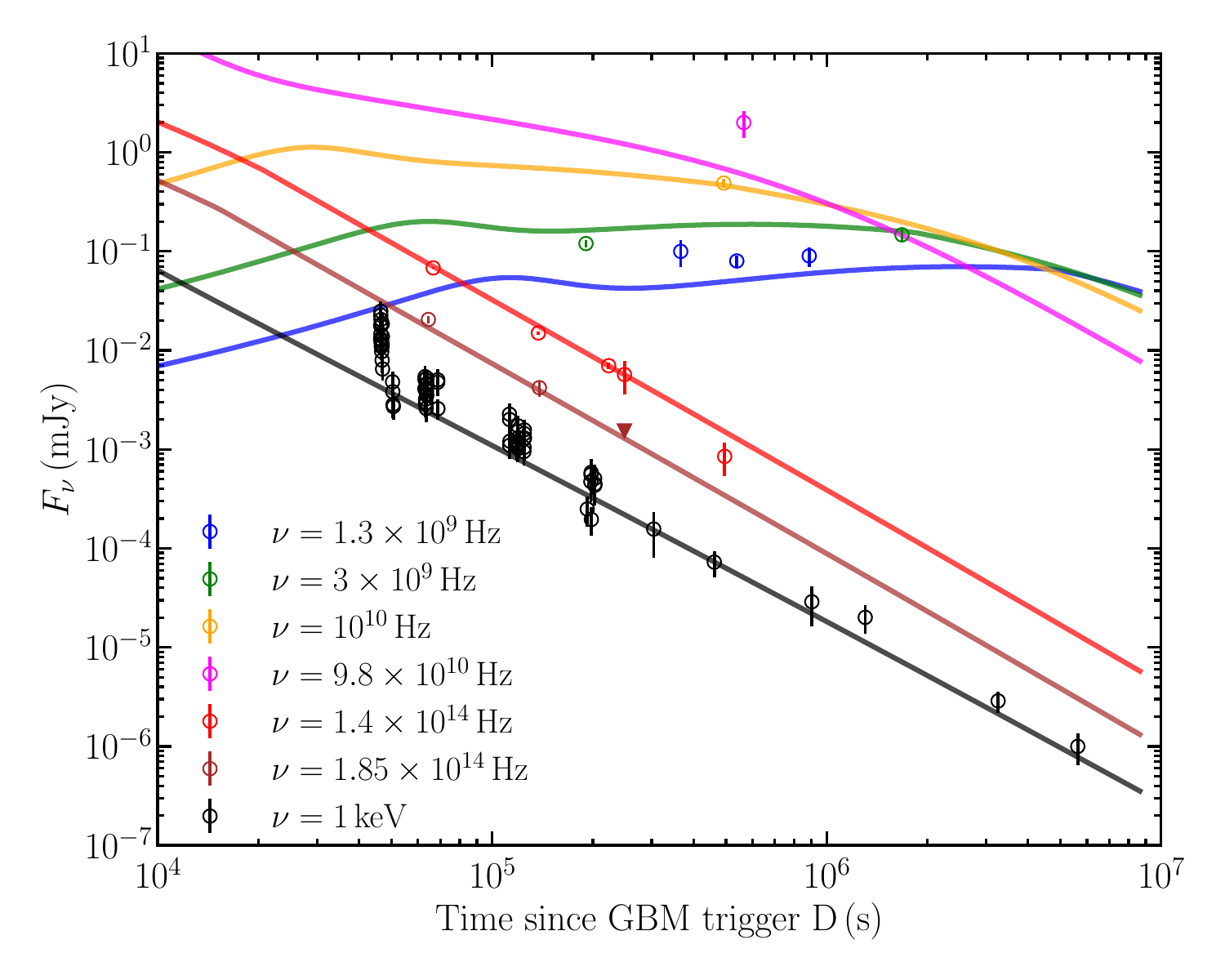}
    \caption{Same as Figure \ref{fig:afterglow} but now using GRB\,250702D as the start time $T_0$. 
    The FS afterglow fit posterior distributions are shown in Fig.\,\ref{fig:afterglowcorner-D}, 
    and the RS emission is obtained now with a different value of $\varepsilon_{B,\rm RS}=2\times10^{-4}$. The description of the data is broadly consistent with the fit in Figure \ref{fig:afterglow}. The lightcurves are obtained for the following model parameters that keep the X-ray afterglow emission below the flaring X-ray emission at early times: $\theta_j=4\times10^{-3}$\,rad, $E_{\rm k,iso}=4\times10^{55}$\,erg, $\Gamma_0=282$, $n_0=1.9$ at $R_0=10^{18}$\,cm, $p=2.24$, $\epsilon_e=0.13$, $\epsilon_B=10^{-4}$, $k=2$.
    }
    \label{fig:afterglow-D}
\end{figure*}

In Figure \ref{fig:afterglowcorner}, we display the corner plot showing the posterior probability distributions for each model parameter based on our broadband afterglow modeling in \S \ref{sec:afterglow} which uses the ``E'' burst as $T_0$. Additional fits using the ``D'' burst as $T_0$ are presented in Figures \ref{fig:afterglowcorner-D} and \ref{fig:afterglow-D}.

\clearpage      

\startlongtable
\begin{deluxetable*}{lccc}
\tablecaption{Log of \textit{Swift}/BAT observations of GRB 250702B between $\delta T$\,$\pm4$ d from GRB 250702D. The reported MJD is the mid-time of the exposure. The fluxes are reported in $14$\,--\,$195$ keV and correspond to upper limits at the $5\sigma$ level. We identify two detections at MJD 60858.6816 and 60858.8170. }
\label{tab:batlimits}
\tablehead{
\colhead{\textbf{MJD}} & \colhead{\textbf{$\mathbf{\delta T}$ (d)}} & \colhead{\textbf{Exposure (s)}} & \colhead{\textbf{Flux (erg cm$\mathbf{^{-2}}$ s$\mathbf{^{-1}}$)}}
}
\startdata
60855.1187 & -3.43 & 489 & $<2.02\times 10^{-8}$ \\
60856.1582 & -2.39 & 494 & $<3.92\times 10^{-8}$ \\
60856.2917 & -2.26 & 778 & $<2.11\times 10^{-8}$ \\
60856.4177 & -2.13 & 996 & $<2.19\times 10^{-8}$ \\
60856.3377 & -2.21 & 764 & $<6.27\times 10^{-9}$ \\
60856.4813 & -2.07 & 116 & $<1.84\times 10^{-8}$ \\
60857.0053 & -1.54 & 981 & $<6.78\times 10^{-9}$ \\
60857.1878 & -1.36 & 334 & $<1.27\times 10^{-8}$ \\
60857.2032 & -1.34 & 508 & $<8.85\times 10^{-9}$ \\
60857.2654 & -1.28 & 891 & $<6.97\times 10^{-9}$ \\
60857.3869 & -1.16 & 1618 & $<5.43\times 10^{-9}$ \\
60857.5171 & -1.03 & 1585 & $<5.42\times 10^{-9}$ \\
60857.5854 & -0.96 & 126 & $<5.77\times 10^{-8}$ \\
60857.5886 & -0.96 & 450 & $<9.48\times 10^{-9}$ \\
60857.6599 & -0.89 & 468 & $<1.32\times 10^{-8}$ \\
60857.7022 & -0.85 & 350 & $<1.28\times 10^{-8}$ \\
60857.7198 & -0.83 & 694 & $<7.61\times 10^{-9}$ \\
60857.7710 & -0.78 & 891 & $<5.43\times 10^{-9}$ \\
60857.7821 & -0.77 & 531 & $<1.03\times 10^{-8}$ \\
60857.9162 & -0.63 & 883 & $<1.06\times 10^{-8}$ \\
60858.2306 & -0.32 & 361 & $<4.22\times 10^{-8}$ \\
60858.3658 & -0.18 & 202 & $<4.59\times 10^{-8}$ \\
60858.3689 & -0.18 & 409 & $<7.39\times 10^{-9}$ \\
60858.6245 & 0.08 & 1065 & $<1.65\times 10^{-8}$ \\
60858.6816 & 0.13 & 856 & $(4.95\pm1.14)\times 10^{-9}$ \\
60858.7446 & 0.20 & 877 & $<3.50\times 10^{-9}$ \\
60858.7557 & 0.21 & 411 & $<1.20\times 10^{-8}$ \\
60858.8170 & 0.27 & 1058 & $(1.08\pm0.39)\times 10^{-8}$ \\
60859.0137 & 0.47 & 681 & $<1.01\times 10^{-8}$ \\
60859.0851 & 0.54 & 702 & $<4.15\times 10^{-9}$ \\
60859.1309 & 0.58 & 375 & $<5.30\times 10^{-9}$ \\
60859.2121 & 0.66 & 519 & $<7.48\times 10^{-9}$ \\
60859.2761 & 0.73 & 1001 & $<3.54\times 10^{-9}$ \\
60859.3432 & 0.80 & 239 & $<6.99\times 10^{-9}$ \\
60859.3471 & 0.80 & 325 & $<6.04\times 10^{-9}$ \\
60859.7278 & 1.18 & 660 & $<9.66\times 10^{-9}$ \\
60859.8519 & 1.30 & 442 & $<4.61\times 10^{-9}$ \\
60859.8610 & 1.31 & 256 & $<1.00\times 10^{-8}$ \\
60859.9193 & 1.37 & 1496 & $<2.99\times 10^{-9}$ \\
60859.9862 & 1.44 & 769 & $<4.05\times 10^{-9}$ \\
60860.1206 & 1.57 & 443 & $<2.70\times 10^{-8}$ \\
60860.3682 & 1.82 & 1725 & $<4.35\times 10^{-9}$ \\
60860.4331 & 1.89 & 266 & $<1.02\times 10^{-8}$ \\
60860.6995 & 2.15 & 827 & $<6.01\times 10^{-9}$ \\
60860.7724 & 2.22 & 558 & $<4.39\times 10^{-9}$ \\
60860.8303 & 2.28 & 1278 & $<3.12\times 10^{-9}$ \\
60860.8910 & 2.34 & 1088 & $<3.42\times 10^{-9}$ \\
60860.9052 & 2.36 & 872 & $<1.86\times 10^{-8}$ \\
60861.0810 & 2.53 & 297 & $<7.83\times 10^{-7}$ \\
60861.0875 & 2.54 & 364 & $<7.71\times 10^{-9}$ \\
60861.2151 & 2.67 & 900 & $<1.48\times 10^{-8}$ \\
60861.4190 & 2.87 & 1200 & $<4.57\times 10^{-9}$ \\
60861.4730 & 2.93 & 256 & $<9.71\times 10^{-9}$ \\
60861.7992 & 3.25 & 976 & $<5.52\times 10^{-9}$ \\
60861.8680 & 3.32 & 368 & $<7.61\times 10^{-9}$ \\
60862.0650 & 3.52 & 571 & $<4.46\times 10^{-9}$ \\
60862.1294 & 3.58 & 641 & $<6.58\times 10^{-9}$ \\
60862.2567 & 3.71 & 863 & $<6.07\times 10^{-9}$ \\
60862.3354 & 3.79 & 525 & $<6.14\times 10^{-9}$ \\
60862.3954 & 3.85 & 900 & $<5.14\times 10^{-9}$ \\
60862.4452 & 3.90 & 358 & $<8.32\times 10^{-9}$ \\
60862.5095 & 3.96 & 1169 & $<4.90\times 10^{-9}$ \\
60862.5240 & 3.98 & 300 & $<6.38\times 10^{-9}$ \\
60862.5746 & 4.03 & 812 & $<5.78\times 10^{-9}$ \\
\enddata
\end{deluxetable*}

\startlongtable
\begin{deluxetable*}{lccc}
\tablecaption{Peak flux of bursts detected in \textit{Swift}/BAT \texttt{GUANO} data. The fluxes are reported in $14$\,--\,$195$ keV. The peak fluxes are for the peak 16.384 s intervals (largest interval tested by \texttt{NITRATES}) within the available \texttt{GUANO} data. }
\label{tab:guanofluxes}
\tablehead{
\colhead{\textbf{Trigger Name}} & 
\colhead{\textbf{MJD}} & 
\colhead{\textbf{Flux (erg cm$\mathbf{^{-2}}$ s$\mathbf{^{-1}}$)}}
}
\startdata
GRB 250702D & 60858.54789 & $(5.08 \pm 0.62)\times 10^{-8}$ \\
GRB 250702C & 60858.61768 & $(3.43 \pm 1.30)\times 10^{-8}$ \\
GRB 250702E & 60858.68161 & $(6.53 \pm 1.17)\times 10^{-8}$ \\
\enddata
\end{deluxetable*}

\bibliography{bib}{}
\bibliographystyle{aasjournal}



\end{document}